\documentclass[twocolumn,traditabstract]{aa}
\usepackage{txfonts}
\usepackage{color}
\usepackage{graphicx}
\usepackage{natbib}
\usepackage{multirow}
\usepackage[colorlinks=true, allcolors=blue]{hyperref}
\newcommand{\PACS}{{\it PACS }}
\newcommand{\SPIRE}{{\it SPIRE }}
\newcommand{\Herschel}{{\it Herschel }}

\begin{document}

   \title{On the underestimation of dust mass in protoplanetary disks: Effects of disk structure and dust properties \thanks{{\it Herschel} is an ESA space observatory with science instruments provided by European-led Principal Investigator consortia and with important participation from NASA.}}
   \author{Yao Liu\inst{1}
          \and
          Hendrik Linz\inst{2}
          \and
		  Min Fang\inst{1}
		  \and
          Thomas Henning\inst{2}
		  \and
          Sebastian Wolf\inst{3}
		  \and
		  Mario Flock\inst{2}
		  \and
		  Giovanni P. Rosotti\inst{4}
		  \and
		  Hongchi Wang\inst{1}
		  \and 
		  Dafa Li\inst{1,5}
          }

   \institute{Purple Mountain Observatory \& Key Laboratory for Radio Astronomy, Chinese Academy of Sciences, 10 Yuanhua Road, Nanjing 210023, China; \\
         \email{yliu@pmo.ac.cn}
         \and
         Max-Planck-Institut f\"ur Astronomie, K\"onigstuhl 17, D-69117 Heidelberg, Germany
		 \and
		 Institut f\"ur Theoretische Physik und Astrophysik, Christian-Albrechts-Universit\"at zu Kiel, Leibnizstr. 15, 24118 Kiel, Germany
		 \and
		 DAMTP, University of Cambridge, CMS, Wilberforce Road, Cambridge CB3 0WA, UK
		 \and
		 School of Astronomy and Space Science, University of Science and Technology of China, 96 Jinzhai Road, Hefei 230026, China}
         \authorrunning{Liu et al.}
         \titlerunning{On the underestimation of protoplanetary disk dust masses}

   \abstract{The total amount of dust grains in protoplanetary disks is one of the key properties that characterize the potential for planet formation. 
   With (sub-)millimeter flux measurements, literature studies usually derive the dust mass using an analytic form under the assumption of optically 
   thin emission, which may lead to substantial underestimation. In this work, we conduct a parameter study with the goal of investigating 
   the effects of disk structure and dust properties on the underestimation through self-consistent radiative transfer models. Different 
   dust models, scattering modes and approaches for dust settling are considered and compared. The influences of disk substructures, such as 
   rings and crescents, on the mass derivation are investigated as well. The results indicate that the traditional analytic method can underestimate 
   the mass by a factor of a few to hundreds, depending on the optical depth along the line of sight set mainly by the true dust mass, disk size and inclination. 
   As an application, we perform a detailed radiative transfer modeling of the spectral energy distribution of DoAr\,33, one of the 
   observed DSHARP disks. When the DSHARP dust opacities are adopted, the most probable dust mass returned from the Bayesian analysis 
   is roughly 7 times higher than the value given by the analytic calculation. Our study demonstrates that estimating disk dust masses from radiative transfer modeling is one solution for alleviating the problem of insufficient mass for planet formation raised in the ALMA era.}

   \keywords{circumstellar matter -- planetary systems: protoplanetary disks -- radiative transfer -- facilities: Herschel}

   \maketitle

\section{Introduction}
Dust and gas material in protoplanetary disks are the building blocks of planetary systems. Within the time scale for disk dispersal, typically 
a few Myrs \citep{Williams2011}, micron-sized dust grains coagulate and grow up by more than ten orders of magnitude until the assembly of large 
entities, such as pebbles and planetesimals \citep{Testi2014,Birnstiel2016,Drazkowska2022}. The gas-phase component either dissipates through a combination of 
complex mechanisms or can be captured by the growing giant planets \citep{Balog2008,Machida2010,Ercolano2017,Picogna2019,Pascucci2022}. The total 
amount of material available in the disk plays a crucial role in determining whether planets can eventually form and what their nature is \citep[e.g.,][]{Mordasini2012}. 
The determination of protoplanetary disk masses can be accomplished through gas line or (sub-)millimeter continuum observations, though limitations 
exist for both diagnoses \citep{Bergin2017,Miotello2022}. 

Determining the disk gas mass is a difficult task, mainly because the most dominant component, ${\rm H_{2}}$, has no 
dipole moment, and it does not emit from the main disk mass reservoir. Carbon monoxide (CO) is an alternative tracer for gas masses. 
However, in addition to the requirement for a CO-${\rm H_{2}}$ abundance ratio appropriate for protoplanetary disk environments, mass determinations 
from CO (and its isotopologues) lines need to be corrected for the effects of CO freeze-out in the cold midplane, which may be more effective than 
previously thought \citep{Powell2022}, and CO photodissociation in the hot surface layers as well as the isotope-selective 
processes \citep{Williams2014,Miotello2014,Miotello2016}. Hydrogen deuteride is believed to be a better probe, but to date its ${\rm J\,{=}\,1–0}$ 
rotational transition line has been detected for only three disks with the Herschel/PACS spectrometer \citep{Bergin2013,McClure2016}. 

Measuring dust continuum emission is less expensive compared to gas line observations. With the advent of the Atacama Large Millimeter/submillimeter 
Array (ALMA), several nearby star formation regions have been surveyed, resulting in a large sample of disks (of the order of 1,000) with flux 
measurements at (sub-)millimeter wavelengths \citep[e.g.,][]{Ansdell2016,Barenfeld2016,Pascucci2016,Cazzoletti2019,Grant2021}. Under optically thin 
conditions, (sub-)millimeter flux densities $F_{\nu}$ can be converted into dust masses via the analytic equation  
\begin{equation}
M_{\rm dust.ana} = \frac{F_{\nu}D^2}{\kappa_{\nu}B_{\nu}(T_{\rm dust})},
\label{eqn:mana}
\end{equation}
where $B_{\nu}(T_{\rm dust})$ refers to the Planck function given at the observed frequency $\nu$ and dust temperature $T_{\rm dust}$, and the distance 
to the object is denoted as $D$. At a reference frequency of $\nu\,{=}\,230\,\rm{GHz}$ (or wavelength $\lambda\,{\sim}\,1.3\,\rm{mm}$), a value 
of $2.3\,{\rm cm^2/g}$ has been frequently adopted for the dust absorption opacity $\kappa_{\nu}$ \citep[e.g.,][]{Beckwith1990,Andrews2013}. For the dust 
temperature, 20\,K is a common choice. However, \citet{Andrews2013} suggested a stellar-luminosity ($L_{\star}$) dependent dust 
temperature $T_{\rm dust}\,{=}\,25\,(L_{\star}/L_{\odot})^{0.25}\,\rm{K}$. The scaling of $T_{\rm dust}$ with $L_{\star}$ was further investigated for 
disks around low-mass stars and brown dwarfs \citep[e.g.,][]{Daemgen2016,vanderPlas2016,Hendler2017}, and generally the relation is found to be 
flatter than the prescription by \citet{Andrews2013}. 

A comparison between the statistic distribution of the analytic dust mass $M_{\rm dust.ana}$ and the minimum-mass solar nebula shows that only few disks have 
enough material to produce our solar system or its counterparts in the exoplanet population \citep[e.g.,][]{Najita2014,Manara2018,Andrews2020,Mulders2021}. 
Explanations for such a discrepancy are generally provided by three different scenarios. The first scenario is that planet formation might be more efficient 
than what has been postulated, namely that in the protoplanetary disk evolutionary stage dust grains have already grown up to pebbles, planetesimals or even planetary 
embryos that are insensitive to (sub-)millimeter observations \citep[e.g.,][]{Tychoniec2020}. Supporting evidence is the prevalence of disk substructures, such as rings and 
gaps \citep[e.g.,][]{alma2015,Long2018,Andrews2018}, even in protostellar disks with relatively young ages \citep{SeguraCox2020,Sheehan2020}. 
These fine-scale features are likely created by (proto-)planets that are still embedded in their natal disk \citep{Kley2012}. 
In the second scenario, the disk acts as a conveyor belt that transports material from the environment to the central star \citep{Manara2018}. 
Therefore, the disk is replenished with material from the surrounding remnant envelope, part of which can contribute to the formation of planets.
The third proposal concerns the reliability of mass estimation from Eq.~\ref{eqn:mana} that is based on an optically thin assumption. Detailed radiative 
transfer analysis of millimeter data have demonstrated that protoplanetary disks, particularly in the inner regions, are not necessarily optically thin at millimeter 
wavelengths \citep[e.g.,][]{Wolf2008,Pinte2016,Liuy2019,Ueda2020}. The measured fluxes only reflect dust material residing above the $\tau\,{=}\,1$ surface. 
Consequently, the calculated $M_{\rm dust.ana}$ is merely a lower limit. Indeed, such an underestimation has been found in previous studies, but mostly on limited 
number of objects \citep[e.g.,][]{Mauco2018,Ballering2019,CarrascoGonzalez2019,Ribas2020,Villenave2020,Macias2021,Guidi2022}. Given the fact that the optical depth 
along the line of sight increases with the viewing angle, the analytic dust mass is expected to further deviate from the true dust mass of disks with high inclinations.

Moreover, deriving dust masses using Eq.~\ref{eqn:mana} does not take into account the dust density distribution in the disk. The accumulating ALMA data show that in 
some cases, dust grains are quite concentrated within substructures like rings and crescents. Examples are the disks around Oph IRS 48 \citep{vanderMarel2013}, 
HD143006 \citep{Perez2018}, HD\,135344B \citep{Cazzoletti2018}, and MWC\,758 \citep{Dong2018}. Whether, and to which level, these configurations affect the 
mass estimation are not clear. In addition to a direct impact of dust opacities (therefore dust emissivities at millimeter wavelengths) on the mass determination, the 
choice for the dust model will also influence the resulting dust temperature, which in turn affects the mass determination. Such parameter coupling is not considered 
in Eq.~\ref{eqn:mana}. In this work, using self-consistent radiative transfer models, we performed a detailed parameter study with the goal of systematically evaluating 
the extent of dust mass underestimation and its dependence on the disk structure and dust properties. The setup of the fiducial model and the method used to quantify 
the mass underestimation are described in Sect.~\ref{sec:fidmodel}. In Sect.~\ref{sec:effect}, effects of various model parameters and assumptions on the results are 
investigated. As an application, we modelled the spectral energy distribution (SED) of the DoAr\,33 disk in Sect.~\ref{sec:appdoar33}. The paper ends up with a 
summary in Sect.~\ref{sec:summary}.

\begin{table}[!t]
\caption{Parameters of the fiducial model.}
\centering
\linespread{1.3}\selectfont
\begin{tabular}{lcc}
\hline 
Parameter                      & Value                    &    Note                      \\
\hline
$M_{\star}\, [M_{\odot}]$      &  0.5                     &    stellar mass              \\
$T_{\rm eff}\, [\rm K]$        &  4000                    &    effective temperature     \\
$L_{\star}\, [L_{\odot}]$      &  0.92                    &    stellar luminosity        \\
$R_{\rm in}\,[\rm{AU}]$        &  0.1                     &    disk inner radius         \\
$R_{\rm out}\,[\rm{AU}]$       &  100                     &    disk outer radius         \\
$\gamma$                       &  1.0                     &    surface density gradient  \\
$\beta$                        &  1.15                    &    flaring index             \\
$H_{100.a_{\rm min}}\, [\rm{AU}]$ &  10                   &    reference scale height    \\ 
$M_{\rm dust}\, [M_{\odot}]$   &  $3\,{\times}\,10^{-4}$  &    disk dust mass            \\
$\xi$                          &  0.14                    &    degree of dust settling   \\
$i\,[\circ]$                   &  41.8                    &    disk inclination          \\
$a_{\rm min}\, [\mu{\rm m}]$   &  0.01                    &    minimum grain size        \\
$a_{\rm max}\, [{\rm mm}]$     &  1                       &    maximum grain size        \\
Dust model                     &  DSHARP                  &    Mie theory                \\
Scattering mode                &  Isotropic               &    RADMC-3D setup            \\
\hline
\end{tabular}
\linespread{1.0}\selectfont
\label{tab:fiducial}
\end{table}

\section{The fiducial disk model}
\label{sec:fidmodel}
In this section, we devise a fiducial radiative tranfer model, and introduce the method used to quantify the degree of underestimation of the dust mass. 
Starting from the fiducial model, we can alter the assumption, and therefore investigate the effects of various parameters on the result. 
The well-tested code \texttt{RADMC-3D} \footnote{https://www.ita.uni-heidelberg.de/~dullemond/software/radmc-3d/.} is invoked to conduct the radiative 
transfer simulation \citep{radmc3d2012}. Table~\ref{tab:fiducial} gives an overview of parameters together with their adopted values used to describe 
the fiducial model.

\subsection{Dust density distribution}
\label{sec:fiddens}

We employ a flared disk surrounding a T Tauri star with stellar luminosity and temperature of $L_{\star}=0.92\,L_{\odot}$ and 
$T_{\rm eff}=4000\,{\rm K}$, respectively. The disk is assumed to be passively heated by stellar irradiation. 
The dust density is assumed to follow a Gaussian vertical profile
\begin{equation}
\rho(R,z,a) = \frac{\Sigma(R,a)}{\sqrt{2\pi}\,h(R,a)}\,\exp\left[-\frac{1}{2}\left(\frac{z}{h(R,a)}\right)^2\right], \\
\label{eqn:dens}
\end{equation}
where $R$ is the radial distance from the central star measured in the disk midplane. The surface density is described as a power law 
\begin{equation}
\Sigma(R,a)\,{=}\,\Sigma_{0}(a)\,\left(\frac{R}{100\,\rm{AU}}\right)^{-\gamma},
\label{eqn:surdens}
\end{equation}
with the proportionality factor $\Sigma_{0}(a)$ determined by normalizing the mass of dust grains in a certain grain size ($a$) bin. 
We set the disk inner radius $R_{\rm{in}}$ to 0.1\,AU which is close to the dust sublimation radius. The disk outer radius 
$R_{\rm{out}}$ is chosen to be 100\,AU, meaning that we truncate the disk at $R_{\rm{out}}$.

\begin{figure}[!t]
\centering
\includegraphics[width=0.4\textwidth]{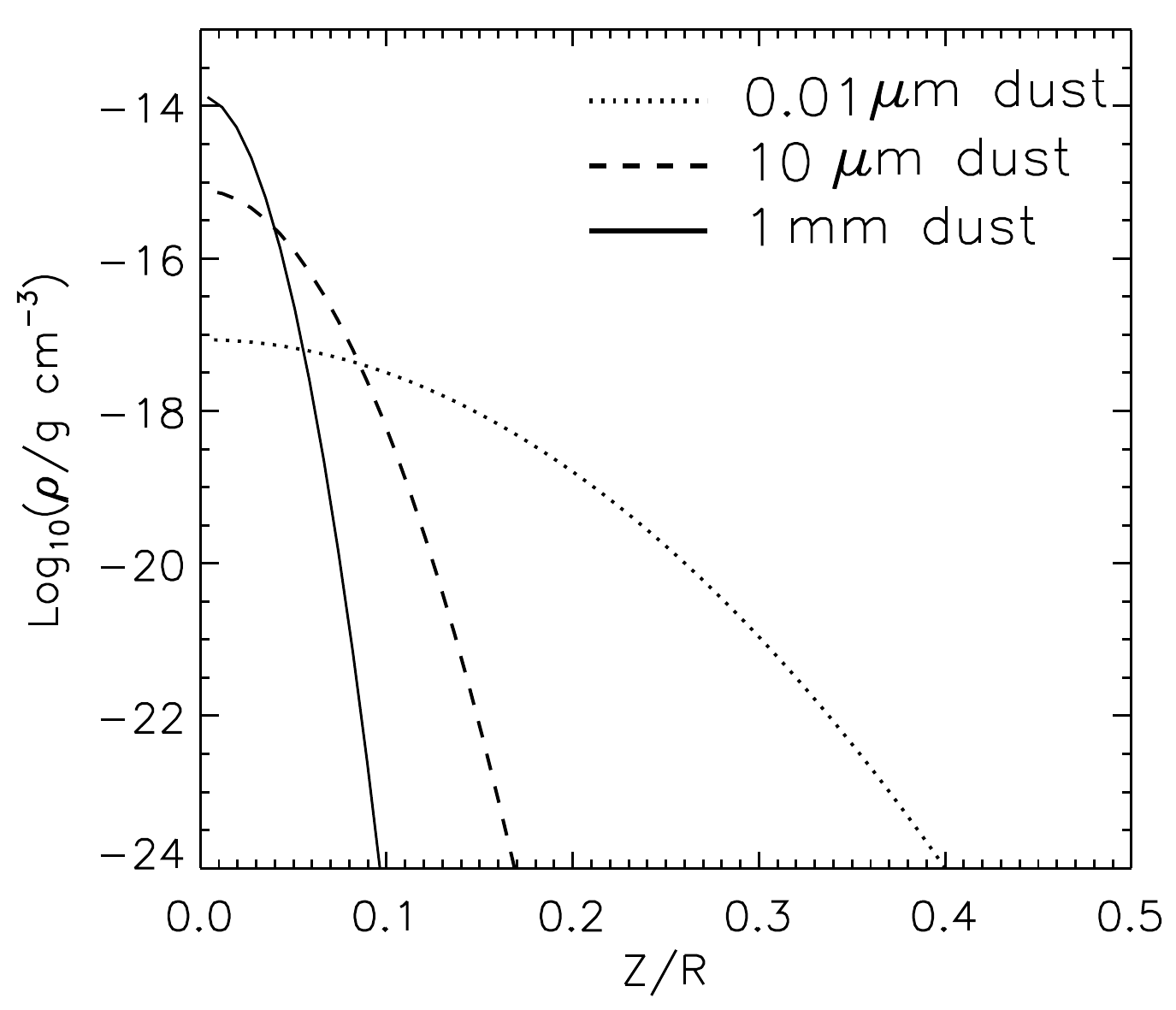}
\caption{Vertical density profile of the fiducial model at a radius of $R\,{=}\,10\,\rm{AU}$. The model includes 32 grains sizes logarithmically spaced from $a_{\rm min}\,{=}\,0.01\,\mu{\rm m}$ to $a_{\rm max}=1\,\rm{mm}$. The results are shown here for dust grains with three representative sizes, i.e. $0.01\,\mu{\rm m}$ (dotted line), $10\,\mu{\rm m}$ (dashed line) and $1\,{\rm mm}$ (solid line).}
\label{fig:fidsketch}
\end{figure}

Theoretical studies show that dust grains gradually decouple from the underlying gas distribution, and tend to settle towards the 
midplane \citep{Dullemond2004}. Such a process has been confirmed by high resolution observations, in which millimeter dust grains 
are found to be confined within scale heights of an order of a few AU \citep[e.g.,][]{Grafe2013,Pinte2016,Villenave2020} which is 
smaller than the typical gas pressure scale height \citep[${>}\,10\,\rm{AU,}$][]{Law2021,Rich2021}. To account for dust settling, we 
assume that the dust scale height $h$ varies with $R$ and the grain size $a$ 
\begin{equation}
h(R,a) = H_{100}(a)\left(\frac{R}{100\,\rm{AU}}\right)^{\beta},
\label{eqn:hra}
\end{equation}
where $\beta$ is the flaring index, and $H_{100}(a)$ stands for the scale height for a certain grain size at $R\,{=}\,100\,\rm{AU}$. 
Using a power law, we parameterize the reference dust scale height as
\begin{equation}
H_{100}(a) = H_{100.a_{\rm min}}\left(\frac{a}{a_{\rm min}}\right)^{-\xi}.
\label{eqn:h100a}
\end{equation}
The parameter $H_{100.a_{\rm min}}$ denotes the scale height of the smallest dust grains that are expected to be well coupled with the gas. 
The quantity $\xi$ is used to characterize the degree of dust settling. This simple prescription of dust settling has been used for multi-wavelength 
modeling of protoplanetary disks \citep{Pinte2008,Liuy2017}. We include 32 grain size bins that are logarithmically distributed from a fixed minimum 
grain size $a_{\rm{min}}\,{=}\,0.01\,\mu{\rm m}$ to a maximum grain size $a_{\rm max}$. The radiative transfer is performed using these 32 grain 
sizes as 32 different dust species, each of which has its own dust opacity (see Sect. \ref{sec:dust}). Moreover, they are assumed to be 
thermally decoupled in \texttt{RADMC-3D}. For the fiducial model, we adopt $a_{\rm max}\,{=}\,1\,\rm{mm}$. The scale height for each of the 32 grain 
sizes is described by Eq.\,\ref{eqn:hra} and \ref{eqn:h100a}. Given $\xi\,{=}\,0.14$ and $H_{100.a_{\rm min}}\,{=}\,10\,\rm{AU}$ 
for the fiducial model, the scale height of the 1\,mm dust grain equals to 2\,AU, which is consistent with observational 
constraints \citep{Pinte2016,Villenave2020,Liuy2022}. When the dust and gas are perfectly mixed, $\xi\,{=}\,0$, and $H_{100}(a)\,{=}\,H_{100.a_{\rm min}}$ becomes 
independent of the grain size and is equal to the gas scale height, the disk turns to be the well-mixed case.  

The fraction of dust mass in each grain size bin is calculated via  
\begin{equation}
 f(a_{j})\,{=}\,\frac{\int_{a_{0}}^{a_{1}}\frac{4\pi}{3}\rho_{\rm grain}n(a)a^3\,\mathrm{d}a}{\int_{a_{\rm lower}}^{a_{\rm upper}}\frac{4\pi}{3}\rho_{\rm grain}n(a)a^3\,\mathrm{d}a},
\end{equation}
where $a_{0}\,{=}\,{\rm exp}\left[{\rm ln}(a_{j})\,{-}\,\Delta/2\right]$, $a_{1}\,{=}\,{\rm exp}\left[{\rm ln}(a_{j})\,{+}\,\Delta/2\right]$, 
$a_{\rm lower}\,{=}\,{\rm exp}\left[{\rm ln}(a_{\rm min})\,{-}\,\Delta/2\right]$, $a_{\rm upper}\,{=}\,{\rm exp}\left[{\rm ln}(a_{\rm max})\,{+}\,\Delta/2\right]$, 
and $\Delta$ is the step width in the logarithmic scale used to sample the 32 grain sizes. The vertically integrated grain size distribution 
is assumed to be $n(a)\,{\propto}\,a^{-3.5}$, and $\rho_{\rm grain}$ is the bulk density of the dust ensemble. Therefore, once the total 
dust mass ($M_{\rm dust}$) in the entire disk is given, we know the dust mass for each grain size bin, therefore can derive the 
proportionality factor $\Sigma_{0}(a)$ in Eq.\,\ref{eqn:surdens}. Figure \ref{fig:fidsketch} shows the vertical density profile of the 
fiducial model at a radius of $R\,{=}\,10\,\rm{AU}$ for three representative grain sizes.

\begin{figure}[!t]
\centering
\includegraphics[width=3.0in]{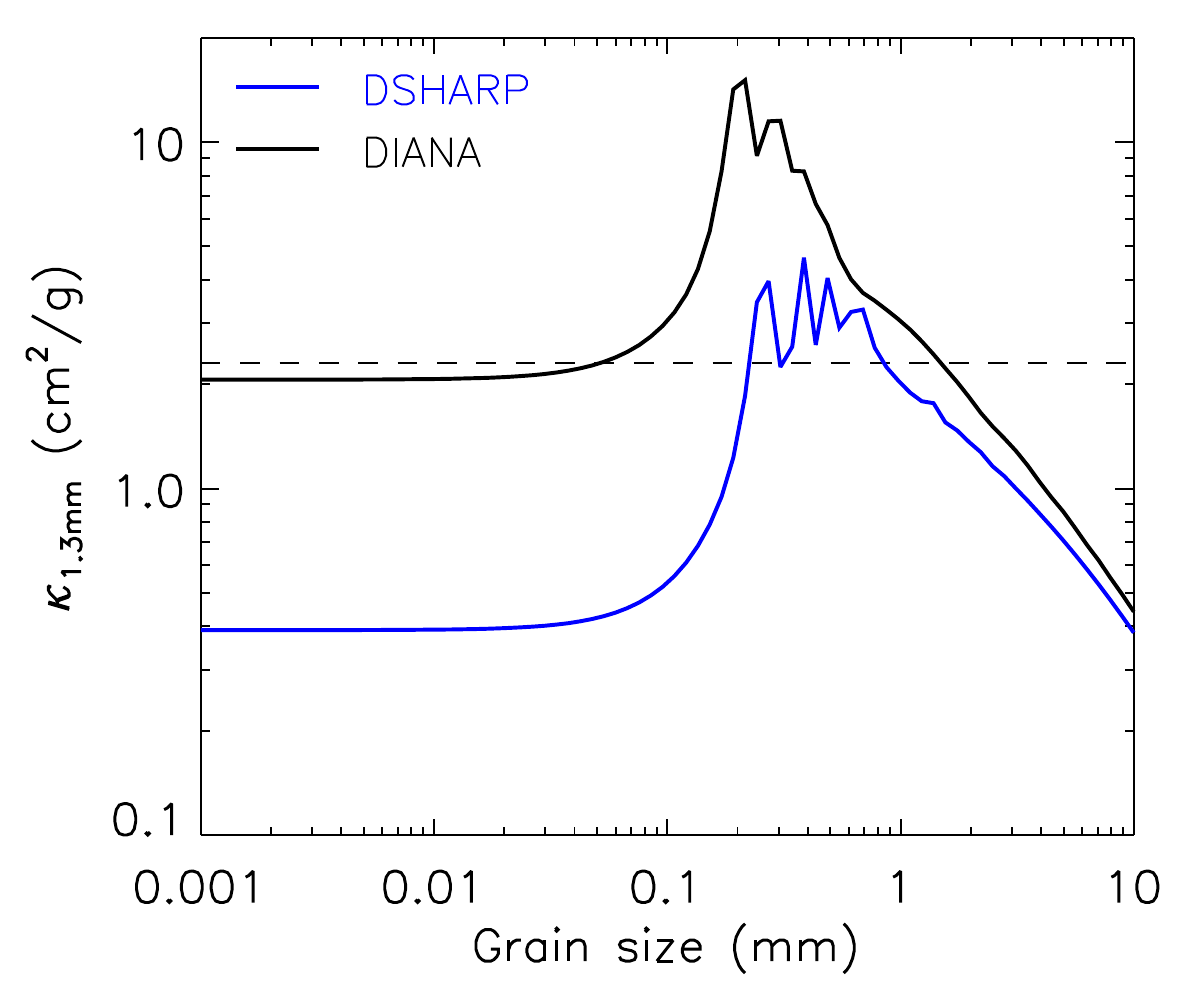}
\caption{Mass absorption coefficient $\kappa_{\nu}$ at $\lambda\,{=}\,1.3\,\rm{mm}$ as a function of grain size $a$. The blue line shows the result 
by using the DSHARP dust model, whereas the case for the DIANA dust model is indicated with the black line. The horizontal dashed line marks the 
value of $2.3\,{\rm cm^2/g}$ that is commonly adopted to derive the analytic dust mass $M_{\rm dust.ana}$ in the literature.}
\label{fig:dustopac}
\end{figure}

\subsection{Dust properties}
\label{sec:dust}

For the dust composition, we directly make use of the model by \citet{Birnstiel2018}, which is composed of water ice, astronomical silicates, 
troilite, and refractory organic material, with volume fractions being 36\%, 17\%, 3\% and 44\%, respectively. The bulk density of the dust mixture 
is $\rho_{\rm grain}\,{=}\,{\rm 1.675\,g\,cm^{-3}}$. This dust model was used for the interpretation of the ALMA data from the Disk Substructures at High 
Angular Resolution Project \citep[DSHARP,][]{Andrews2018}. We use the Mie theory to calculate the mass absorption/scattering coefficients for 
the sampled 32 grain sizes. The blue solid line in Figure~\ref{fig:dustopac} shows the absorption coefficient at $\lambda\,{=}\,1.3\,\rm{mm}$ 
as a function of grain size. There are clear fluctuations for $a\,{\sim}\,0.2-0.7\,\rm{mm}$ due to the resonances when the size of the particle is 
comparable with $\lambda/2\pi$. The horizontal dashed line indicates the value of $2.3\,{\rm cm^2/g}$.

\begin{figure}[!t]
\centering
\includegraphics[width=3.2in]{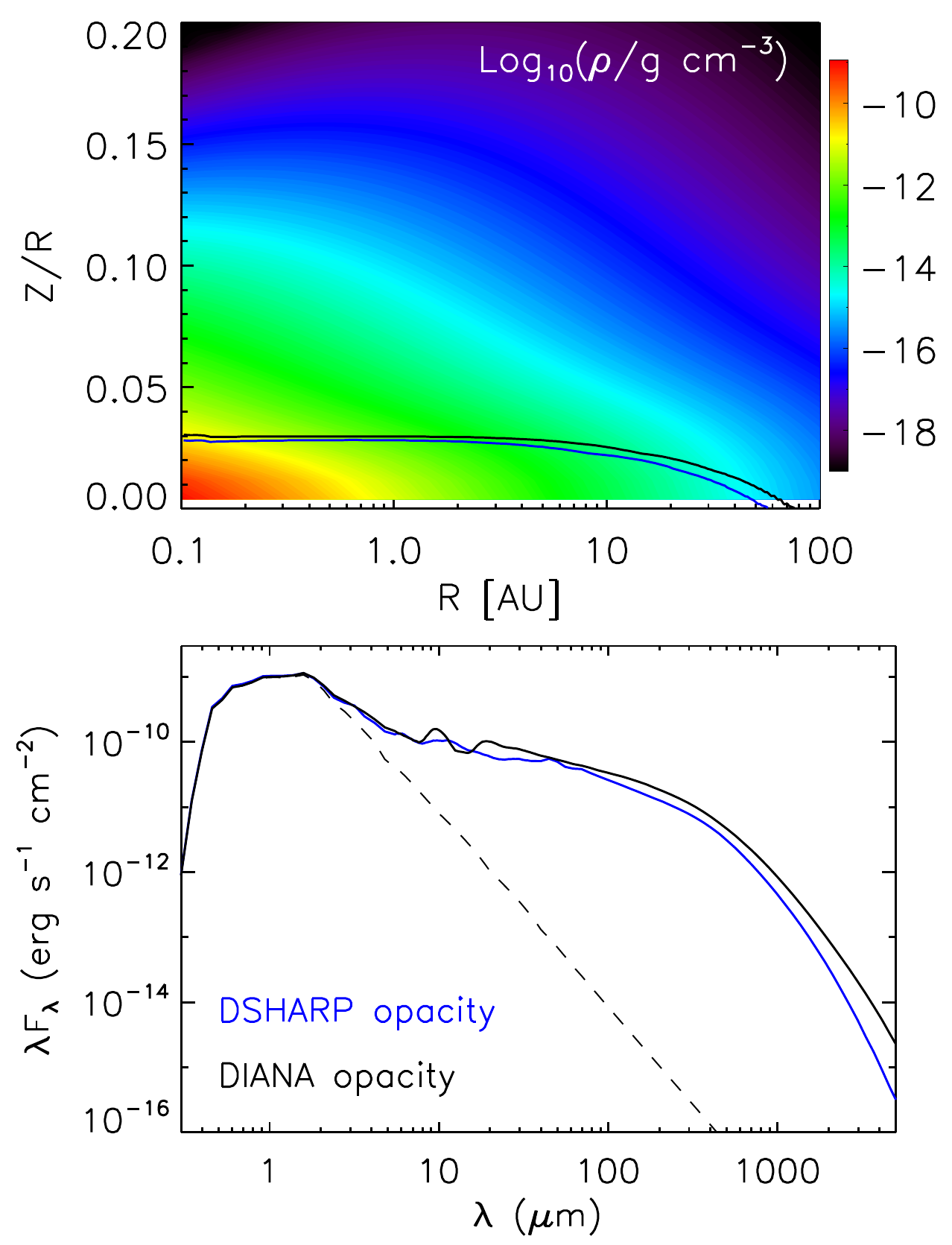}
\caption{{\it Upper panel:} 2D density distribution of the fiducial model with parameters given in Table~\ref{tab:fiducial}. 
The blue and black lines draw the contour of $\tau_{1.3\,\rm{mm}}\,{=}\,1$ when the disk is viewed at $i\,{=}\,41.8^{\circ}$, and 
includes the DSHARP and DIANA dust opacitites, respectively. {\it Bottom panel:} SEDs of the fiducial model when the DSHARP (blue line) 
or DIANA (black line) opacities are adopted in the simulation. The dashed curve stands for the input photospheric spectrum 
with $T_{\rm eff}\,{=}\,4000\,\rm{K}$, assuming ${\rm log}\,g\,{=}\,3.5$ and solar metallicity \citep{Kurucz1994}.}
\label{fig:fiddens}
\end{figure} 

Dust scattering is expected to significantly affect the emission level and millimeter spectral index of optically thick disks \citep{Zhu2019,Liuh2019}. 
Consequently, whether or not this mechanism is taken into account will influence the inferred dust temperature, optical depth, dust mass and grain 
size, see Appendix C of \citet{CarrascoGonzalez2019} for instance. We switch on isotropic scattering in both the thermal Monte Carlo simulation for computing the 
dust temperature and the subsequent step for SED generation. A more realistic treatment for dust scattering is considered and compared in Sect.~\ref{sec:effsca}.

Multi-wavelength observations have shown that millimeter spectral slopes differ among disks even in the same star formation 
region \citep[e.g.,][]{Ricci2010,Pinilla2017,Andrews2020}. The scatter of the observed millimeter spectral slopes can be attributed to  
different levels of dust processing or to different optical depths. Without a detailed analysis of multi-wavelength data, the grain size distribution 
is not known, which could have a direct impact on the mass determination with Eq.~\ref{eqn:mana}. Moreover, the composition of dust grains is also 
found to vary for different disks \citep[e.g.,][]{Juhasz2010}. Therefore, we also consider another dust model that is introduced in the DiscAnalysis 
project \citep[DIANA,][]{Woitke2016}, and investigate the influence of a different choice for dust properties on the result, see Sect.~\ref{sec:effopac}.

\subsection{Properties of the fiducial model}
\label{sec:fidprop}

The upper panel of Figure~\ref{fig:fiddens} shows the dust density distribution of the fiducial model. The blue curve indicates the location 
of the $\tau_{1.3\,\rm{mm}}\,{=}\,1$ surface considering that the disk is viewed at $i\,{=}\,41.8^{\circ}$. 
Within $R\,{\lesssim}\,40\,\rm{AU}$, disk interior layers are optically thick. We run the thermal 
Monte Carlo simulation using $3\times10^{7}$ photons in order to obtain a smooth temperature structure. The SED of the fiducial 
model is simulated and scaled to the distance of 140\,pc, which is indicated as the blue line in the bottom panel of Figure~\ref{fig:fiddens}. 

The midplane dust temperature ranges from ${\sim}\,1300\,\rm{K}$ to ${\sim}\,10\,\rm{K}$, roughly following a power law $T_{\rm midplane}\,{\propto}\,R^{-0.5}$. 
To give a representative dust temperature characterizing millimeter continuum emission, we calculate the 
mass-averaged dust temperature via 
\begin{equation}
 T_{\rm dust}\,{=}\,\frac{1}{M_{\rm dust}} \sum_{i=1}^{N_{\rm cells}}\sum_{j=1}^{N=32}m_{\rm dust}(i,j)\,t_{\rm dust}(i,j),
\end{equation}
where $m_{\rm dust}(i,j)$ and $t_{\rm dust}(i,j)$ are the dust mass and temperature in each cell ($i$) for each grain size ($j$), respectively. 
Similarly, we derive the mass-averaged dust opacity from 
\begin{equation}
 \overline{\kappa_{\nu}}\,{=}\,\frac{1}{M_{\rm dust}} \sum_{i=1}^{N_{\rm cells}}\sum_{j=1}^{N=32}m_{\rm dust}(i,j)\,\kappa_{\nu}(i,j),
\end{equation}
where $\kappa_{\nu}(i,j)$ are the dust opacity in each cell for each grain size. The fiducial model features 
a $T_{\rm dust}\,{=}\,16.3\,\rm{K}$ (see Table~\ref{tab:scatcom}) and $\overline{\kappa_{\rm 1.3mm}}\,{=}\,1.9\,\rm{cm^2/g}$. 

The simulated flux density at $\lambda\,{=}\,1.3\,\rm{mm}$ is $F_{\nu}\,{=}\,80.7\,\rm{mJy}$. Because $F_{\nu}$, $T_{\rm dust}$ and $\overline{\kappa_{\nu}}$ of 
the fiducial model are exactly known, we can use Eq.~\ref{eqn:mana} to derive a corresponding analytic dust mass required to produce the same amount of continuum 
in the optically thin regime. This gives us $M_{\rm dust.ana}\,{=}\,2.2\,{\times}\,10^{-4}\,M_{\odot}$. The true dust mass imported into the fiducial model 
is $M_{\rm dust}\,{=}\,3\,{\times}\,10^{-4}\,M_{\odot}$. We define the degree of mass underestimation as 
\begin{equation}
\Lambda_{1.3\rm{mm}} = \frac{M_{\rm dust}}{M_{\rm dust.ana}} = 1.4,
\end{equation}
where the subscript is used to state that the calculation is based on 1.3\,mm flux densities. Similar factors can be derived  
using data points at other wavelengths. The difference between $M_{\rm dust}$ and $M_{\rm dust.ana}$ basically reflects dust 
material hidden below the $\tau\,{=}\,1$ surface (see Figure~\ref{fig:fiddens}), because uncertainties on the dust temperature 
and opacity are eliminated in our definition.


\begin{table}[!t]
\caption{Comparison of results between models with isotropic and anisotropic scattering.}
\centering
\linespread{1.3}\selectfont
\begin{tabular}{lcccc}
\hline
$M_{\rm dust}$ $[M_{\odot}]$ & $i$ $[\circ]$  &  $F_{\rm 1.3\,mm}$ [mJy]    &    $T_{\rm dust}$ [K]  &   $\Lambda_{\rm 1.3\,mm}$    \\
\hline
$3\,{\times}\,10^{-5}$   & 41.8   &    12.2 (12.6)    &   17.5 (18.0)       & 1.00 (1.00)            \\
$3\,{\times}\,10^{-4}$   & 41.8   &    80.7 (86.7)    &   16.3 (16.9)       & 1.38 (1.35)            \\
$3\,{\times}\,10^{-3}$   & 41.8   &    168.7 (195.5)  &   15.7 (16.2)       & 6.25 (5.62)            \\
\hline
$3\,{\times}\,10^{-4}$   & 8.1            &  101.3 (105.9)           &   16.3 (16.9)         &  1.10 (1.10)        \\
$3\,{\times}\,10^{-4}$   & 63.1           &  51.5 (57.0)             &   16.3 (16.9)         &  2.16 (2.04)        \\
$3\,{\times}\,10^{-4}$   & 79.3           &  21.6 (24.0)             &   16.3 (16.9)         &  5.16 (4.86)        \\
\hline
\end{tabular}
\linespread{1.0}\selectfont
\tablefoot{Numbers given outside parentheses refer to the result when isotropic scattering is assumed, while 
the ones in the parentheses are the case when full scattering matrices are taken into account.}
\label{tab:scatcom}
\end{table}

\begin{figure*}[!h]
\centering
\includegraphics[width=0.75\textwidth]{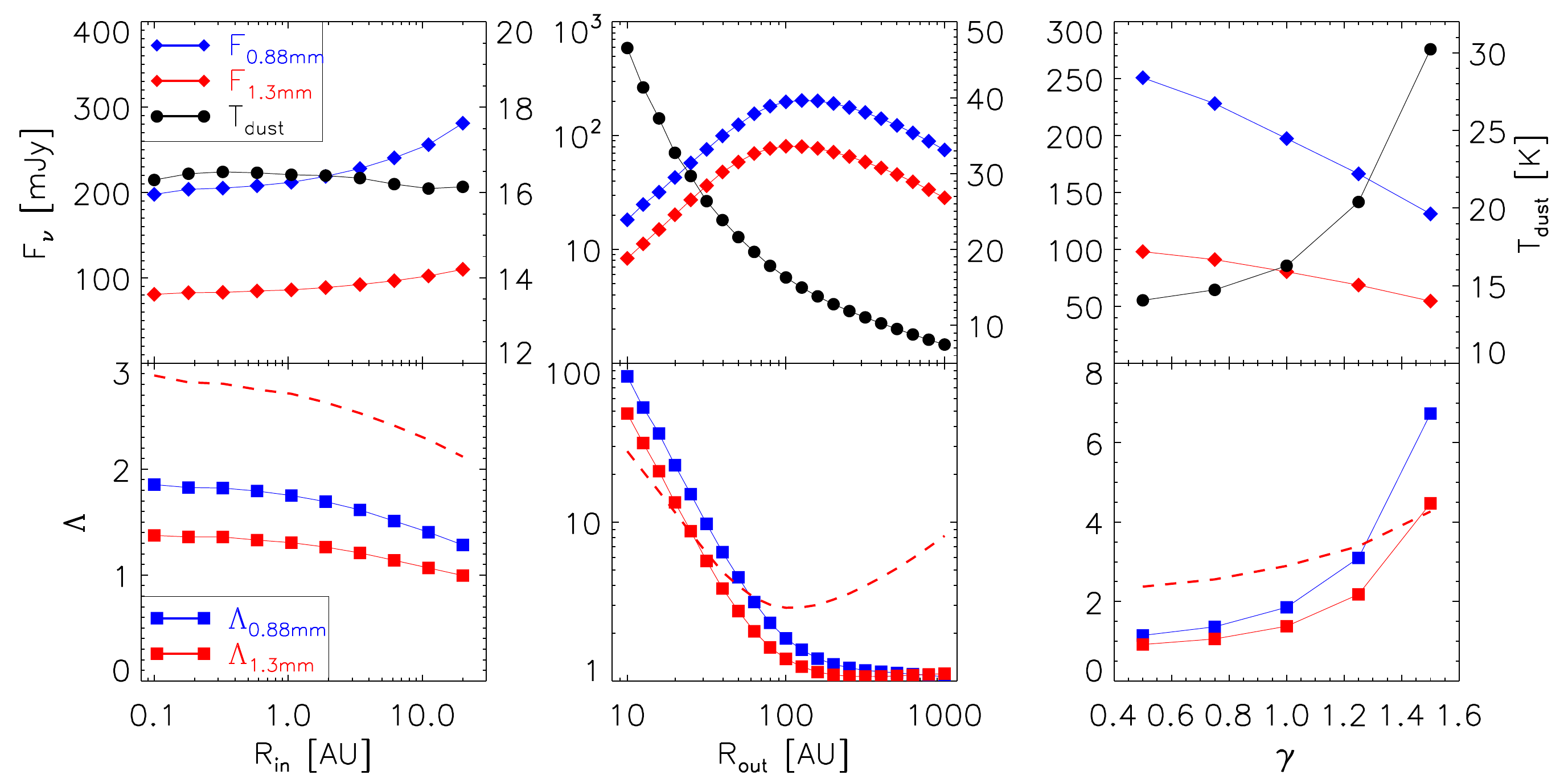}
\includegraphics[width=0.75\textwidth]{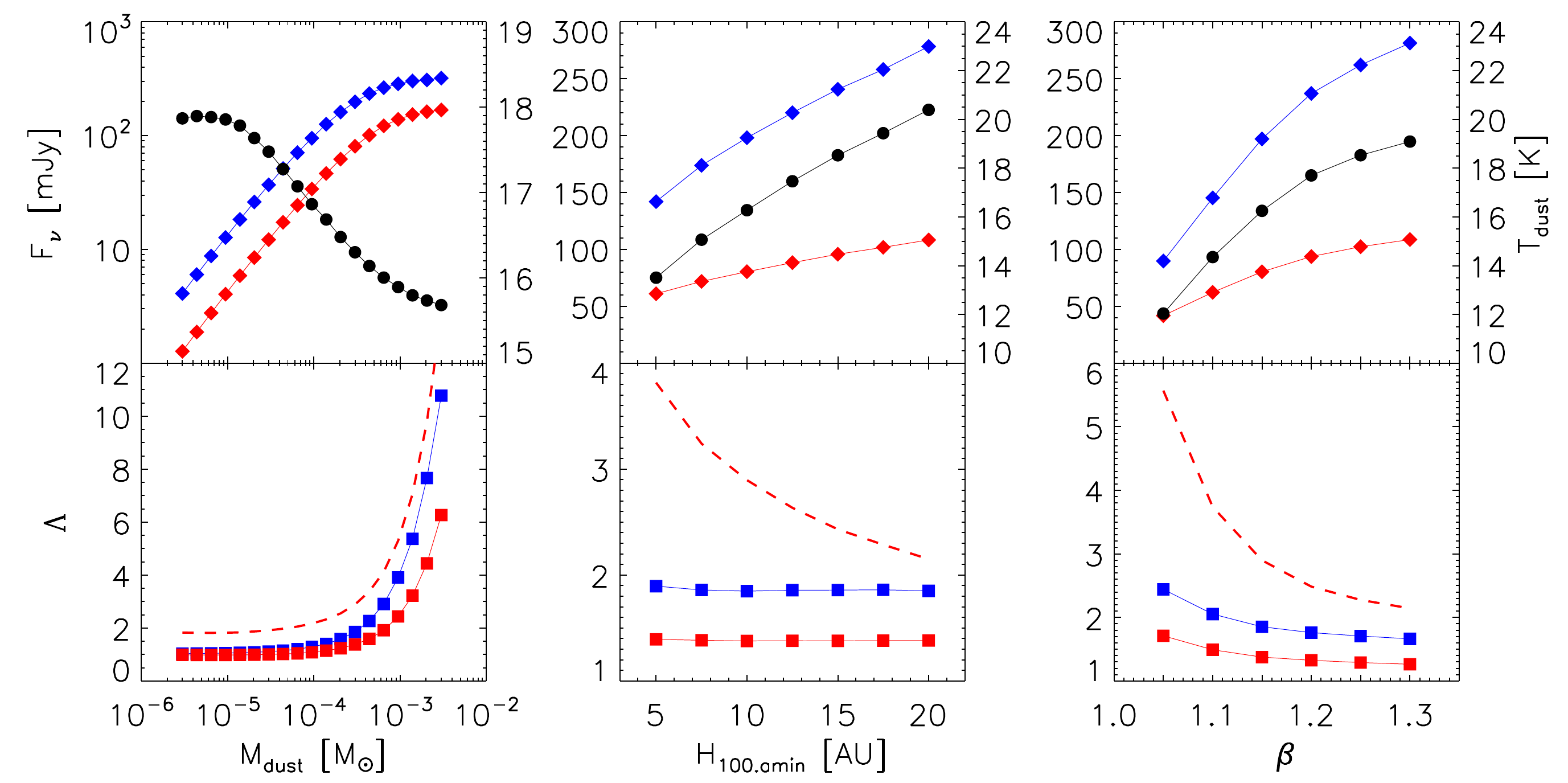}
\includegraphics[width=0.75\textwidth]{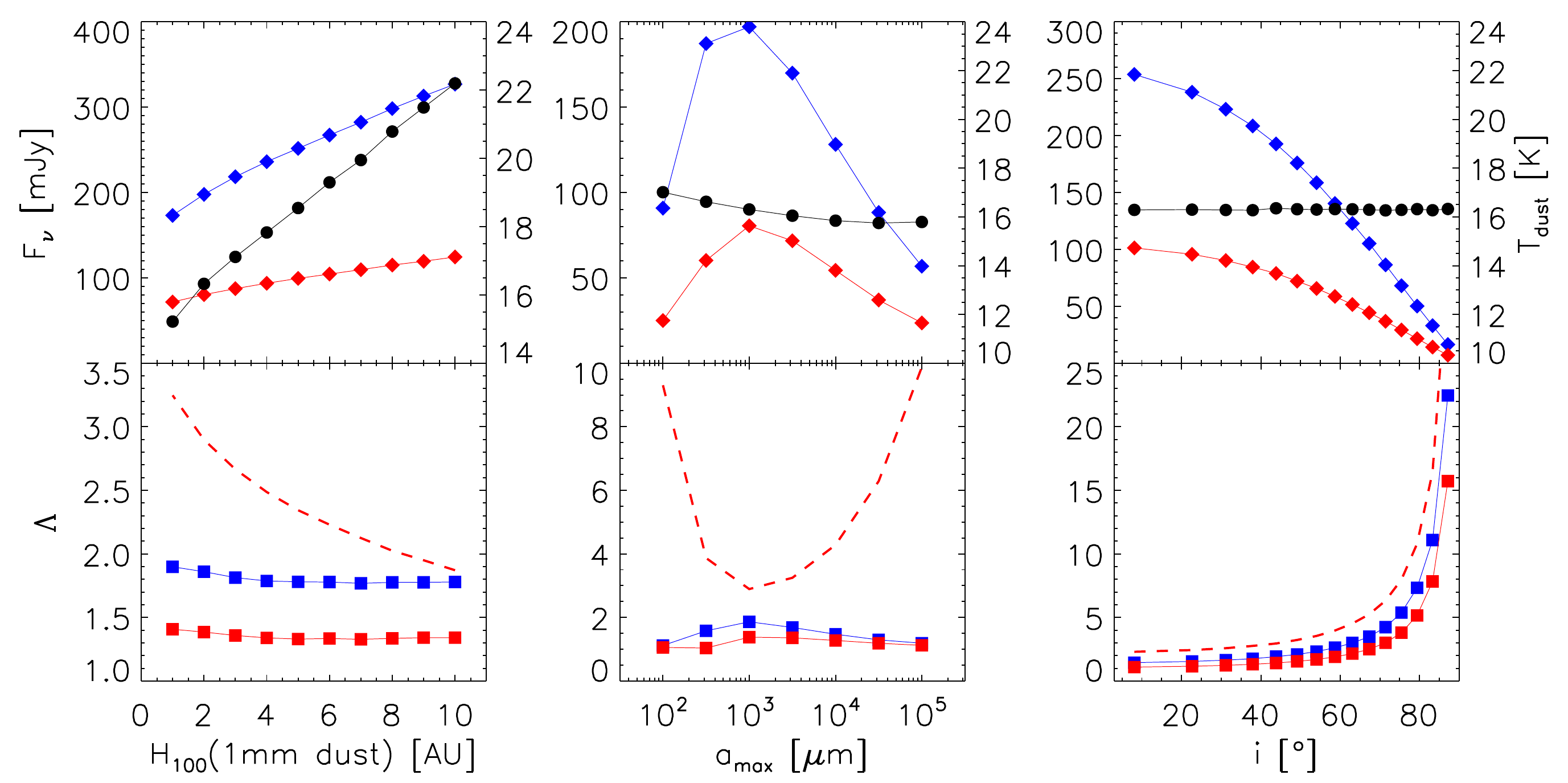}
\caption{The effect of model parameters on the flux densities at 0.88 and 1.3\,mm ($F_{\rm 0.88mm}$, $F_{\rm 1.3mm}$, top row on the left Y axis), mass-averaged dust 
temperature ($T_{\rm dust}$, top row on the right Y axis), and underestimation factors ($\Lambda_{\rm 0.88mm}$, $\Lambda_{\rm 1.3mm}$, bottom row). When exploring one 
particular parameter, all the rest parameters are fixed to their fiducial values listed in Table \ref{tab:fiducial}. The red dashed lines show the underestimation
factors $\Lambda_{\rm 1.3mm}$ that are calculated by assuming a constant $T_{\rm dust}\,{=}\,25\,(L_{\star}/L_{\odot})^{0.25}\,{=}\,24.5\,\rm{K}$ 
and $\kappa_{1.3\rm{mm}}\,{=}\,2.3\,\rm{cm^2/g}$.}
\label{fig:paraeff}
\end{figure*}

\section{Effects of model parameters and assumptions on the results}
\label{sec:effect}

\subsection{Effects of model parameters}
\label{sec:effpara}

With the methodology used to quantify the underestimation factor outlined in Sect.~\ref{sec:fidprop}, we then change the 
values for each model parameter to investigate their effects on the results. Parameters considered in the exploration 
are $R_{\rm in}$, $R_{\rm out}$, $\gamma$, $M_{\rm dust}$, $H_{100.a_{\rm min}}$, $\beta$, $H_{100}({\rm 1mm\,\,dust})$ or equivalent to $\xi$, 
$a_{\rm max}$, and $i$. Calculating the underestimation factor always utilizes the mass-averaged dust opacity and temperature of 
the corresponding model. Figure~\ref{fig:paraeff} shows the flux densities at 0.88\,mm and 1.3\,mm, mass-averaged temperature and 
underestimation factors.

As can be seen, although the stellar properties are fixed, there is a broad range for the mass-averaged temperature. Particularly, the variation in $T_{\rm dust}$ with 
$R_{\rm out}$ can be up to a factor of 4, from ${\sim}\,32\,\rm{K}$ for compact disks with $R_{\rm out}\,{=}\,20\,\rm{AU}$ to ${\sim}\,8\,\rm{K}$ for large disks 
with $R_{\rm out}\,{=}\,800\,\rm{AU}$. Of the nine parameters, only $R_{\rm in}$, $a_{\rm amax}$ and $i$ have minor impacts on $T_{\rm dust}$. 
Therefore, a simple presumption on the temperature according to $T_{\rm dust}\,{=}\,25\,(L_{\star}/L_{\odot})^{0.25}\,{=}\,24.5\,\rm{K}$ will lead to differences 
in the determination of $M_{\rm dust.ana}$. 

The underestimation factor $\Lambda$ ranges from a few to hundreds. Higher dust masses, inclinations, or smaller disk sizes result in the greatest underestimation in the 
dust mass, because they work together to achieve a higher optical depth along the line of sight. Other parameters that aggravate the underestimation include smaller $R_{\rm in}$, 
steeper $\gamma$, lower $\beta$ and $a_{\rm amax}$ that yields a higher $\kappa$. As the optical depth decreases with increasing wavelength, underestimations
calculated based on 1.3\,mm fluxes are systematically lower than those derived using the 0.88\,mm fluxes. Therefore, long wavelength observations are required to 
better probe the dust mass in protoplanetary disks. The red dashed lines in Figure~\ref{fig:paraeff} show the underestimation factors $\Lambda_{\rm 1.3mm}$ 
that are calculated by assuming a constant $\kappa_{1.3\rm{mm}}\,{=}\,2.3\,\rm{cm^2/g}$ and $T_{\rm dust}\,{=}\,25\,(L_{\star}/L_{\odot})^{0.25}\,{=}\,24.5\,\rm{K}$
(hotter than most of the models). As can be seen, they are generally larger than those obtained by using the mass-averaged dust opacity and temperature of 
the models, directly demonstrating the importance of taking disk structure and dust properties into account in the task of mass estimation.

Figure~\ref{fig:lambdafaceon} and \ref{fig:lambdaedgeon} show how $\Lambda$ varies with different model parameters when the disk is viewed at two extreme 
inclinations, i.e., $i\,{=}\,0^{\circ}$ (face on) and $i\,{=}\,90^{\circ}$ (edge-on). The behavior of face-on disks is quite similar to the fiducial 
inclined disks. For edge-on disks, the underestimation factor becomes more sensitive to most of the explored parameters. This tendency can be clearly 
identified from the $\Lambda\,{-}\,R_{\rm out}$, $\Lambda\,{-}\,M_{\rm dust}$, $\Lambda\,{-}\,H_{100}({\rm 1\,mm\,\,dust})$ and $\Lambda\,{-}\,a_{\rm max}$ profiles. 
As a consequence, mass determination for edge-on disks suffers more difficulties than for disks with low inclinations, because the accuracy of the result highly 
depends on how well many other parameters are constrained.

\subsection{Effects of dust scattering mode}
\label{sec:effsca}

When dust grains grow up to millimeter sizes, scattering coefficients at millimeter wavelengths become significantly larger than absorption coefficients, resulting in 
a very high albedo. For instance, the albedo of the fiducial dust model is 0.9 at 1.3\,mm. In this case, scattering of thermal reemission radiation on the SED should be 
taken into account. Theoretical studies have found that dust scattering can considerably reduce the emission from optically thick regions, with the reduction level 
depending on the inclination \citep[e.g.,][]{Zhu2019,Sierra2020}. Thus, an optically thick disk with scattering can be misidentified as an optically thin disk if 
scattering is ignored. 

We compared the models when different modes of dust scattering are turned on. The results are summarized in Table~\ref{tab:scatcom}. The difference in dust temperature 
between isotropic and anisotropic scattering models is merely ${\sim}\,0.5\,\rm{K}$, regardless of the total dust mass (hence, intrinsic optical thickness). Anisotropic 
scattering always yields stronger emission. However, the flux discrepancy between the two modes does not exceed 16\%, with larger differences generally for more massive 
or more inclined disks. Therefore, once dust scattering is included, this does not have a significant impact on the underestimation factor.

\begin{figure}[!t]
\centering
\includegraphics[width=3.0in]{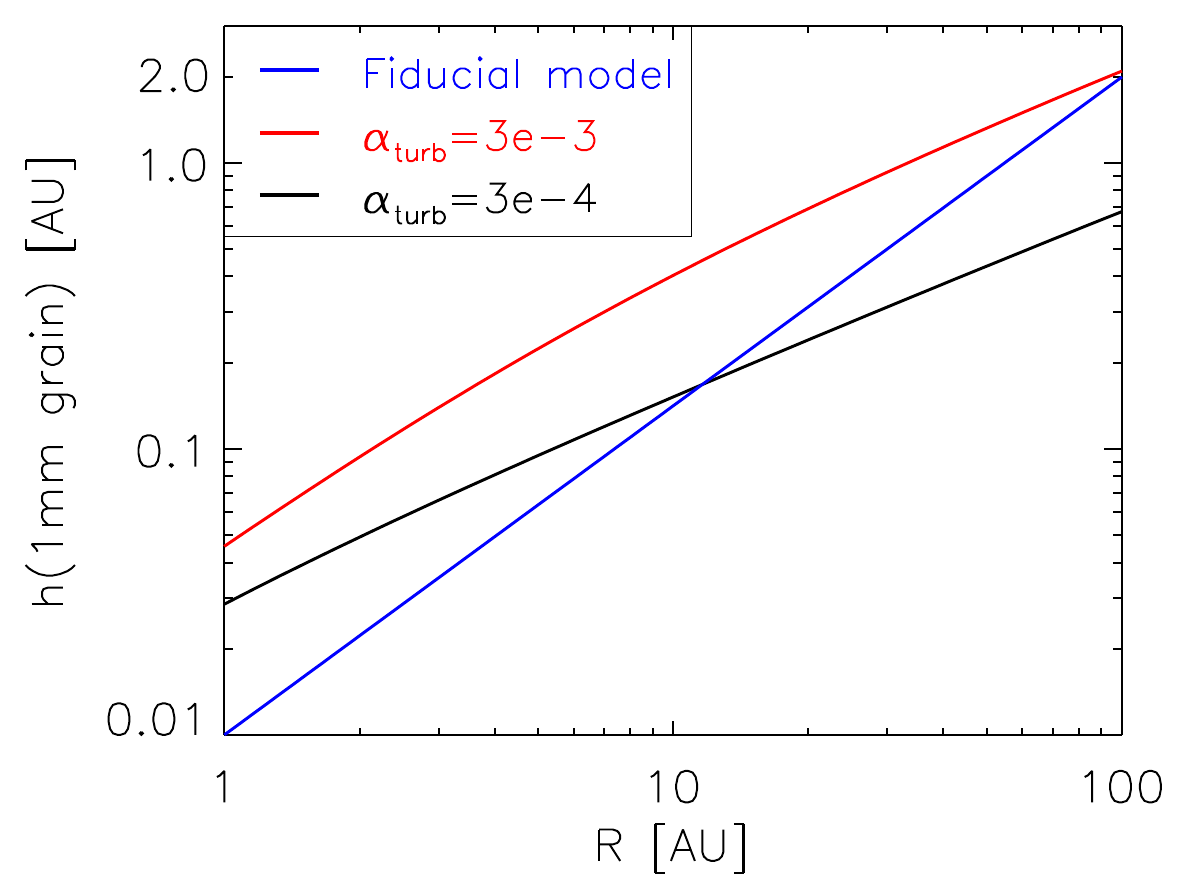}
\caption{Scale height of 1\,mm dust as a function of $R$. The blue line shows the fiducial model with dust settling prescribed by Eq.~\ref{eqn:h100a}. The black and 
red lines stand for models when dust settling is parameterized according to Eq.~\ref{eqn:hr2a} under different turbulence strengths $\alpha_{\rm turb}$.}
\label{fig:dustheight}
\end{figure}

\begin{table}[!t]
\caption{Comparison of models using different prescriptions for dust settling.}
\centering
\linespread{1.3}\selectfont
\begin{tabular}{lcccc}
\hline
Settling approach &  $\xi$ or $\alpha_{\rm turb}$  &  $F_{\rm 1.3\,mm}$  &  $T_{\rm dust}$ &  $\Lambda_{\rm 1.3\,mm}$   \\
                  &                                &   [Jy]  &  [K]  &   \\

\hline
Fiducial model    & $\xi\,{=}\,0.14$             &   80.7                    &  16.3           &    1.38                     \\
Well-mixed model                                 &    $\xi\,{=}\,0$                & 124.4           &    22.2        &  1.34    \\
\multirow{2}{*}{Settling by Eq.~\ref{eqn:hr2a}}  &  $\alpha_{\rm turb}\,{=}\,3{\rm e-}3$ & 108.5           &    20.6        &  1.40    \\
                                                 & $\alpha_{\rm turb}\,{=}\,3{\rm e-}4$  & 86.6            &    18.1        &  1.47    \\
\hline
\end{tabular}
\linespread{1.0}\selectfont
\tablefoot{Other parameters of these models are fixed to the fiducial values.} 
\label{tab:reshcom}
\end{table}

\subsection{Effects of dust settling approach}

We also consider different approaches for dust settling that basically determines the millimeter dust scale height. In the fiducial model, we use only three free parameters
($H_{100.a_{\rm min}}$, $\xi$ and $\beta$) to parameterize the scale height, see Eq.~\ref{eqn:hra} and \ref{eqn:h100a}. The scale height of 1\,mm dust grain is indicated as 
the blue line in Figure~\ref{fig:dustheight}. Theoretically, the vertical distribution of dust grains is a consequence of equilibrium between dust settling and vertical 
stirring induced by turbulent motions. Therefore, the scale heights of dust grains with different sizes are related to the turbulence strength $\alpha_{\rm turb}$
\begin{equation}
 h(R,a) = H_{\rm gas}\left(1+\frac{\rm St}{\alpha_{\rm turb}} \frac{\rm 1+2\,St}{\rm 1+St}\right)^{-1/2},
 \label{eqn:hr2a}
\end{equation}
where ${\rm St} = \frac{\pi}{2} \frac{\rho_{\rm grain}\,a}{\Sigma_{\rm gas}(R)}$ is the Stokes number \citep[e.g.,][]{Dubrulle1995,Schrapler2004,Woitke2016,Birnstiel2016}.
Since the smallest dust grain ($0.01\,\mu{\rm m}$ in our case) is expected to be well-mixed with the gas, we assume that the gas pressure scale height $H_{\rm gas}$ follows 
\begin{equation}
 H_{\rm gas}=H_{100.a_{\rm min}}\,\left(\frac{R}{100\,\rm AU}\right)^{\beta}.
 \label{eqn:hr2b}
\end{equation}
This settling approach has four free parameters: $H_{100.a_{\rm min}}$, $\alpha_{\rm turb}$, $\beta$ and gas surface density $\Sigma_{\rm gas}$. For $\Sigma_{\rm gas}$, we simply 
scaled the dust surface density by a constant gas-to-dust mass ratio of 100. For $\alpha_{\rm turb}$, we consider $3\,{\times}\,10^{-4}$ and $3\,{\times}\,10^{-3}$ that are consistent 
with observational constraints from gas line observations \citep[e.g.,][]{Flaherty2017,Teague2018,Flaherty2020}. The $\alpha_{\rm turb}/{\rm St}$ ratios for the 0.2\,mm dust 
grain calculated at $R\,{\sim}\,80\,\rm{AU}$ are ${\sim}\,0.03$ and ${\sim}\,0.3$ for $\alpha_{\rm turb}\,{=}\,3\,{\times}\,10^{-4}$ and $3\,{\times}\,10^{-3}$, respectively. 
These ratios are comparable with the results derived from analyzing the widths of dust rings in several DSHARP disks \citep{Dullemond2018}. The dust scale heights are shown with 
the black and red lines in Figure~\ref{fig:dustheight}. The model with $\alpha_{\rm turb}\,{=}\,3\,{\times}\,10^{-3}$ features a larger scale height than the fiducial setup across 
the entire radius. Consequently, it has a higher temperature and emits more continuum (see Table~\ref{tab:reshcom}). A weaker turbulence of $3\,{\times}\,10^{-4}$ makes the disk 
cooler and fainter, and now $T_{\rm dust}$ and $F_{1.3\,\rm{mm}}$ are comparable with those of the fiducial model. 

In Table~\ref{tab:reshcom}, we also provide the results when dust grains and gas are well coupled, which can be considered as a limiting case of the fiducial settled disk. When the 
settling degree $\xi$ (see Eq.~\ref{eqn:h100a}) tends towards zero, the settled disk model turns into the well-mixed case, and becomes the brightest at millimeter wavelengths among 
the models under comparison. It is naturally understood that using well-mixed radiative transfer models to fit observed SEDs, one would need the least amount of dust grains. 
Hence, dust mass determinations from radiative transfer analysis based on the well-mixed assumption are probably still lower limits. 

\begin{figure}[!t]
\centering
\includegraphics[width=3.0in]{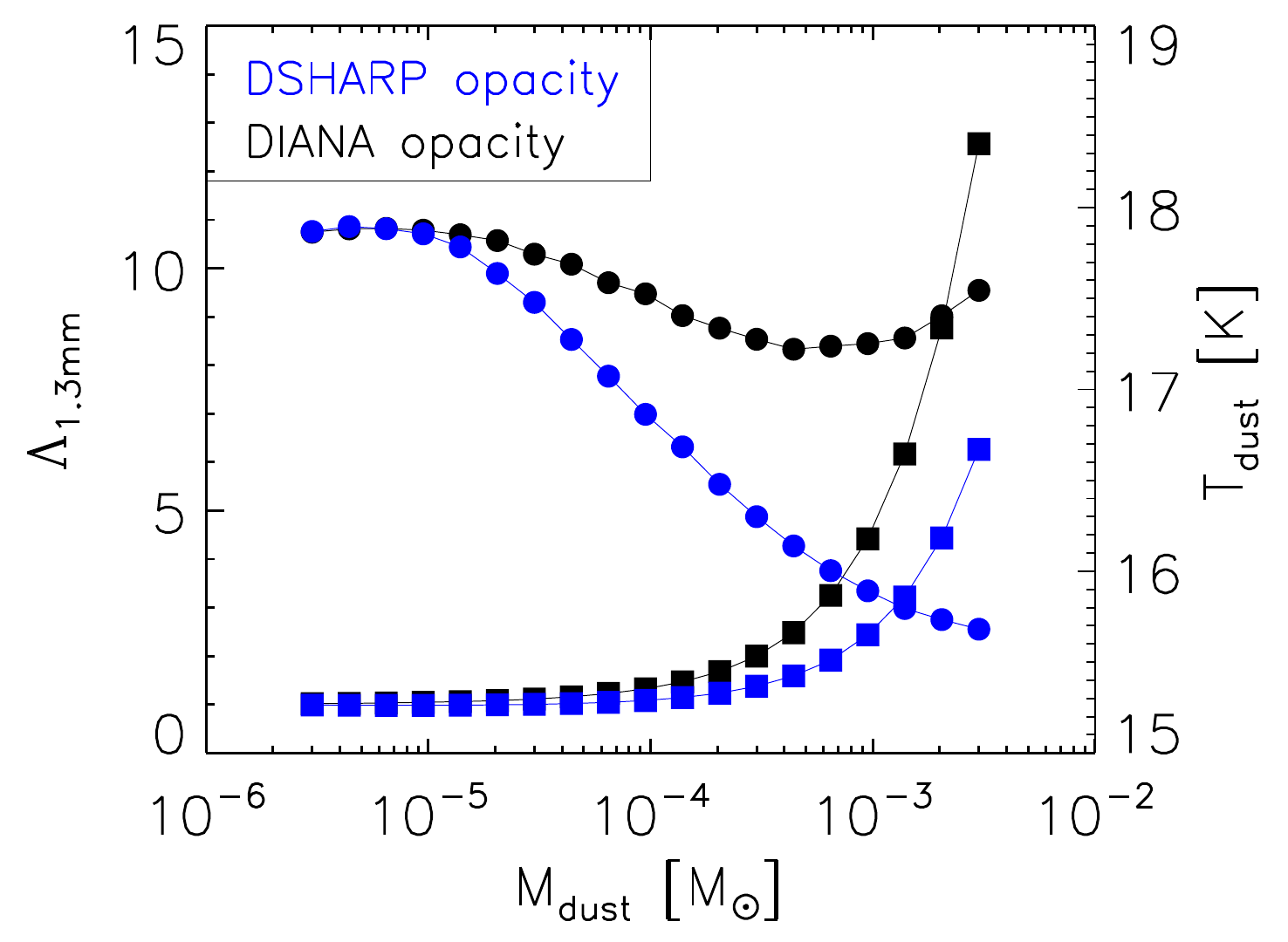}
\caption{A comparison of $\Lambda_{1.3\rm{mm}}$ (squares) and $T_{\rm dust}$ (filled circles) between models that use the DSHARP (blue symbols) and DIANA (black symbols) dust opacities in the simulation, respectively. Details can be found in Sect.~\ref{sec:effopac}.}
\label{fig:ratiodiana}
\end{figure}

\subsection{Effects of dust properties}
\label{sec:effopac}
To investigate how the dust properties affect the results, we make use of the standard dust properties introduced by the DIANA project \citep{Woitke2016}.
The dust grains consist of 60\% silicate \citep[$\rm{Mg_{0.7}Fe_{0.3}SiO_{3}}$,][]{Dorschner1995}, 15\% amorphous carbon \citep[BE$-$sample,][]{Zubko1996}, and 25\% porosity. 
These percentages are volume fractions that are used to derive the effective refractory indices by applying the Bruggeman mixing rule \citep{Bruggeman1935}. 
Porous grains are considered because porosity, and its evolution during the collisional growth, helps to overcome the radial-drift 
barrier \citep[e.g.,][]{Ormel2007,Garcia2020}. We assumed a distribution of hollow spheres with a maximum hollow volume ratio of 0.8, and calculated 
the opacities using the \texttt{OpacityTool} package \citep{Toon1981,Min2005}. 

The black line in Figure~\ref{fig:dustopac} shows $\kappa_{1.3\,\rm{mm}}$ as a function of grain size. The DIANA dust absorption coefficients are systematically 
larger than the DSHARP values. The choice for the material termed ``organic'' in the DSHARP model has a significant impact on the absorption coefficient, see 
Appendix B in \citet{Birnstiel2018}. When using the amorphous carbon \citep[BE$-$sample,][]{Zubko1996} instead, the absorption coefficients increase by 
a factor of ${\sim}\,5$. The SEDs using different dust opacities are compared in the lower panel of Figure~\ref{fig:fiddens}, with $F_{1.3\rm{mm}}\,{=}\,165.2\,\rm{mJy}$
and $80.7\,\rm{mJy}$ for the DIANA and DSHARP dust being imported, respectively. Models using the DIANA opacity are generally hotter, but the differences 
in the mass-averaged temperature are within ${\sim}\,2\,\rm{K}$, even for very massive disks, see Figure~\ref{fig:ratiodiana}. Given the disk is more 
optically thick when adopting the DIANA dust model, correspondingly, the underestimation factors are systematically higher, as shown 
in Figure~\ref{fig:ratiodiana}.

\begin{table}[!t]
 \centering
   \caption{Ring and crescent parameters.}
     \linespread{1.3}\selectfont
     \begin{tabular}[h]{lcccc}
      \hline
	  Parameter                                       &  1st ring   &   2nd ring  &   3rd ring  &  Crescent   \\
	  \hline
      $R_{\rm ring}$ or $R_{\rm cres}$ [AU]           &   10        &   35        &   65        &  70       \\
      $\sigma_{\rm ring}$ or $\sigma_{R_{\rm cres}}$ [AU]  &   5         &   5         &   5         &  5        \\
      $\theta_{\rm cres}$ $[\circ]$                   &   $-$       &  $-$        & $-$         &  45       \\
      $\sigma_{\theta_{\rm cres}}$  $[\circ]$         &   $-$       &  $-$        & $-$         &  20       \\
      \hline
    \end{tabular}
\label{tab:structures}
\end{table}

\begin{figure*}[!t]
 \centering
 \includegraphics[width=0.9\textwidth]{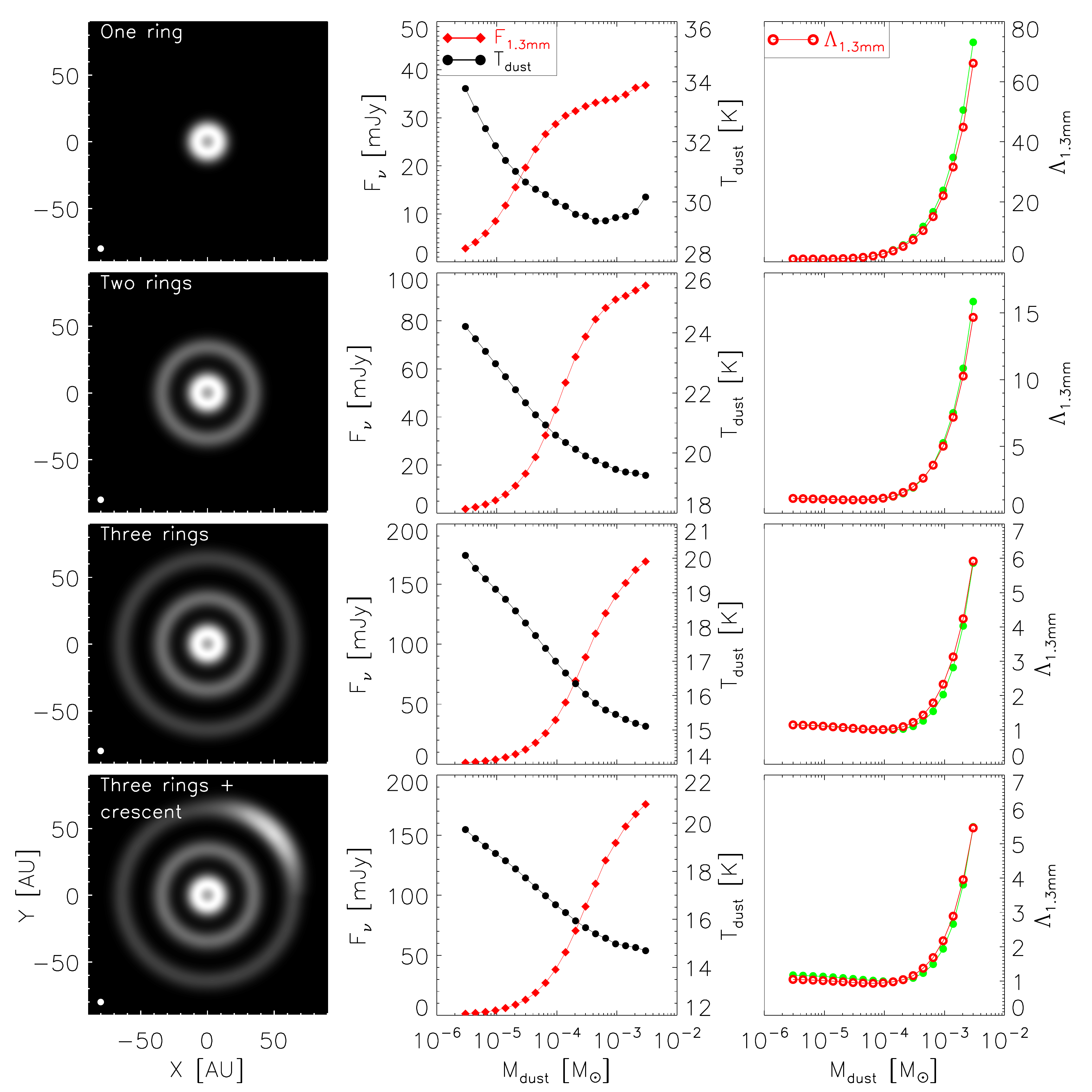}
 \caption{Effects of disk substructures on the mass underestimation. {\it Left column:} representative 1.3\,mm dust continuum image for each scenario of disk substructures. 
 The raw radiative transfer images are convolved with a Gaussian beam of 35\,mas in size (${\sim}\,5\,\rm{AU}$ at the distance of 140\,pc, indicated with a white filled 
 circle in the lower left corner of each image). The disk is seen face-on merely for a better presentation of the disk configuration. {\it Middle column:} 1.3\,mm flux 
 density ($F_{\rm 1.3mm}$) and mass-averaged temperature ($T_{\rm dust}$) as a function of total dust mass ($M_{\rm dust}$). When simulating the SED, the fiducial 
 inclination of $41.8^{\circ}$ is adopted. {\it Right column:} underestimation factor as a function of $M_{\rm dust}$. The green filled circles refer to the 
 underestimation factors for smooth disks that have an equivalent $R_{\rm out}$ to that of the disks with substructures, see Sect.~\ref{sec:substructure}.}
 \label{fig:ringres}
\end{figure*}

\subsection{Effects of the presence of disk substructures}
\label{sec:substructure}
In the above modeling procedure, dust surface densities are assumed to follow a smooth power law, without sharp density 
fluctuations, see Eq.~\ref{eqn:surdens}. However, high resolution multi-wavelength observations have revealed that many 
protoplanetary disks possess rings and crescents \citep[e.g.,][]{vanBoekel2017,Avenhaus2018,Fedele2018,Long2018,Andrews2018,Cieza2021}, 
which are believed to be the favorable place for planet formation \citep{Andrews2020}. In some cases, these substructures, even located 
at large radial distances from the host central star, appear very bright in the millimeter continuum, indicating that a large amount 
of dust grains are concentrated \citep[e.g.,][]{alma2015,Perez2018,Dong2018,Cazzoletti2018}.  

The overdensities in disk substructures will increase the optical depth locally, which in principle can affect the mass determination. Detailed radiative transfer analysis 
confirms that some substructures are optically thick at ALMA wavelengths \citep[e.g.,][]{Pinte2016,Liuy2017,Ohashi2019,Sierra2021,Liuy2022}. In order to investigate how the 
redistribition of dust grains in disks with substructures affect the mass determination, we consider four scenarios: one ring, two rings, three rings, and three rings plus 
a crescent. The surface densities of the substructures are defined as 
\begin{equation}
 \Sigma_{\rm ring}\,{=}\,\Sigma_{0}{\rm exp}^{-(R-R_{\rm ring})^2/2\sigma_{\rm ring}^2},
 \label{eqn:ringsurdens}
\end{equation}
\begin{equation}
 \Sigma_{\rm crescent}\,{=}\,3\Sigma_{0}\,{\rm exp}^{-(R-R_{\rm cres})^2/2\sigma_{R_{\rm cres}}^2} {\rm exp}^{-(\theta-\theta_{\rm cres})^2/2\sigma_{\theta_{\rm cres}}^2}, 
 \label{eqn:cressurdens}
\end{equation}
where $R_{\rm ring}$ or $R_{\rm cres}$ refers to the radial location of the feature, and $\sigma_{\rm ring}$ or $\sigma_{R_{\rm cres}}$ stands for the radial width. For 
the crescent, $\theta_{\rm cres}$ and $\sigma_{\theta_{\rm cres}}$ are used to control the location and width in the azimuthal direction. The proportionality constant 
is set to be 3 times larger than that for the rings, under which the crescent can feature a high millimeter flux contrast to its neighboring ring. Table~\ref{tab:structures} 
lists the adopted parameter values. These values are chosen to be representative among the results that are derived by analyzing the brightness profiles of 
rings/crescents revealed by ALMA \citep[e.g.,][]{Long2018,Huang2018,Perez2018}. If all the features are included, integrating Eq.~\ref{eqn:ringsurdens} 
and \ref{eqn:cressurdens} results in the mass fractions of 8\%, 28\%, 50\%, and 14\% for the 1st, 2nd, 3rd ring and crescent, respectively. 

We gradually increase $M_{\rm dust}$, therefore to let the substructures turn to be optically thick. Dust opacities and settling approach are kept to the fiducial assumptions. 
The results for each scenario are presented in Figure~\ref{fig:ringres}. We found that the tendencies of $F_{1.3\rm{mm}}$, $T_{\rm dust}$, and the overall level of $\Lambda_{1.3\rm{mm}}$  
are generally similar to those for smooth disks. The maximum underestimation is achieved by only introducing the innermost ring. In this case, dust grains are confined within the 
smallest disk radius. When the outer substructures are added, the underestimation becomes less severe. Such an effect is mainly driven by the actual $R_{\rm out}$. 
In other words, $R_{\rm out}$ has a stronger influence on the mass underestimation than the presence/absence of substructures. In order to testify the hypothesis, 
we run additional simulations for smooth disks. The surface densities for $R\,{\le}\,R_f$ are all fixed to the surface density at $R\,{=}\,R_f$, where $R_f$ is the location 
of the outermost substructure (e.g., $R_f\,{=}\,65\,\rm{AU}$ for the case of three rings, see Table~\ref{tab:structures}), and for $R\,{>}\,R_f$ we set the surface densities 
according to Eq.~\ref{eqn:ringsurdens}. In the case of three rings plus a crescent, we set the surface densities for $R\,{>}\,R_f$ according to an azimuthally averaged profile 
described by a combination of Eq.~\ref{eqn:ringsurdens} and Eq.~\ref{eqn:cressurdens}. Under such a circumstance, the dust surface density does not vary within 
$R_f$ (i.e., no rings/gaps/crescent), and follows a same profile to that of the disk with substructures in the outermost region, which would lead to an equivalent $R_{\rm out}$. 
The green filled circles shown in the right column of Figure~\ref{fig:ringres} refer to the underestimation factors, which are similar to those of the disks with substructures.

\section{Application to the DoAr\,33 disk}
\label{sec:appdoar33}

We have shown that due to the optical depth effect, calculating disk dust masses using the analytic approach can lead to substantial underestimation, even if the 
dust temperature ($T_{\rm dust}$) and opacity ($\kappa$) are known. The disk outer radius ($R_{\rm out}$), inclination ($i$) and true dust mass ($M_{\rm dust}$) 
are most important to create such optically thick regions. Fortunately, $R_{\rm out}$ and $i$ can be derived from high resolution (sub-)millimeter images that 
are nowadays available for a large number of disks thanks to ALMA. In this section, we conduct a detailed radiative transfer modeling of the SED for DoAr\,33 
in order to constrain the total dust mass in the disk, and then compare the result with the analytic dust mass.  

\begin{table}[!t]
 \centering
 \caption{The photometry of DoAr\,33.}	
     \begin{tabular}[h]{lccc}
     \hline
     $\lambda$ [$\mu{\rm m}$]  &  $F_{\nu}$ [mJy]   &  Instrument / Filter &  Reference \\
     \hline
      0.505     &  4.5    $\pm$ 0.039    & GAIA/bp     &  1    \\
      0.623     &  16.1   $\pm$ 0.039    & GAIA/g      &  1    \\
      0.773     &  38.8   $\pm$ 0.239    & GAIA/rp     &  1    \\
      0.64      &  15.5   $\pm$ 1.548    & R           &  2    \\
      0.79      &  37.7   $\pm$ 3.765    & I           &  2    \\
      1.235     &  175    $\pm$ 3.88     & 2MASS/J     &  3    \\
      1.662     &  332    $\pm$ 11       & 2MASS/H     &  3    \\
      2.159     &  348    $\pm$ 5.5      & 2MASS/K     &  3    \\
      3.55      &  212    $\pm$ 15.3     & Spitzer/IRAC\,1 &  3    \\
      4.493     &  169    $\pm$ 12.3     & Spitzer/IRAC\,2 &  3    \\
      5.731     &  193    $\pm$ 11.9     & Spitzer/IRAC\,3 &  3    \\
      7.872     &  222    $\pm$ 11.7     & Spitzer/IRAC\,4 &  3    \\
      12        &  200.4  $\pm$ 3.8      & WISE\,3         &  4    \\
      22        &  225    $\pm$ 9.5      & WISE\,4         &  4    \\
      23.68     &  211    $\pm$ 19.5     & Spitzer/MIPS\,1 &  3    \\
      70        &  469    $\pm$ 33       & Herschel/PACS   &  this work  \\
      100       &  537    $\pm$ 38       & Herschel/PACS   &  this work  \\
      160       &  539    $\pm$ 38       & Herschel/PACS   &  this work  \\
      250       &  416    $\pm$ 23       & Herschel/SPIRE  &  this work  \\
      350       &  260    $\pm$ 14       & Herschel/SPIRE  &  this work  \\
      350       &  212    $\pm$ 21.2     & CSO/SHARC\,II   &  this work  \\
      850       &  79     $\pm$ 7        & JCMT            &  5  \\
      880       &  80     $\pm$ 2        & SMA             &  6  \\
      870       &  76.4   $\pm$ 1.4      & ALMA            &  7  \\
      1254      &  35     $\pm$ 0.017    & ALMA            &  8  \\
      1300      &  33.6   $\pm$ 0.47     & ALMA            &  3  \\
      3300      &  3.7    $\pm$ 0.3      & ATCA            &  9  \\
      \hline
    \end{tabular}
	\tablefoot{\small References: (1) \citet{Gaia2018}; (2) \citet{Wilking2005}; (3) \citet{Cieza2019}; (4) \citet{Cutri2013};  (5) \citet{Andrews2007}; (6) \citet{Andrews2010}; (7) \citet{Cox2017}; (8) \citet{Andrews2018}; (9) \citet{Ricci2010}}
\label{tab:csoflux}
\end{table}

\subsection{DoAr\,33 and its SED}

\begin{figure}[!t]
\centering
  \includegraphics[width=3.0in]{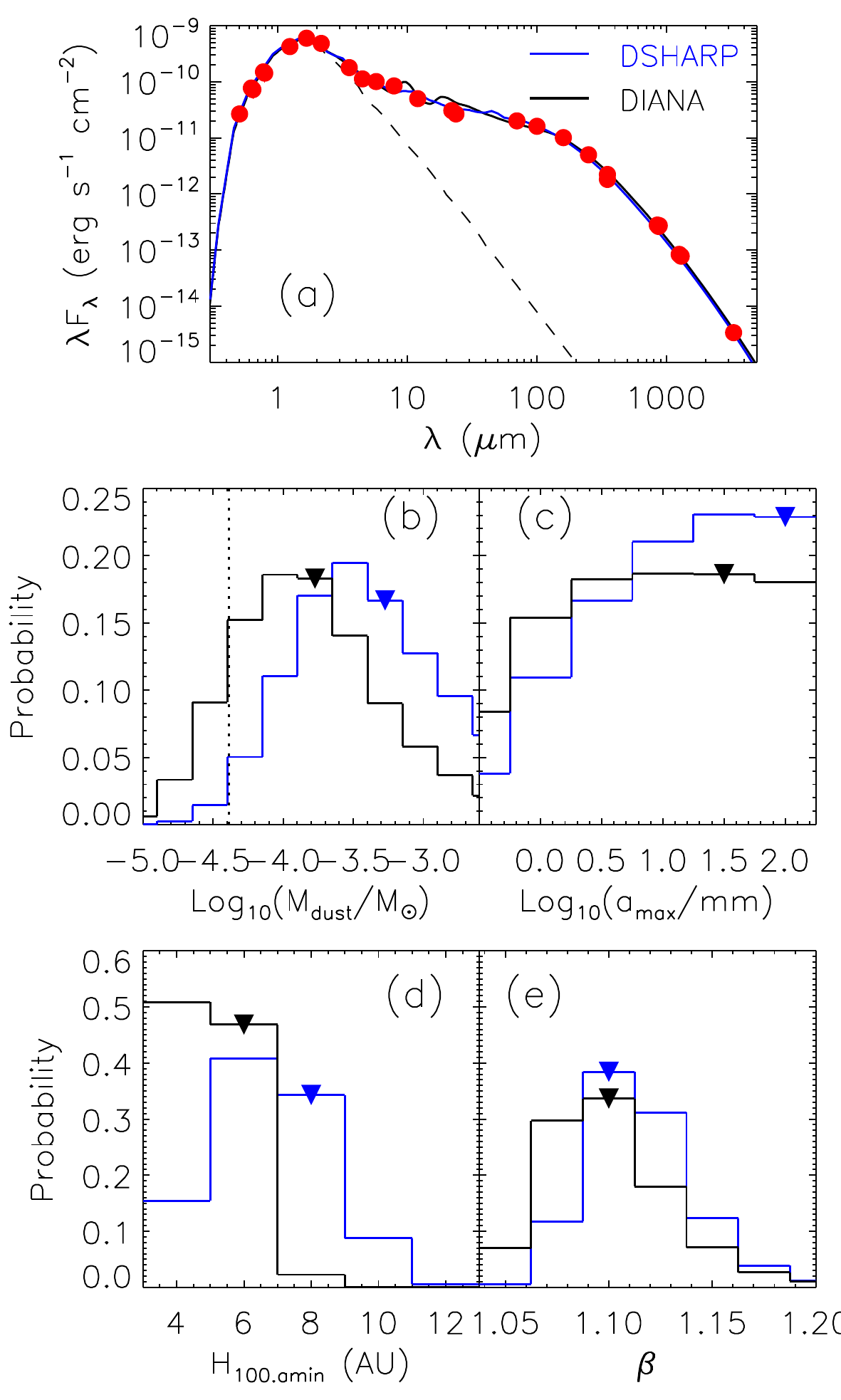}
  \caption{{\it Panel (a):} SEDs of DoAr\,33. The best-fit radiative transfer models using the DSHARP and DIANA opacities 
          are shown with blue and black solid lines, respectively. The dashed line refers to the input photospheric spectrum, while observational 
		  data points are overlaid with red dots. {\it Panels (b)$-$(e):} Bayesian probability distributions for ${\rm Log}_{10}(M_{\rm dust}/M_{\odot})$, 
		  ${\rm Log}_{10}(a_{\rm max}/{\rm mm})$, $H_{100.a_{\rm min}}$ and $\beta$. The triangles indicate the parameter values of the best-fit radiative 
		  transfer model. The vertical dotted line in panel (b) marks the analytic dust mass $M_{\rm dust.ana}$ that is derived by using the flux density 
		  measured at 1.3\,mm, $T_{\rm dust}\,{=}\,24.8\,\rm{K}$, and $\kappa_{1.3\rm{mm}}\,{=}\,2.3\,\rm{cm^2/g}$, see Sect.~\ref{sec:doar33res}.}
\label{fig:doar33sed}
\end{figure}

DoAr\,33 is a T Tauri star located in the Ophiuchus star formation region at a distance of 140\,pc \citep{Gaia2018}. The spectral type K5.5 derived 
by \citet{Wilking2005} translates to an effective temperature of $4160\,\rm{K}$ using the $\rm{SpT-T_{\rm{eff}}}$ conversion reported by \citet{Herczeg2014}. 
To compile the SED, on one hand we collected data points at various wavelengths from the literature. On the other hand, through the Chinese Telescope 
Access Program, we conducted new CSO/SHARC\,II observations at $350\,\mu{\rm m}$, see Sect.~\ref{sec:cso} in the Appendix for details about the observations and 
data reduction. Herschel photometry for DoAr\,33 in the literature show a surprisingly steep decline toward submillimeter wavelengths \citep{2015A&A...581A..30R}. 
Therefore, we re-reduced the Herschel/PACS and SPIRE data, and performed the photometry based on a point-spread function fitting, see Sect.~\ref{sec:herschel}.
These efforts allow us to build the SED with an excellent wavelength coverage. The measurements from optical to millimeter domains are summarized 
in Table~\ref{tab:csoflux}, and shown with red dots in panel (a) of Figure~\ref{fig:doar33sed}. Fitting the optical photometry with the \texttt{Kurucz} 
atmosphere models yields a stellar luminosity of $L_{\star}\,{=}\,0.97\,L_{\odot}$.

\subsection{A large grid of radiative transfer model SEDs}
\label{sec:doar33grid} 
 
As one of the 20 disks selected in the DSHARP project, DoAr\,33 was observed by ALMA with an angular resolution of $\sim$\,35\,mas \citep{Andrews2018}. The disk is 
inclined by $i\,{=}\,41.8^{\circ}$, and appears smooth in the 1.3\,mm continuum image. \citet{Huang2018} defined the outer boundary of the millimeter disk to be the 
radius at which the enclosed flux is equal to 95\% of the total flux. That is 27\,AU, making DoAr\,33 to be the smallest disk among the 18 single-star disk systems 
targeted by DSHARP. We checked the $^{12}{\rm CO}\,J\,{=}\,2{-}1$ line data, and found that the emission of $^{12}{\rm CO}$ is more extended, with a 95\% enclosing 
flux radius of $70\,\rm{AU}$. A more compact millimeter dust disk than the CO disk has also been found for other targets \citep[e.g.,][]{Ansdell2018,Long2022}, and 
can be naturally explained by radial drift of dust particles \citep{Trapman2019,Toci2021}.

We built a large grid of radiative transfer models for DoAr\,33. The model configuration is a slight variant of the fiducial setup. Dust grains with sizes ranging 
from $a_{\rm min}\,{=}\,0.01\,\mu\rm{m}$ to $10\,\mu\rm{m}$ are assumed to be distributed from 0.1\,AU to 70\,AU. Larger dust grains (i.e., $a\,{>}\,10\,\mu{\rm m}$) 
have already drifted to the location where the DSHARP continuum data probes, and therefore we truncated them at $R\,{=}\,27\,\rm{AU}$. For the most interesting 
parameter $M_{\rm dust}$, we sampled 13 grid points that are logarithmically distributed from $3\,{\times}\,10^{-6}$ to $3\,{\times}\,10^{-3}\,M_{\odot}$. 
For $a_{\rm max}$, we took 7 logarithmically spaced points from 0.1\,mm to 100\,mm. There are 8 points for $H_{100.a_{\rm min}}$ linearly spaced between 4 and 18\,AU, 
and 9 linear points for $\beta$ from 1.05 to 1.25. For the power-law index ($\gamma$) of the surface density, we chose a value of $0.5$ which is close to the slope 
of a power law used to fit the azimuthally-averaged surface brightness of the DSHARP image. Parameters not mentioned here were fixed to the fiducial values, see 
Table~\ref{tab:fiducial}. In total, there are 6,552 models in the grid, and we separately ran the simulation with the DSHARP and DIANA dust opacities.   

\subsection{Result and discussion}
\label{sec:doar33res}
The best-fit model SEDs, identified as the ones with the lowest reduced $\chi^2$, are shown in panel (a) of Figure~\ref{fig:doar33sed}. Following the procedure  
described in \citet{Pinte2008}, a Bayesian analysis is conducted by using the 6,552 SED models and assuming flat priors because there is no preliminary information 
about the parameters. The resulting marginalized probability distributions are presented in panels (b)$-$(e) of Figure~\ref{fig:doar33sed}. The triangles indicate 
the parameter values of the best-fit radiative transfer model, all of which are less than one bin away from the most probable values. The relatively wide distribution of 
the probabilities, especially for $a_{\rm max}$, implies that there are degeneracies between model parameters in fitting the SED. The most probable dust masses are $M_{\rm dust.RT}\,{=}\,3\,{\times}\,10^{-4}\,M_{\odot}$ ($9.5\,{\times}\,10^{-5}$, $1.7\,{\times}\,10^{-3}$) and $9.5\,{\times}\,10^{-5}\,M_{\odot}$ ($5.3\,{\times}\,10^{-5}$, $5.3\,{\times}\,10^{-4}$) when the DSHARP and DIANA dust opacities are used, respectively, where the numbers given in parenthesis correspond to the 68\% confidence interval. 
To calculate the analytic dust mass $M_{\rm dust.ana}$, we adopted $\kappa_{1.3\rm{mm}}\,{=}\,2.3\,\rm{cm^2/g}$ and $T_{\rm dust}\,{=}\,25\,(L_{\star}/L_{\odot})^{0.25}\,{=}\,24.8\,\rm{K}$ for consistency with literature studies. We took the integrated flux of $35\,\rm{mJy}$ from the DSHARP project (see Table~\ref{tab:csoflux}), and derived $M_{\rm dust.ana}\,{=}\,4.1\,{\times}\,10^{-5}\,M_{\odot}$ that is indicated with the vertical dotted line in panel (b) of Figure~\ref{fig:doar33sed}. Dust masses from 
radiative transfer modeling are 7.3 and 2.3 times higher than the analytic dust mass, when the DSHARP and DIANA dust opacities are adopted in the simulation, respectively. 

\begin{figure}[!t]
\centering
  \includegraphics[width=0.47\textwidth]{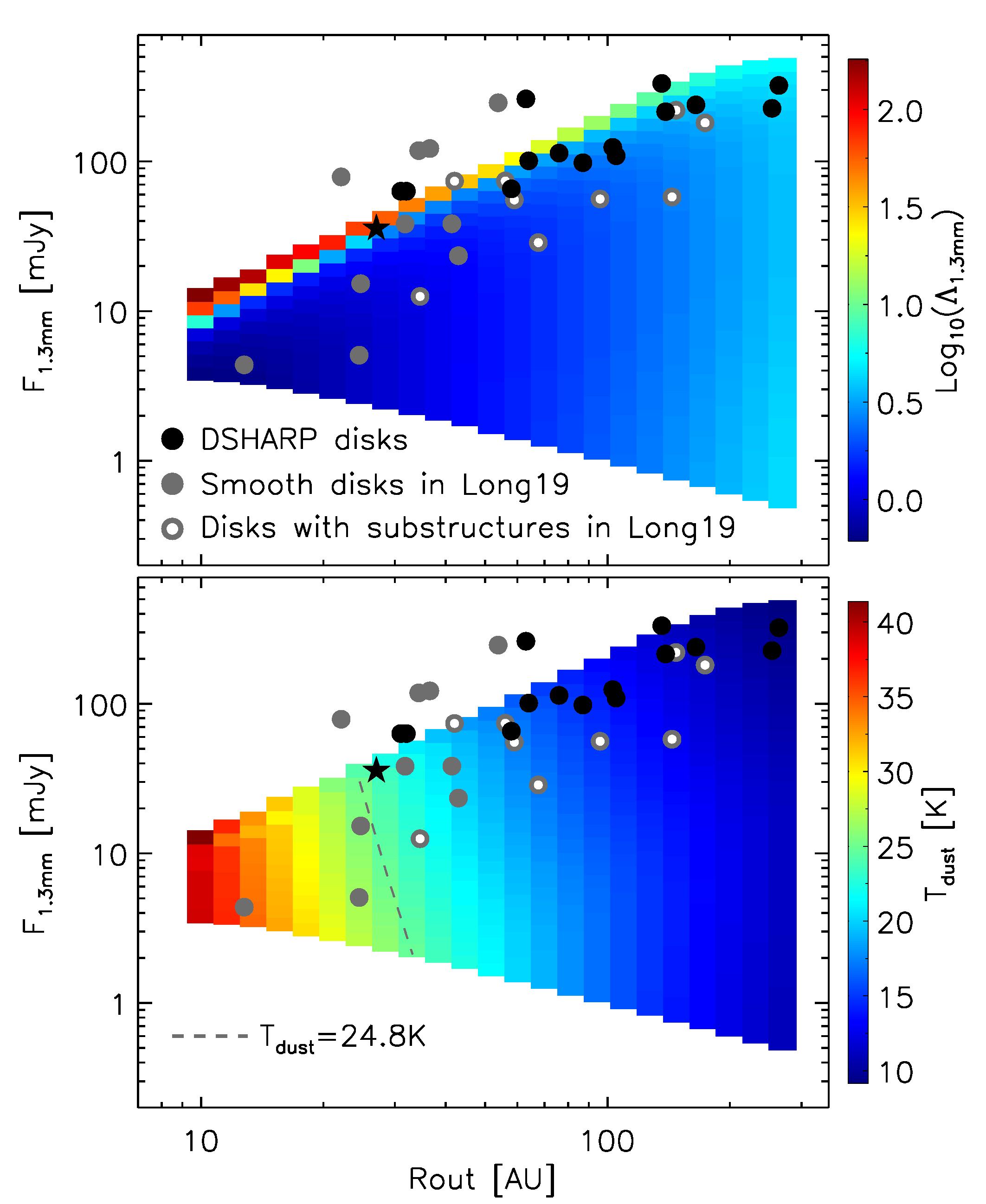}
  \caption{ {\it Upper panel:} $\Lambda_{\rm 1.3mm}$ as a function of $F_{\rm 1.3mm}$ and $R_{\rm out}$. Black dots stand for the DSHARP disks. 
   The black star represents the DoAr\,33 disk. Smooth disks and disks with substructures observed in \citet{Long2019} are indicated with grey dots and 
   grey rings, respectively. All the flux densities are scaled to the distance of 140\,pc. {\it Bottom panel: } $T_{\rm dust}$ distribution in 
   the $F_{\rm 1.3mm}\,{-}\,R_{\rm out}$ space. The dashed line draws the contour level of 24.8\,K, the analytic temperature simply given 
   by $25\,(L_{\star}/L_{\odot})^{0.25}\,\rm{K}$.}
\label{fig:ratioapp}
\end{figure}

Taking the most probable model with the inclusion of the DSHARP opacity as an illustration, regions below the $\tau_{1.3\rm{mm}}\,{=}\,1$ surface occupy ${\sim}\,70$\% of the total mass, which partly explains the large difference between $M_{\rm dust.ana}$ and $M_{\rm dust.RT}$. The mass-averaged dust temperature is $T_{\rm dust}\,{=}\,19.6\,\rm{K}$ that is cooler than the analytic dust temperature 24.8\,K. This also contributes to the mass difference. To reproduce the shallow millimeter spectral slope $\alpha_{\rm mm}\,{=}\,2.1$, the most probable maximum grain size achieves $a_{\rm max}\,{=}\,32\,\rm{mm}$, yielding an absorption coefficient of $\overline{\kappa_{1.3\rm{mm}}}\,{=}\,0.7\,\rm{cm^2/g}$. The discrepancy in $\kappa_{1.3\rm{mm}}$ accounts for a factor of $2.3/0.7\,{=}\,3.3$ difference in the dust mass. The millimeter spectral slope of DoAr\,33 is typically found in other disks around both 
brown dwarfs (BDs) and T Tauri/Herbig stars \citep[e.g.,][]{Ricci2010,Ricci2014,Pinilla2017}, implying that dust grains have grown to millimeter sizes at least. 
A multi-wavelength analysis, such as self-consistent radiative transfer modeling implemented in this study, is necessary to recover the grain size distribution, which is 
very important to characterize dust emissivities, and therefore obtain a reliable determination of dust masses. This information is basically missing in the 
traditionally analytic approach.  

The large underestimation of the dust mass obtained for DoAr\,33 may not be a peculiarity of that source.. With a thorough analysis of millimeter 
observations, \citet{Macias2021} recovered $4.5\,{-}\,5.9$ times more mass for the TW\,Hydrae disk than with the usual analytic methodology. 
Similar underestimation factors have also been found for some disks in the Taurus star formation region \citep{Ballering2019, Ribas2020}. 
To give a rough assessment on how frequently a large underestimation would be encountered, Figure~\ref{fig:ratioapp} shows $\Lambda_{\rm 1.3mm}$ 
and $T_{\rm dust}$ as a function of $F_{\rm 1.3mm}$ and $R_{\rm out}$. When making the plot, we include models in which only $M_{\rm dust}$ and 
$R_{\rm out}$ are varied, because as discussed in Sect.~\ref{sec:effpara} they have the most significant impact on the underestimation 
factor. Other parameters including the disk inclination are fixed to their fiducial values given in Table \ref{tab:fiducial}.
Disks from the DSHARP program and ALMA survey conducted by \citet{Long2019} are overlaid. We excluded disks in multiple systems and disks around Herbig stars. 
For simplicity, we directly took the 95\% enclosing flux radius from the literature to be $R_{\rm out}$. A majority of disks are located in parameter spaces that 
yield a relatively large underestimation (e.g., $\Lambda_{\rm 1.3mm}\,{\gtrsim}\,3$). There are few outliers even brighter than model predictions at the highest level. 
The maximum dust mass sampled in our grid is $M_{\rm dust}\,{=}\,3\,{\times}\,10^{-3}\,M_{\odot}$. Under the 2D density distribution describe 
in Sect.~\ref{sec:fiddens}, such a large amount of material concentrated in a narrow radius range makes the disk quite optically thick, any further increase 
in $M_{\rm dust}$ does not contribute much more to the emergent flux. However, stellar properties, power-law index for the dust surface density, flaring index,  
scale height, and degree of dust settling are all fixed in generating the plot. A combination of adjustments to (some of) these assumptions is able to induce 
a higher dust temperature $T_{\rm dust}$. Alternatively, lowering the disk inclination would increase the solid angle towards the observer. Both possibilities 
can bring the millimeter flux up to the observed level.

Correcting for the underestimation on disk dust masses will certainly ease the mass budget problem for planet formation \citep[e.g.,][]{Ueda2020,Andrews2020}. 
It is also expected to have an impact on the $M_{\rm dust}\,{-}\,M_{\star}$ relation. Millimeter surveys for protoplanetary disks have found that the measured 
fluxes strongly correlate to the stellar masses $F_{\rm mm}\,{\propto}\,M_{\star}^{\,{\sim}1.7}$ \citep[e.g.,][]{Andrews2013,Pascucci2016,Andrews2020}. 
A correlation between the dust mass and stellar mass will be naturally identified, when converting $F_{\rm mm}(s)$ to dust masses with Eq.~\ref{eqn:mana} and 
using a constant absorption coefficient $\kappa_{\rm mm}$ and dust temperature $T_{\rm dust}$. If the measured fluxes fully reflect the total amount of dust 
material (i.e., not considering the effects of $\kappa_{\rm mm}$ and $T_{\rm dust}$), T Tauri/Herbig disks are expected to be more massive than BD disks, because 
of the $F_{\rm mm}\,{-}\,M_{\star}$ relation. This means that the optical depth of T Tauri/Herbig disks is systematically higher than for BD disks if all disks 
have similar sizes. In such a circumstance, corrections for true dust masses are larger in disks around higher mass stars, which in principle 
will steepen the $M_{\rm dust}\,{-}\,M_{\star}$ relation. However, the size of the millimeter disk $R_{\rm disk.mm}$ has been found to scale with the stellar 
mass $R_{\rm disk.mm}\,{\propto}\,M_{\star}^{0.6}$, though such a relation is less pronounced \citep[][]{Andrews2018b,Andrews2020}, and whether or not it  
extends down to the BD regime needs more spatially resolved observations for further confirmation \citep[e.g.,][]{Ricci2013,Ricci2014,Testi2016}. As shown in 
Sect~\ref{sec:effpara}, the disk size significantly affects the mass derivation, with more severe underestimation for smaller disks. In this regard, correction factors 
for the analytic dust masses of BD disks are higher than those for their higher mass counterparts, which conversely will shallow the $M_{\rm dust}\,{-}\,M_{\star}$ relation. 
Moreover, there is a large scatter on the millimeter spectral slope $\alpha_{\rm mm}$ for each individual stellar mass bin (see for instance Figure 2 in \citet{Pinilla2017} 
and Figure 9 in \citet{Andrews2020}), pointing to inhomogeneities of the dust properties and/or a dispersion in optical depths. Given the complexity and parameter 
coupling, a dedicated investigation on the relation between $M_{\rm dust}$ and $M_{\star}$, and its evolution with time \citep{Pascucci2016,Rilinger2021}, requires a 
homogeneous radiative transfer analysis. 

Spatially resolved observations at various wavelengths provide the radial and vertical brightness distributions, which are directly linked to the product of the 
dust emissivity and density distribution at different locations in the disk. Therefore, spatially resolved data, when available, should be taken into 
account \citep[e.g.,][]{Grafe2013,Menu2014,Pinte2016,Liuy2017}. If the disk is completely optically thick at all the observed millimeter wavelengths, having 
multi-wavelength observations would still not be enough to provide a good measurement of the dust mass. Observations at longer wavelengths (even with the Very
Large Array) are also necessary in this situation \citep[e.g.,][]{Macias2018,CarrascoGonzalez2019,Macias2021,Guidi2022}.
 
\section{Summary}
\label{sec:summary}

Dust mass in protoplanetary disks is an important parameter characterizing the potential for planet formation. Literature studies usually derive dust masses using an
analytic approach based on the optically thin assumption. Statistic analysis of dust masses obtained in this way reveals that only few disks are able to form our solar 
system or its analogs. However, protoplanetary disks, particularly in the inner regions, are probably optically thick, which might lead to substantial underestimation 
on the true dust mass. 

Using self-consistent radiative transfer models, we conducted a detailed parameter study to investigate to which degree dust masses can be underestimated, and how the 
underestimation is influenced by disk and dust properties. Our results show that mass underestimations can be a few times up to hundreds depending on the optical depth 
along the line of sight. The most significant impacts on the underestimation are produced by the disk outer radius, inclination and the true dust mass. We also compared 
models with different dust settling approaches that control the millimeter dust scale height. The results show that the underestimation is not significantly affected, but the 
well-mixed model produces the strongest millimeter emisson, and consequently it needs the lowest $M_{\rm dust}$ to fit the data. Different dust scattering modes are also 
tested, but the differences in the millimeter flux, dust temperature, and therefore the underestimation are generally small. Given the prevalence of disk substructures 
revealed by ALMA, we also included these small-scale features in the models. The results show that their impacts on the mass underestimation is weaker than that induced 
by the disk outer radius and true dust mass. 

As an application, we also conducted a detailed SED modeling for $\rm{DoAr\,33}$ that is one of the 20 disks observed by the DSHARP project. In the modeling 
procedure, we fixed the disk outer radius and disk inclination to the constraints set by the ALMA observations. The most probable dust masses are 7.3 and 2.3 times higher 
than the analytic dust mass, when the DSHARP and DIANA dust opacities are adopted in the radiative transfer simulation, respectively. 

A homogeneous radiative transfer modeling is a more appropriate way to determine the disk dust mass, and investigate its dependence on the host stellar properties. 
In addition, in the analysis, multi-wavelength spatially resolved data are required for better reliability.

\begin{acknowledgements}
We thank the anonymous referee for the constructive comments that highly improved the manuscript. Y.L. acknowledges the financial support by the National Natural Science 
Foundation of China (Grant No. 11973090), and the science research grants from the China Manned Space Project with NO. CMS-CSST-2021-B06. T.H. acknowledges support 
from the European Research Council under the Horizon 2020 Framework Program via the ERC Advanced Grant Origins 832428. M.F. acknowledges funding from the European 
Research Council (ERC) under the European Union's Horizon 2020 research and innovation program (grant agreement No. 757957). G.R. acknowledges support from the 
Netherlands Organisation for Scientific Research (NWO, program number 016.Veni.192.233) and from an STFC Ernest Rutherford Fellowship (grant number ST/T003855/1). 
H.W. acknowledges the support by NSFC grant 11973091. 

This publication makes use of data products from the {\it Wide-field Infrared Survey Explorer}, which is a joint project of the University of California, Los 
Angeles, and the Jet Propulsion Laboratory/California Institute of Technology, funded by the National Aeronautics and Space Administration.

{\it PACS} has been developed by a consortium of institutes led by MPE (Germany) and including UVIE (Austria); KU Leuven, CSL, IMEC (Belgium); CEA, LAM (France); MPIA (Germany); INAF-IFSI/OAA/OAP/OAT, LENS, SISSA (Italy); IAC (Spain). This development has been supported by the funding agencies BMVIT (Austria), ESA-PRODEX (Belgium), CEA/CNES (France), DLR (Germany), ASI/INAF (Italy), and CICYT/MCYT (Spain). {\it SPIRE} has been developed by a consortium of institutes led by Cardiff University (UK) and including Univ. Lethbridge (Canada); NAOC (China); CEA, LAM (France); IFSI, Univ. Padua (Italy); IAC (Spain); Stockholm Observatory (Sweden); Imperial College London, RAL, UCL-MSSL, UKATC, Univ. Sussex (UK); and Caltech, JPL, NHSC, Univ. Colorado (USA). This development has been supported by national funding agencies: CSA (Canada); NAOC (China); CEA, CNES, CNRS (France); ASI (Italy); MCINN (Spain); SNSB (Sweden); STFC, UKSA (UK); and NASA (USA).

ALMA is a partnership of ESO (representing its member states), NSF (USA), and NINS (Japan), together with 
NRC (Canada), MOST and ASIAA (Taiwan), and KASI (Republic of Korea), in cooperation with the Republic of Chile. The Joint ALMA Observatory is operated 
by ESO, AUI/NRAO, and NAOJ.
\end{acknowledgements}

\bibliographystyle{aa}
\bibliography{dustmass.bib}

\begin{thebibliography}{124}
\expandafter\ifx\csname natexlab\endcsname\relax\def\natexlab#1{#1}\fi

\bibitem[{{ALMA Partnership} {et~al.}(2015){ALMA Partnership}, {Brogan},
  {P{\'e}rez}, {Hunter}, {Dent}, {Hales}, {Hills}, {Corder}, {Fomalont},
  {Vlahakis}, {Asaki}, {Barkats}, {Hirota}, {Hodge}, {Impellizzeri}, {Kneissl},
  {Liuzzo}, {Lucas}, {Marcelino}, {Matsushita}, {Nakanishi}, {Phillips},
  {Richards}, {Toledo}, {Aladro}, {Broguiere}, {Cortes}, {Cortes}, {Espada},
  {Galarza}, {Garcia-Appadoo}, {Guzman-Ramirez}, {Humphreys}, {Jung}, {Kameno},
  {Laing}, {Leon}, {Marconi}, {Mignano}, {Nikolic}, {Nyman}, {Radiszcz},
  {Remijan}, {Rod{\'o}n}, {Sawada}, {Takahashi}, {Tilanus}, {Vila Vilaro},
  {Watson}, {Wiklind}, {Akiyama}, {Chapillon}, {de Gregorio-Monsalvo}, {Di
  Francesco}, {Gueth}, {Kawamura}, {Lee}, {Nguyen Luong}, {Mangum}, {Pietu},
  {Sanhueza}, {Saigo}, {Takakuwa}, {Ubach}, {van Kempen}, {Wootten},
  {Castro-Carrizo}, {Francke}, {Gallardo}, {Garcia}, {Gonzalez}, {Hill},
  {Kaminski}, {Kurono}, {Liu}, {Lopez}, {Morales}, {Plarre}, {Schieven},
  {Testi}, {Videla}, {Villard}, {Andreani}, {Hibbard}, \&
  {Tatematsu}}]{alma2015}
{ALMA Partnership}, {Brogan}, C.~L., {P{\'e}rez}, L.~M., {et~al.} 2015, \apjl,
  808, L3

\bibitem[{{Andr{\'e}} {et~al.}(2010){Andr{\'e}}, {Men'shchikov}, {Bontemps},
  {K{\"o}nyves}, {Motte}, {Schneider}, {Didelon}, {Minier}, {Saraceno},
  {Ward-Thompson}, {di Francesco}, {White}, {Molinari}, {Testi}, {Abergel},
  {Griffin}, {Henning}, {Royer}, {Mer{\'{\i}}n}, {Vavrek}, {Attard},
  {Arzoumanian}, {Wilson}, {Ade}, {Aussel}, {Baluteau}, {Benedettini},
  {Bernard}, {Blommaert}, {Cambr{\'e}sy}, {Cox}, {di Giorgio}, {Hargrave},
  {Hennemann}, {Huang}, {Kirk}, {Krause}, {Launhardt}, {Leeks}, {Le Pennec},
  {Li}, {Martin}, {Maury}, {Olofsson}, {Omont}, {Peretto}, {Pezzuto}, {Prusti},
  {Roussel}, {Russeil}, {Sauvage}, {Sibthorpe}, {Sicilia-Aguilar}, {Spinoglio},
  {Waelkens}, {Woodcraft}, \& {Zavagno}}]{andre2010}
{Andr{\'e}}, P., {Men'shchikov}, A., {Bontemps}, S., {et~al.} 2010, \aap, 518,
  L102

\bibitem[{{Andrews}(2020)}]{Andrews2020}
{Andrews}, S.~M. 2020, \araa, 58, 483

\bibitem[{{Andrews} {et~al.}(2018{\natexlab{a}}){Andrews}, {Huang},
  {P{\'e}rez}, {Isella}, {Dullemond}, {Kurtovic}, {Guzm{\'a}n}, {Carpenter},
  {Wilner}, {Zhang}, {Zhu}, {Birnstiel}, {Bai}, {Benisty}, {Hughes},
  {{\"O}berg}, \& {Ricci}}]{Andrews2018}
{Andrews}, S.~M., {Huang}, J., {P{\'e}rez}, L.~M., {et~al.} 2018{\natexlab{a}},
  \apjl, 869, L41

\bibitem[{{Andrews} {et~al.}(2013){Andrews}, {Rosenfeld}, {Kraus}, \&
  {Wilner}}]{Andrews2013}
{Andrews}, S.~M., {Rosenfeld}, K.~A., {Kraus}, A.~L., \& {Wilner}, D.~J. 2013,
  \apj, 771, 129

\bibitem[{{Andrews} {et~al.}(2018{\natexlab{b}}){Andrews}, {Terrell},
  {Tripathi}, {Ansdell}, {Williams}, \& {Wilner}}]{Andrews2018b}
{Andrews}, S.~M., {Terrell}, M., {Tripathi}, A., {et~al.} 2018{\natexlab{b}},
  \apj, 865, 157

\bibitem[{{Andrews} \& {Williams}(2007)}]{Andrews2007}
{Andrews}, S.~M. \& {Williams}, J.~P. 2007, \apj, 671, 1800

\bibitem[{{Andrews} {et~al.}(2010){Andrews}, {Wilner}, {Hughes}, {Qi}, \&
  {Dullemond}}]{Andrews2010}
{Andrews}, S.~M., {Wilner}, D.~J., {Hughes}, A.~M., {Qi}, C., \& {Dullemond},
  C.~P. 2010, \apj, 723, 1241

\bibitem[{{Ansdell} {et~al.}(2018){Ansdell}, {Williams}, {Trapman}, {van
  Terwisga}, {Facchini}, {Manara}, {van der Marel}, {Miotello}, {Tazzari},
  {Hogerheijde}, {Guidi}, {Testi}, \& {van Dishoeck}}]{Ansdell2018}
{Ansdell}, M., {Williams}, J.~P., {Trapman}, L., {et~al.} 2018, \apj, 859, 21

\bibitem[{{Ansdell} {et~al.}(2016){Ansdell}, {Williams}, {van der Marel},
  {Carpenter}, {Guidi}, {Hogerheijde}, {Mathews}, {Manara}, {Miotello},
  {Natta}, {Oliveira}, {Tazzari}, {Testi}, {van Dishoeck}, \& {van
  Terwisga}}]{Ansdell2016}
{Ansdell}, M., {Williams}, J.~P., {van der Marel}, N., {et~al.} 2016, \apj,
  828, 46

\bibitem[{{Avenhaus} {et~al.}(2018){Avenhaus}, {Quanz}, {Garufi}, {Perez},
  {Casassus}, {Pinte}, {Bertrang}, {Caceres}, {Benisty}, \&
  {Dominik}}]{Avenhaus2018}
{Avenhaus}, H., {Quanz}, S.~P., {Garufi}, A., {et~al.} 2018, \apj, 863, 44

\bibitem[{{Ballering} \& {Eisner}(2019)}]{Ballering2019}
{Ballering}, N.~P. \& {Eisner}, J.~A. 2019, \aj, 157, 144

\bibitem[{{Balog} {et~al.}(2014){Balog}, {M{\"u}ller}, {Nielbock}, {Altieri},
  {Klaas}, {Blommaert}, {Linz}, {Lutz}, {Mo{\'o}r}, {Billot}, {Sauvage}, \&
  {Okumura}}]{2014ExA....37..129B}
{Balog}, Z., {M{\"u}ller}, T., {Nielbock}, M., {et~al.} 2014, Experimental
  Astronomy, 37, 129

\bibitem[{{Balog} {et~al.}(2008){Balog}, {Rieke}, {Muzerolle}, {Bally}, {Su},
  {Misselt}, \& {G{\'a}sp{\'a}r}}]{Balog2008}
{Balog}, Z., {Rieke}, G.~H., {Muzerolle}, J., {et~al.} 2008, \apj, 688, 408

\bibitem[{{Barenfeld} {et~al.}(2016){Barenfeld}, {Carpenter}, {Ricci}, \&
  {Isella}}]{Barenfeld2016}
{Barenfeld}, S.~A., {Carpenter}, J.~M., {Ricci}, L., \& {Isella}, A. 2016,
  \apj, 827, 142

\bibitem[{{Beckwith} {et~al.}(1990){Beckwith}, {Sargent}, {Chini}, \&
  {Guesten}}]{Beckwith1990}
{Beckwith}, S. V.~W., {Sargent}, A.~I., {Chini}, R.~S., \& {Guesten}, R. 1990,
  \aj, 99, 924

\bibitem[{{Bergin} {et~al.}(2013){Bergin}, {Cleeves}, {Gorti}, {Zhang},
  {Blake}, {Green}, {Andrews}, {Evans}, {Henning}, {{\"O}berg}, {Pontoppidan},
  {Qi}, {Salyk}, \& {van Dishoeck}}]{Bergin2013}
{Bergin}, E.~A., {Cleeves}, L.~I., {Gorti}, U., {et~al.} 2013, \nat, 493, 644

\bibitem[{{Bergin} \& {Williams}(2017)}]{Bergin2017}
{Bergin}, E.~A. \& {Williams}, J.~P. 2017, in Astrophysics and Space Science
  Library, Vol. 445, Formation, Evolution, and Dynamics of Young Solar Systems,
  ed. M.~{Pessah} \& O.~{Gressel}, 1

\bibitem[{{Beuther} \& {Steinacker}(2007)}]{2007ApJ...656L..85B}
{Beuther}, H. \& {Steinacker}, J. 2007, \apjl, 656, L85

\bibitem[{{Birnstiel} {et~al.}(2018){Birnstiel}, {Dullemond}, {Zhu}, {Andrews},
  {Bai}, {Wilner}, {Carpenter}, {Huang}, {Isella}, {Benisty}, {P{\'e}rez}, \&
  {Zhang}}]{Birnstiel2018}
{Birnstiel}, T., {Dullemond}, C.~P., {Zhu}, Z., {et~al.} 2018, \apjl, 869, L45

\bibitem[{{Birnstiel} {et~al.}(2016){Birnstiel}, {Fang}, \&
  {Johansen}}]{Birnstiel2016}
{Birnstiel}, T., {Fang}, M., \& {Johansen}, A. 2016, \ssr, 205, 41

\bibitem[{{Bruggeman}(1935)}]{Bruggeman1935}
{Bruggeman}, D. 1935, Ann. Phys., 416, 636

\bibitem[{{Carrasco-Gonz{\'a}lez} {et~al.}(2019){Carrasco-Gonz{\'a}lez},
  {Sierra}, {Flock}, {Zhu}, {Henning}, {Chandler}, {Galv{\'a}n-Madrid},
  {Mac{\'\i}as}, {Anglada}, {Linz}, {Osorio}, {Rodr{\'\i}guez}, {Testi},
  {Torrelles}, {P{\'e}rez}, \& {Liu}}]{CarrascoGonzalez2019}
{Carrasco-Gonz{\'a}lez}, C., {Sierra}, A., {Flock}, M., {et~al.} 2019, \apj,
  883, 71

\bibitem[{{Cazzoletti} {et~al.}(2019){Cazzoletti}, {Manara}, {Liu}, {van
  Dishoeck}, {Facchini}, {Alcal{\`a}}, {Ansdell}, {Testi}, {Williams},
  {Carrasco-Gonz{\'a}lez}, {Dong}, {Forbrich}, {Fukagawa}, {Galv{\'a}n-Madrid},
  {Hirano}, {Hogerheijde}, {Hasegawa}, {Muto}, {Pinilla}, {Takami}, {Tamura},
  {Tazzari}, \& {Wisniewski}}]{Cazzoletti2019}
{Cazzoletti}, P., {Manara}, C.~F., {Liu}, H.~B., {et~al.} 2019, \aap, 626, A11

\bibitem[{{Cazzoletti} {et~al.}(2018){Cazzoletti}, {van Dishoeck}, {Pinilla},
  {Tazzari}, {Facchini}, {van der Marel}, {Benisty}, {Garufi}, \&
  {P{\'e}rez}}]{Cazzoletti2018}
{Cazzoletti}, P., {van Dishoeck}, E.~F., {Pinilla}, P., {et~al.} 2018, \aap,
  619, A161

\bibitem[{{Cieza} {et~al.}(2021){Cieza}, {Gonz{\'a}lez-Ruilova}, {Hales},
  {Pinilla}, {Ru{\'\i}z-Rodr{\'\i}guez}, {Zurlo}, {Casassus}, {P{\'e}rez},
  {C{\'a}novas}, {Arce-Tord}, {Flock}, {Kurtovic}, {Marino}, {Nogueira},
  {Perez}, {Price}, {Principe}, \& {Williams}}]{Cieza2021}
{Cieza}, L.~A., {Gonz{\'a}lez-Ruilova}, C., {Hales}, A.~S., {et~al.} 2021,
  \mnras, 501, 2934

\bibitem[{{Cieza} {et~al.}(2019){Cieza}, {Ru{\'\i}z-Rodr{\'\i}guez}, {Hales},
  {Casassus}, {P{\'e}rez}, {Gonzalez-Ruilova}, {C{\'a}novas}, {Williams},
  {Zurlo}, {Ansdell}, {Avenhaus}, {Bayo}, {Bertrang}, {Christiaens}, {Dent},
  {Ferrero}, {Gamen}, {Olofsson}, {Orcajo}, {Pe{\~n}a Ram{\'\i}rez},
  {Principe}, {Schreiber}, \& {van der Plas}}]{Cieza2019}
{Cieza}, L.~A., {Ru{\'\i}z-Rodr{\'\i}guez}, D., {Hales}, A., {et~al.} 2019,
  \mnras, 482, 698

\bibitem[{{Cox} {et~al.}(2017){Cox}, {Harris}, {Looney}, {Chiang}, {Chandler},
  {Kratter}, {Li}, {Perez}, \& {Tobin}}]{Cox2017}
{Cox}, E.~G., {Harris}, R.~J., {Looney}, L.~W., {et~al.} 2017, \apj, 851, 83

\bibitem[{{Cutri} \& {et al.}(2013)}]{Cutri2013}
{Cutri}, R.~M. \& {et al.} 2013, VizieR Online Data Catalog, 2328

\bibitem[{{Daemgen} {et~al.}(2016){Daemgen}, {Natta}, {Scholz}, {Testi},
  {Jayawardhana}, {Greaves}, \& {Eastwood}}]{Daemgen2016}
{Daemgen}, S., {Natta}, A., {Scholz}, A., {et~al.} 2016, \aap, 594, A83

\bibitem[{{Diolaiti} {et~al.}(2000){Diolaiti}, {Bendinelli}, {Bonaccini},
  {Close}, {Currie}, \& {Parmeggiani}}]{2000A&AS..147..335D}
{Diolaiti}, E., {Bendinelli}, O., {Bonaccini}, D., {et~al.} 2000, \aaps, 147,
  335

\bibitem[{{Dong} {et~al.}(2018){Dong}, {Liu}, {Eisner}, {Andrews}, {Fung},
  {Zhu}, {Chiang}, {Hashimoto}, {Liu}, {Casassus}, {Esposito}, {Hasegawa},
  {Muto}, {Pavlyuchenkov}, {Wilner}, {Akiyama}, {Tamura}, \&
  {Wisniewski}}]{Dong2018}
{Dong}, R., {Liu}, S.-y., {Eisner}, J., {et~al.} 2018, \apj, 860, 124

\bibitem[{{Dorschner} {et~al.}(1995){Dorschner}, {Begemann}, {Henning},
  {Jaeger}, \& {Mutschke}}]{Dorschner1995}
{Dorschner}, J., {Begemann}, B., {Henning}, T., {Jaeger}, C., \& {Mutschke}, H.
  1995, \aap, 300, 503

\bibitem[{{Dowell} {et~al.}(2003){Dowell}, {Allen}, {Babu}, {Freund},
  {Gardner}, {Groseth}, {Jhabvala}, {Kovacs}, {Lis}, {Moseley}, {Phillips},
  {Silverberg}, {Voellmer}, \& {Yoshida}}]{dowell2003}
{Dowell}, C.~D., {Allen}, C.~A., {Babu}, R.~S., {et~al.} 2003, in \procspie,
  Vol. 4855, Millimeter and Submillimeter Detectors for Astronomy, ed. T.~G.
  {Phillips} \& J.~{Zmuidzinas}, 73--87

\bibitem[{{Drazkowska} {et~al.}(2022){Drazkowska}, {Bitsch}, {Lambrechts},
  {Mulders}, {Harsono}, {Vazan}, {Liu}, {Ormel}, {Kretke}, \&
  {Morbidelli}}]{Drazkowska2022}
{Drazkowska}, J., {Bitsch}, B., {Lambrechts}, M., {et~al.} 2022, arXiv
  e-prints, arXiv:2203.09759

\bibitem[{{Dubrulle} {et~al.}(1995){Dubrulle}, {Morfill}, \&
  {Sterzik}}]{Dubrulle1995}
{Dubrulle}, B., {Morfill}, G., \& {Sterzik}, M. 1995, \icarus, 114, 237

\bibitem[{{Dullemond} {et~al.}(2018){Dullemond}, {Birnstiel}, {Huang},
  {Kurtovic}, {Andrews}, {Guzm{\'a}n}, {P{\'e}rez}, {Isella}, {Zhu}, {Benisty},
  {Wilner}, {Bai}, {Carpenter}, {Zhang}, \& {Ricci}}]{Dullemond2018}
{Dullemond}, C.~P., {Birnstiel}, T., {Huang}, J., {et~al.} 2018, \apjl, 869,
  L46

\bibitem[{{Dullemond} \& {Dominik}(2004)}]{Dullemond2004}
{Dullemond}, C.~P. \& {Dominik}, C. 2004, \aap, 421, 1075

\bibitem[{{Dullemond} {et~al.}(2012){Dullemond}, {Juhasz}, {Pohl}, {Sereshti},
  {Shetty}, {Peters}, {Commercon}, \& {Flock}}]{radmc3d2012}
{Dullemond}, C.~P., {Juhasz}, A., {Pohl}, A., {et~al.} 2012, {RADMC-3D: A
  multi-purpose radiative transfer tool}, Astrophysics Source Code Library

\bibitem[{{Ercolano} \& {Pascucci}(2017)}]{Ercolano2017}
{Ercolano}, B. \& {Pascucci}, I. 2017, Royal Society Open Science, 4, 170114

\bibitem[{{Fedele} {et~al.}(2018){Fedele}, {Tazzari}, {Booth}, {Testi},
  {Clarke}, {Pascucci}, {Kospal}, {Semenov}, {Bruderer}, {Henning}, \&
  {Teague}}]{Fedele2018}
{Fedele}, D., {Tazzari}, M., {Booth}, R., {et~al.} 2018, \aap, 610, A24

\bibitem[{{Flaherty} {et~al.}(2020){Flaherty}, {Hughes}, {Simon}, {Qi}, {Bai},
  {Bulatek}, {Andrews}, {Wilner}, \& {K{\'o}sp{\'a}l}}]{Flaherty2020}
{Flaherty}, K., {Hughes}, A.~M., {Simon}, J.~B., {et~al.} 2020, \apj, 895, 109

\bibitem[{{Flaherty} {et~al.}(2017){Flaherty}, {Hughes}, {Rose}, {Simon}, {Qi},
  {Andrews}, {K{\'o}sp{\'a}l}, {Wilner}, {Chiang}, {Armitage}, \&
  {Bai}}]{Flaherty2017}
{Flaherty}, K.~M., {Hughes}, A.~M., {Rose}, S.~C., {et~al.} 2017, \apj, 843,
  150

\bibitem[{{Fruchter} \& {Hook}(2002)}]{2002PASP..114..144F}
{Fruchter}, A.~S. \& {Hook}, R.~N. 2002, \pasp, 114, 144

\bibitem[{{Gaia Collaboration} {et~al.}(2018){Gaia Collaboration}, {Brown},
  {Vallenari}, {Prusti}, {de Bruijne}, {Babusiaux}, {Bailer-Jones}, {Biermann},
  {Evans}, {Eyer}, {Jansen}, {Jordi}, {Klioner}, {Lammers}, {Lindegren},
  {Luri}, {Mignard}, {Panem}, {Pourbaix}, {Randich}, {Sartoretti}, {Siddiqui},
  {Soubiran}, {van Leeuwen}, {Walton}, {Arenou}, {Bastian}, {Cropper},
  {Drimmel}, {Katz}, {Lattanzi}, {Bakker}, {Cacciari}, {Casta{\~n}eda},
  {Chaoul}, {Cheek}, {De Angeli}, {Fabricius}, {Guerra}, {Holl}, {Masana},
  {Messineo}, {Mowlavi}, {Nienartowicz}, {Panuzzo}, {Portell}, {Riello},
  {Seabroke}, {Tanga}, {Th{\'e}venin}, {Gracia-Abril}, {Comoretto},
  {Garcia-Reinaldos}, {Teyssier}, {Altmann}, {Andrae}, {Audard},
  {Bellas-Velidis}, {Benson}, {Berthier}, {Blomme}, {Burgess}, {Busso},
  {Carry}, {Cellino}, {Clementini}, {Clotet}, {Creevey}, {Davidson}, {De
  Ridder}, {Delchambre}, {Dell'Oro}, {Ducourant},
  {Fern{\'a}ndez-Hern{\'a}ndez}, {Fouesneau}, {Fr{\'e}mat}, {Galluccio},
  {Garc{\'\i}a-Torres}, {Gonz{\'a}lez-N{\'u}{\~n}ez}, {Gonz{\'a}lez-Vidal},
  {Gosset}, {Guy}, {Halbwachs}, {Hambly}, {Harrison}, {Hern{\'a}ndez},
  {Hestroffer}, {Hodgkin}, {Hutton}, {Jasniewicz}, {Jean-Antoine-Piccolo},
  {Jordan}, {Korn}, {Krone-Martins}, {Lanzafame}, {Lebzelter}, {L{\"o}ffler},
  {Manteiga}, {Marrese}, {Mart{\'\i}n-Fleitas}, {Moitinho}, {Mora}, {Muinonen},
  {Osinde}, {Pancino}, {Pauwels}, {Petit}, {Recio-Blanco}, {Richards},
  {Rimoldini}, {Robin}, {Sarro}, {Siopis}, {Smith}, {Sozzetti}, {S{\"u}veges},
  {Torra}, {van Reeven}, {Abbas}, {Abreu Aramburu}, {Accart}, {Aerts},
  {Altavilla}, {{\'A}lvarez}, {Alvarez}, {Alves}, {Anderson}, {Andrei},
  {Anglada Varela}, {Antiche}, {Antoja}, {Arcay}, {Astraatmadja}, {Bach},
  {Baker}, {Balaguer-N{\'u}{\~n}ez}, {Balm}, {Barache}, {Barata}, {Barbato},
  {Barblan}, {Barklem}, {Barrado}, {Barros}, {Barstow}, {Bartholom{\'e}
  Mu{\~n}oz}, {Bassilana}, {Becciani}, {Bellazzini}, {Berihuete}, {Bertone},
  {Bianchi}, {Bienaym{\'e}}, {Blanco-Cuaresma}, {Boch}, {Boeche}, {Bombrun},
  {Borrachero}, {Bossini}, {Bouquillon}, {Bourda}, {Bragaglia}, {Bramante},
  {Breddels}, {Bressan}, {Brouillet}, {Br{\"u}semeister}, {Brugaletta},
  {Bucciarelli}, {Burlacu}, {Busonero}, {Butkevich}, {Buzzi}, {Caffau},
  {Cancelliere}, {Cannizzaro}, {Cantat-Gaudin}, {Carballo}, {Carlucci},
  {Carrasco}, {Casamiquela}, {Castellani}, {Castro-Ginard}, {Charlot},
  {Chemin}, {Chiavassa}, {Cocozza}, {Costigan}, {Cowell}, {Crifo}, {Crosta},
  {Crowley}, {Cuypers}, {Dafonte}, {Damerdji}, {Dapergolas}, {David}, {David},
  {de Laverny}, {De Luise}, {De March}, {de Martino}, {de Souza}, {de Torres},
  {Debosscher}, {del Pozo}, {Delbo}, {Delgado}, {Delgado}, {Di Matteo},
  {Diakite}, {Diener}, {Distefano}, {Dolding}, {Drazinos}, {Dur{\'a}n},
  {Edvardsson}, {Enke}, {Eriksson}, {Esquej}, {Eynard Bontemps}, {Fabre},
  {Fabrizio}, {Faigler}, {Falc{\~a}o}, {Farr{\`a}s Casas}, {Federici},
  {Fedorets}, {Fernique}, {Figueras}, {Filippi}, {Findeisen}, {Fonti},
  {Fraile}, {Fraser}, {Fr{\'e}zouls}, {Gai}, {Galleti}, {Garabato},
  {Garc{\'\i}a-Sedano}, {Garofalo}, {Garralda}, {Gavel}, {Gavras}, {Gerssen},
  {Geyer}, {Giacobbe}, {Gilmore}, {Girona}, {Giuffrida}, {Glass}, {Gomes},
  {Granvik}, {Gueguen}, {Guerrier}, {Guiraud}, {Guti{\'e}rrez-S{\'a}nchez},
  {Haigron}, {Hatzidimitriou}, {Hauser}, {Haywood}, {Heiter}, {Helmi}, {Heu},
  {Hilger}, {Hobbs}, {Hofmann}, {Holland}, {Huckle}, {Hypki}, {Icardi},
  {Jan{\ss}en}, {Jevardat de Fombelle}, {Jonker}, {Juh{\'a}sz}, {Julbe},
  {Karampelas}, {Kewley}, {Klar}, {Kochoska}, {Kohley}, {Kolenberg},
  {Kontizas}, {Kontizas}, {Koposov}, {Kordopatis}, {Kostrzewa-Rutkowska},
  {Koubsky}, {Lambert}, {Lanza}, {Lasne}, {Lavigne}, {Le Fustec}, {Le
  Poncin-Lafitte}, {Lebreton}, {Leccia}, {Leclerc}, {Lecoeur-Taibi},
  {Lenhardt}, {Leroux}, {Liao}, {Licata}, {Lindstr{\o}m}, {Lister}, {Livanou},
  {Lobel}, {L{\'o}pez}, {Managau}, {Mann}, {Mantelet}, {Marchal}, {Marchant},
  {Marconi}, {Marinoni}, {Marschalk{\'o}}, {Marshall}, {Martino}, {Marton},
  {Mary}, {Massari}, {Matijevi{\v{c}}}, {Mazeh}, {McMillan}, {Messina},
  {Michalik}, {Millar}, {Molina}, {Molinaro}, {Moln{\'a}r}, {Montegriffo},
  {Mor}, {Morbidelli}, {Morel}, {Morris}, {Mulone}, {Muraveva}, {Musella},
  {Nelemans}, {Nicastro}, {Noval}, {O'Mullane}, {Ord{\'e}novic},
  {Ord{\'o}{\~n}ez-Blanco}, {Osborne}, {Pagani}, {Pagano}, {Pailler},
  {Palacin}, {Palaversa}, {Panahi}, {Pawlak}, {Piersimoni}, {Pineau}, {Plachy},
  {Plum}, {Poggio}, {Poujoulet}, {Pr{\v{s}}a}, {Pulone}, {Racero}, {Ragaini},
  {Rambaux}, {Ramos-Lerate}, {Regibo}, {Reyl{\'e}}, {Riclet}, {Ripepi}, {Riva},
  {Rivard}, {Rixon}, {Roegiers}, {Roelens}, {Romero-G{\'o}mez}, {Rowell},
  {Royer}, {Ruiz-Dern}, {Sadowski}, {Sagrist{\`a} Sell{\'e}s}, {Sahlmann},
  {Salgado}, {Salguero}, {Sanna}, {Santana-Ros}, {Sarasso}, {Savietto},
  {Schultheis}, {Sciacca}, {Segol}, {Segovia}, {S{\'e}gransan}, {Shih},
  {Siltala}, {Silva}, {Smart}, {Smith}, {Solano}, {Solitro}, {Sordo}, {Soria
  Nieto}, {Souchay}, {Spagna}, {Spoto}, {Stampa}, {Steele},
  {Steidelm{\"u}ller}, {Stephenson}, {Stoev}, {Suess}, {Surdej}, {Szabados},
  {Szegedi-Elek}, {Tapiador}, {Taris}, {Tauran}, {Taylor}, {Teixeira},
  {Terrett}, {Teyssandier}, {Thuillot}, {Titarenko}, {Torra Clotet}, {Turon},
  {Ulla}, {Utrilla}, {Uzzi}, {Vaillant}, {Valentini}, {Valette}, {van Elteren},
  {Van Hemelryck}, {van Leeuwen}, {Vaschetto}, {Vecchiato}, {Veljanoski},
  {Viala}, {Vicente}, {Vogt}, {von Essen}, {Voss}, {Votruba}, {Voutsinas},
  {Walmsley}, {Weiler}, {Wertz}, {Wevers}, {Wyrzykowski}, {Yoldas},
  {{\v{Z}}erjal}, {Ziaeepour}, {Zorec}, {Zschocke}, {Zucker}, {Zurbach}, \&
  {Zwitter}}]{Gaia2018}
{Gaia Collaboration}, {Brown}, A.~G.~A., {Vallenari}, A., {et~al.} 2018, \aap,
  616, A1

\bibitem[{{Garcia} \& {Gonzalez}(2020)}]{Garcia2020}
{Garcia}, A. J.~L. \& {Gonzalez}, J.-F. 2020, \mnras, 493, 1788

\bibitem[{{Graci{\'a}-Carpio} {et~al.}(2015){Graci{\'a}-Carpio}, {Wetzstein},
  \& {Roussel}}]{2015arXiv151203252G}
{Graci{\'a}-Carpio}, J., {Wetzstein}, M., \& {Roussel}, H. 2015, ArXiv e-prints

\bibitem[{{Gr{\"a}fe} {et~al.}(2013){Gr{\"a}fe}, {Wolf}, {Guilloteau},
  {Dutrey}, {Stapelfeldt}, {Pontoppidan}, \& {Sauter}}]{Grafe2013}
{Gr{\"a}fe}, C., {Wolf}, S., {Guilloteau}, S., {et~al.} 2013, \aap, 553, A69

\bibitem[{{Grant} {et~al.}(2021){Grant}, {Espaillat}, {Wendeborn}, {Tobin},
  {Mac{\'\i}as}, {Rilinger}, {Ribas}, {Megeath}, {Fischer}, {Calvet}, \& {Hee
  Kim}}]{Grant2021}
{Grant}, S.~L., {Espaillat}, C.~C., {Wendeborn}, J., {et~al.} 2021, \apj, 913,
  123

\bibitem[{{Griffin} {et~al.}(2010){Griffin}, {Abergel}, {Abreu}, {Ade},
  {Andr{\'e}}, {Augueres}, {Babbedge}, {Bae}, {Baillie}, {Baluteau}, {Barlow},
  {Bendo}, {Benielli}, {Bock}, {Bonhomme}, {Brisbin}, {Brockley-Blatt},
  {Caldwell}, {Cara}, {Castro-Rodriguez}, {Cerulli}, {Chanial}, {Chen},
  {Clark}, {Clements}, {Clerc}, {Coker}, {Communal}, {Conversi}, {Cox},
  {Crumb}, {Cunningham}, {Daly}, {Davis}, {de Antoni}, {Delderfield}, {Devin},
  {di Giorgio}, {Didschuns}, {Dohlen}, {Donati}, {Dowell}, {Dowell}, {Duband},
  {Dumaye}, {Emery}, {Ferlet}, {Ferrand}, {Fontignie}, {Fox}, {Franceschini},
  {Frerking}, {Fulton}, {Garcia}, {Gastaud}, {Gear}, {Glenn}, {Goizel},
  {Griffin}, {Grundy}, {Guest}, {Guillemet}, {Hargrave}, {Harwit}, {Hastings},
  {Hatziminaoglou}, {Herman}, {Hinde}, {Hristov}, {Huang}, {Imhof}, {Isaak},
  {Israelsson}, {Ivison}, {Jennings}, {Kiernan}, {King}, {Lange}, {Latter},
  {Laurent}, {Laurent}, {Leeks}, {Lellouch}, {Levenson}, {Li}, {Li},
  {Lilienthal}, {Lim}, {Liu}, {Lu}, {Madden}, {Mainetti}, {Marliani}, {McKay},
  {Mercier}, {Molinari}, {Morris}, {Moseley}, {Mulder}, {Mur}, {Naylor},
  {Nguyen}, {O'Halloran}, {Oliver}, {Olofsson}, {Olofsson}, {Orfei}, {Page},
  {Pain}, {Panuzzo}, {Papageorgiou}, {Parks}, {Parr-Burman}, {Pearce},
  {Pearson}, {P{\'e}rez-Fournon}, {Pinsard}, {Pisano}, {Podosek}, {Pohlen},
  {Polehampton}, {Pouliquen}, {Rigopoulou}, {Rizzo}, {Roseboom}, {Roussel},
  {Rowan-Robinson}, {Rownd}, {Saraceno}, {Sauvage}, {Savage}, {Savini},
  {Sawyer}, {Scharmberg}, {Schmitt}, {Schneider}, {Schulz}, {Schwartz},
  {Shafer}, {Shupe}, {Sibthorpe}, {Sidher}, {Smith}, {Smith}, {Smith},
  {Spencer}, {Stobie}, {Sudiwala}, {Sukhatme}, {Surace}, {Stevens}, {Swinyard},
  {Trichas}, {Tourette}, {Triou}, {Tseng}, {Tucker}, {Turner}, {Vaccari},
  {Valtchanov}, {Vigroux}, {Virique}, {Voellmer}, {Walker}, {Ward}, {Waskett},
  {Weilert}, {Wesson}, {White}, {Whitehouse}, {Wilson}, {Winter}, {Woodcraft},
  {Wright}, {Xu}, {Zavagno}, {Zemcov}, {Zhang}, \&
  {Zonca}}]{2010A&A...518L...3G}
{Griffin}, M.~J., {Abergel}, A., {Abreu}, A., {et~al.} 2010, \aap, 518, L3

\bibitem[{{Guidi} {et~al.}(2022){Guidi}, {Isella}, {Testi}, {Chandler}, {Liu},
  {Schmid}, {Rosotti}, {Meng}, {Jennings}, {Williams}, {Carpenter}, {de
  Gregorio-Monsalvo}, {Li}, {Liu}, {Ortolani}, {Quanz}, {Ricci}, \&
  {Tazzari}}]{Guidi2022}
{Guidi}, G., {Isella}, A., {Testi}, L., {et~al.} 2022, arXiv e-prints,
  arXiv:2207.01496

\bibitem[{{Hendler} {et~al.}(2017){Hendler}, {Mulders}, {Pascucci},
  {Greenwood}, {Kamp}, {Henning}, {M{\'e}nard}, {Dent}, \&
  {Evans}}]{Hendler2017}
{Hendler}, N.~P., {Mulders}, G.~D., {Pascucci}, I., {et~al.} 2017, \apj, 841,
  116

\bibitem[{{Herczeg} \& {Hillenbrand}(2014)}]{Herczeg2014}
{Herczeg}, G.~J. \& {Hillenbrand}, L.~A. 2014, \apj, 786, 97

\bibitem[{{Huang} {et~al.}(2018){Huang}, {Andrews}, {Dullemond}, {Isella},
  {P{\'e}rez}, {Guzm{\'a}n}, {{\"O}berg}, {Zhu}, {Zhang}, {Bai}, {Benisty},
  {Birnstiel}, {Carpenter}, {Hughes}, {Ricci}, {Weaver}, \&
  {Wilner}}]{Huang2018}
{Huang}, J., {Andrews}, S.~M., {Dullemond}, C.~P., {et~al.} 2018, \apjl, 869,
  L42

\bibitem[{{Juh{\'a}sz} {et~al.}(2010){Juh{\'a}sz}, {Bouwman}, {Henning},
  {Acke}, {van den Ancker}, {Meeus}, {Dominik}, {Min}, {Tielens}, \&
  {Waters}}]{Juhasz2010}
{Juh{\'a}sz}, A., {Bouwman}, J., {Henning}, T., {et~al.} 2010, \apj, 721, 431

\bibitem[{{Kley} \& {Nelson}(2012)}]{Kley2012}
{Kley}, W. \& {Nelson}, R.~P. 2012, \araa, 50, 211

\bibitem[{{Kov{\'a}cs}(2008)}]{kovacs2008}
{Kov{\'a}cs}, A. 2008, in \procspie, Vol. 7020, Millimeter and Submillimeter
  Detectors and Instrumentation for Astronomy IV, 70201S

\bibitem[{{Kurucz}(1994)}]{Kurucz1994}
{Kurucz}, R. 1994, Solar abundance model atmospheres for 0,1,2,4,8 km/s.~Kurucz
  CD-ROM No.~19.~ Cambridge, Mass.: Smithsonian Astrophysical Observatory,
  1994., 19

\bibitem[{{Law} {et~al.}(2021){Law}, {Teague}, {Loomis}, {Bae}, {{\"O}berg},
  {Czekala}, {Andrews}, {Aikawa}, {Alarc{\'o}n}, {Bergin}, {Bergner}, {Booth},
  {Bosman}, {Calahan}, {Cataldi}, {Cleeves}, {Furuya}, {Guzm{\'a}n}, {Huang},
  {Ilee}, {Le Gal}, {Liu}, {Long}, {M{\'e}nard}, {Nomura}, {P{\'e}rez}, {Qi},
  {Schwarz}, {Soto}, {Tsukagoshi}, {Yamato}, {van't Hoff}, {Walsh}, {Wilner},
  \& {Zhang}}]{Law2021}
{Law}, C.~J., {Teague}, R., {Loomis}, R.~A., {et~al.} 2021, \apjs, 257, 4

\bibitem[{{Liu}(2019)}]{Liuh2019}
{Liu}, H.~B. 2019, \apjl, 877, L22

\bibitem[{{Liu} {et~al.}(2022){Liu}, {Bertrang}, {Flock}, {Rosotti}, {van
  Dishoeck}, {Boehler}, {Facchini}, {Cui}, {Wolf}, \& {Fang}}]{Liuy2022}
{Liu}, Y., {Bertrang}, G. H.~M., {Flock}, M., {et~al.} 2022, arXiv e-prints,
  arXiv:2208.09230

\bibitem[{{Liu} {et~al.}(2019){Liu}, {Dipierro}, {Ragusa}, {Lodato}, {Herczeg},
  {Long}, {Harsono}, {Boehler}, {Menard}, {Johnstone}, {Pascucci}, {Pinilla},
  {Salyk}, {van der Plas}, {Cabrit}, {Fischer}, {Hendler}, {Manara}, {Nisini},
  {Rigliaco}, {Avenhaus}, {Banzatti}, \& {Gully-Santiago}}]{Liuy2019}
{Liu}, Y., {Dipierro}, G., {Ragusa}, E., {et~al.} 2019, \aap, 622, A75

\bibitem[{{Liu} {et~al.}(2017){Liu}, {Henning}, {Carrasco-Gonz{\'a}lez},
  {Chandler}, {Linz}, {Birnstiel}, {van Boekel}, {P{\'e}rez}, {Flock}, {Testi},
  {Rodr{\'\i}guez}, \& {Galv{\'a}n-Madrid}}]{Liuy2017}
{Liu}, Y., {Henning}, T., {Carrasco-Gonz{\'a}lez}, C., {et~al.} 2017, \aap,
  607, A74

\bibitem[{{Long} {et~al.}(2022){Long}, {Andrews}, {Rosotti}, {Harsono},
  {Pinilla}, {Wilner}, {{\"O}berg}, {Teague}, {Trapman}, \&
  {Tabone}}]{Long2022}
{Long}, F., {Andrews}, S.~M., {Rosotti}, G., {et~al.} 2022, \apj, 931, 6

\bibitem[{{Long} {et~al.}(2019){Long}, {Herczeg}, {Harsono}, {Pinilla},
  {Tazzari}, {Manara}, {Pascucci}, {Cabrit}, {Nisini}, {Johnstone}, {Edwards},
  {Salyk}, {Menard}, {Lodato}, {Boehler}, {Mace}, {Liu}, {Mulders}, {Hendler},
  {Ragusa}, {Fischer}, {Banzatti}, {Rigliaco}, {van de Plas}, {Dipierro},
  {Gully-Santiago}, \& {Lopez-Valdivia}}]{Long2019}
{Long}, F., {Herczeg}, G.~J., {Harsono}, D., {et~al.} 2019, \apj, 882, 49

\bibitem[{{Long} {et~al.}(2018){Long}, {Pinilla}, {Herczeg}, {Harsono},
  {Dipierro}, {Pascucci}, {Hendler}, {Tazzari}, {Ragusa}, {Salyk}, {Edwards},
  {Lodato}, {van de Plas}, {Johnstone}, {Liu}, {Boehler}, {Cabrit}, {Manara},
  {Menard}, {Mulders}, {Nisini}, {Fischer}, {Rigliaco}, {Banzatti}, {Avenhaus},
  \& {Gully-Santiago}}]{Long2018}
{Long}, F., {Pinilla}, P., {Herczeg}, G.~J., {et~al.} 2018, \apj, 869, 17

\bibitem[{{Machida} {et~al.}(2010){Machida}, {Kokubo}, {Inutsuka}, \&
  {Matsumoto}}]{Machida2010}
{Machida}, M.~N., {Kokubo}, E., {Inutsuka}, S.-I., \& {Matsumoto}, T. 2010,
  \mnras, 405, 1227

\bibitem[{{Mac{\'\i}as} {et~al.}(2018){Mac{\'\i}as}, {Espaillat}, {Ribas},
  {Schwarz}, {Anglada}, {Osorio}, {Carrasco-Gonz{\'a}lez}, {G{\'o}mez}, \&
  {Robinson}}]{Macias2018}
{Mac{\'\i}as}, E., {Espaillat}, C.~C., {Ribas}, {\'A}., {et~al.} 2018, \apj,
  865, 37

\bibitem[{{Mac{\'\i}as} {et~al.}(2021){Mac{\'\i}as}, {Guerra-Alvarado},
  {Carrasco-Gonz{\'a}lez}, {Ribas}, {Espaillat}, {Huang}, \&
  {Andrews}}]{Macias2021}
{Mac{\'\i}as}, E., {Guerra-Alvarado}, O., {Carrasco-Gonz{\'a}lez}, C., {et~al.}
  2021, \aap, 648, A33

\bibitem[{{Manara} {et~al.}(2018){Manara}, {Morbidelli}, \&
  {Guillot}}]{Manara2018}
{Manara}, C.~F., {Morbidelli}, A., \& {Guillot}, T. 2018, \aap, 618, L3

\bibitem[{{Mauc{\'o}} {et~al.}(2018){Mauc{\'o}}, {Brice{\~n}o}, {Calvet},
  {Hern{\'a}ndez}, {Ballesteros-Paredes}, {Gonz{\'a}lez}, {Espaillat}, {Li},
  {Telesco}, {Downes}, {Mac{\'\i}as}, {Qi}, {Michel}, {D'Alessio}, \&
  {Ali}}]{Mauco2018}
{Mauc{\'o}}, K., {Brice{\~n}o}, C., {Calvet}, N., {et~al.} 2018, \apj, 859, 1

\bibitem[{{McClure} {et~al.}(2016){McClure}, {Bergin}, {Cleeves}, {van
  Dishoeck}, {Blake}, {Evans}, {Green}, {Henning}, {{\"O}berg}, {Pontoppidan},
  \& {Salyk}}]{McClure2016}
{McClure}, M.~K., {Bergin}, E.~A., {Cleeves}, L.~I., {et~al.} 2016, \apj, 831,
  167

\bibitem[{{Menu} {et~al.}(2014){Menu}, {van Boekel}, {Henning}, {Chandler},
  {Linz}, {Benisty}, {Lacour}, {Min}, {Waelkens}, {Andrews}, {Calvet},
  {Carpenter}, {Corder}, {Deller}, {Greaves}, {Harris}, {Isella}, {Kwon},
  {Lazio}, {Le Bouquin}, {M{\'e}nard}, {Mundy}, {P{\'e}rez}, {Ricci},
  {Sargent}, {Storm}, {Testi}, \& {Wilner}}]{Menu2014}
{Menu}, J., {van Boekel}, R., {Henning}, T., {et~al.} 2014, \aap, 564, A93

\bibitem[{{Min} {et~al.}(2005){Min}, {Hovenier}, \& {de Koter}}]{Min2005}
{Min}, M., {Hovenier}, J.~W., \& {de Koter}, A. 2005, \aap, 432, 909

\bibitem[{{Miotello} {et~al.}(2014){Miotello}, {Bruderer}, \& {van
  Dishoeck}}]{Miotello2014}
{Miotello}, A., {Bruderer}, S., \& {van Dishoeck}, E.~F. 2014, \aap, 572, A96

\bibitem[{{Miotello} {et~al.}(2022){Miotello}, {Kamp}, {Birnstiel}, {Cleeves},
  \& {Kataoka}}]{Miotello2022}
{Miotello}, A., {Kamp}, I., {Birnstiel}, T., {Cleeves}, L.~I., \& {Kataoka}, A.
  2022, arXiv e-prints, arXiv:2203.09818

\bibitem[{{Miotello} {et~al.}(2016){Miotello}, {van Dishoeck}, {Kama}, \&
  {Bruderer}}]{Miotello2016}
{Miotello}, A., {van Dishoeck}, E.~F., {Kama}, M., \& {Bruderer}, S. 2016,
  \aap, 594, A85

\bibitem[{{Mordasini} {et~al.}(2012){Mordasini}, {Alibert}, {Klahr}, \&
  {Henning}}]{Mordasini2012}
{Mordasini}, C., {Alibert}, Y., {Klahr}, H., \& {Henning}, T. 2012, \aap, 547,
  A111

\bibitem[{{Mulders} {et~al.}(2021){Mulders}, {Pascucci}, {Ciesla}, \&
  {Fernandes}}]{Mulders2021}
{Mulders}, G.~D., {Pascucci}, I., {Ciesla}, F.~J., \& {Fernandes}, R.~B. 2021,
  \apj, 920, 66

\bibitem[{{Najita} \& {Kenyon}(2014)}]{Najita2014}
{Najita}, J.~R. \& {Kenyon}, S.~J. 2014, \mnras, 445, 3315

\bibitem[{{Ohashi} \& {Kataoka}(2019)}]{Ohashi2019}
{Ohashi}, S. \& {Kataoka}, A. 2019, \apj, 886, 103

\bibitem[{{Ormel} {et~al.}(2007){Ormel}, {Spaans}, \& {Tielens}}]{Ormel2007}
{Ormel}, C.~W., {Spaans}, M., \& {Tielens}, A.~G.~G.~M. 2007, \aap, 461, 215

\bibitem[{{Pascucci} {et~al.}(2022){Pascucci}, {Cabrit}, {Edwards}, {Gorti},
  {Gressel}, \& {Suzuki}}]{Pascucci2022}
{Pascucci}, I., {Cabrit}, S., {Edwards}, S., {et~al.} 2022, arXiv e-prints,
  arXiv:2203.10068

\bibitem[{{Pascucci} {et~al.}(2016){Pascucci}, {Testi}, {Herczeg}, {Long},
  {Manara}, {Hendler}, {Mulders}, {Krijt}, {Ciesla}, {Henning}, {Mohanty},
  {Drabek-Maunder}, {Apai}, {Sz{\H{u}}cs}, {Sacco}, \&
  {Olofsson}}]{Pascucci2016}
{Pascucci}, I., {Testi}, L., {Herczeg}, G.~J., {et~al.} 2016, \apj, 831, 125

\bibitem[{{Pearson} {et~al.}(2014){Pearson}, {Lim}, {North}, {Bendo},
  {Conversi}, {Dowell}, {Griffin}, {Jin}, {Laporte}, {Papageorgiou}, {Schulz},
  {Shupe}, {Smith}, \& {Xu}}]{2014ExA....37..175P}
{Pearson}, C., {Lim}, T., {North}, C., {et~al.} 2014, Experimental Astronomy,
  37, 175

\bibitem[{{P{\'e}rez} {et~al.}(2018){P{\'e}rez}, {Benisty}, {Andrews},
  {Isella}, {Dullemond}, {Huang}, {Kurtovic}, {Guzm{\'a}n}, {Zhu}, {Birnstiel},
  {Zhang}, {Carpenter}, {Wilner}, {Ricci}, {Bai}, {Weaver}, \&
  {{\"O}berg}}]{Perez2018}
{P{\'e}rez}, L.~M., {Benisty}, M., {Andrews}, S.~M., {et~al.} 2018, \apjl, 869,
  L50

\bibitem[{{Picogna} {et~al.}(2019){Picogna}, {Ercolano}, {Owen}, \&
  {Weber}}]{Picogna2019}
{Picogna}, G., {Ercolano}, B., {Owen}, J.~E., \& {Weber}, M.~L. 2019, \mnras,
  487, 691

\bibitem[{{Pinilla} {et~al.}(2017){Pinilla}, {Quiroga-Nu{\~n}ez}, {Benisty},
  {Natta}, {Ricci}, {Henning}, {van der Plas}, {Birnstiel}, {Testi}, \&
  {Ward-Duong}}]{Pinilla2017}
{Pinilla}, P., {Quiroga-Nu{\~n}ez}, L.~H., {Benisty}, M., {et~al.} 2017, \apj,
  846, 70

\bibitem[{{Pinte} {et~al.}(2016){Pinte}, {Dent}, {M{\'e}nard}, {Hales}, {Hill},
  {Cortes}, \& {de Gregorio-Monsalvo}}]{Pinte2016}
{Pinte}, C., {Dent}, W.~R.~F., {M{\'e}nard}, F., {et~al.} 2016, \apj, 816, 25

\bibitem[{{Pinte} {et~al.}(2008){Pinte}, {Padgett}, {M{\'e}nard},
  {Stapelfeldt}, {Schneider}, {Olofsson}, {Pani{\'c}}, {Augereau},
  {Duch{\^e}ne}, {Krist}, {Pontoppidan}, {Perrin}, {Grady}, {Kessler-Silacci},
  {van Dishoeck}, {Lommen}, {Silverstone}, {Hines}, {Wolf}, {Blake}, {Henning},
  \& {Stecklum}}]{Pinte2008}
{Pinte}, C., {Padgett}, D.~L., {M{\'e}nard}, F., {et~al.} 2008, \aap, 489, 633

\bibitem[{{Powell} {et~al.}(2022){Powell}, {Gao}, {Murray-Clay}, \&
  {Zhang}}]{Powell2022}
{Powell}, D., {Gao}, P., {Murray-Clay}, R., \& {Zhang}, X. 2022, Nature
  Astronomy

\bibitem[{{Ragan} {et~al.}(2012){Ragan}, {Henning}, {Krause}, {Pitann},
  {Beuther}, {Linz}, {Tackenberg}, {Balog}, {Hennemann}, {Launhardt}, {Lippok},
  {Nielbock}, {Schmiedeke}, {Schuller}, {Steinacker}, {Stutz}, \&
  {Vasyunina}}]{2012A&A...547A..49R}
{Ragan}, S., {Henning}, T., {Krause}, O., {et~al.} 2012, \aap, 547, A49

\bibitem[{{Rebollido} {et~al.}(2015){Rebollido}, {Mer{\'{\i}}n}, {Ribas},
  {Bustamante}, {Bouy}, {Riviere-Marichalar}, {Prusti}, {Pilbratt},
  {Andr{\'e}}, \& {{\'A}brah{\'a}m}}]{2015A&A...581A..30R}
{Rebollido}, I., {Mer{\'{\i}}n}, B., {Ribas}, {\'A}., {et~al.} 2015, \aap, 581,
  A30

\bibitem[{{Ribas} {et~al.}(2020){Ribas}, {Espaillat}, {Mac{\'\i}as}, \&
  {Sarro}}]{Ribas2020}
{Ribas}, {\'A}., {Espaillat}, C.~C., {Mac{\'\i}as}, E., \& {Sarro}, L.~M. 2020,
  \aap, 642, A171

\bibitem[{{Ricci} {et~al.}(2013){Ricci}, {Isella}, {Carpenter}, \&
  {Testi}}]{Ricci2013}
{Ricci}, L., {Isella}, A., {Carpenter}, J.~M., \& {Testi}, L. 2013, \apjl, 764,
  L27

\bibitem[{{Ricci} {et~al.}(2010){Ricci}, {Testi}, {Natta}, {Neri}, {Cabrit}, \&
  {Herczeg}}]{Ricci2010}
{Ricci}, L., {Testi}, L., {Natta}, A., {et~al.} 2010, \aap, 512, A15

\bibitem[{{Ricci} {et~al.}(2014){Ricci}, {Testi}, {Natta}, {Scholz}, {de
  Gregorio-Monsalvo}, \& {Isella}}]{Ricci2014}
{Ricci}, L., {Testi}, L., {Natta}, A., {et~al.} 2014, \apj, 791, 20

\bibitem[{{Rich} {et~al.}(2021){Rich}, {Teague}, {Monnier}, {Davies}, {Bosman},
  {Harries}, {Calvet}, {Adams}, {Wilner}, \& {Zhu}}]{Rich2021}
{Rich}, E.~A., {Teague}, R., {Monnier}, J.~D., {et~al.} 2021, \apj, 913, 138

\bibitem[{{Rilinger} \& {Espaillat}(2021)}]{Rilinger2021}
{Rilinger}, A.~M. \& {Espaillat}, C.~C. 2021, \apj, 921, 182

\bibitem[{{Roussel}(2013)}]{2013PASP..125.1126R}
{Roussel}, H. 2013, \pasp, 125, 1126

\bibitem[{{Schr{\"a}pler} \& {Henning}(2004)}]{Schrapler2004}
{Schr{\"a}pler}, R. \& {Henning}, T. 2004, \apj, 614, 960

\bibitem[{{Segura-Cox} {et~al.}(2020){Segura-Cox}, {Schmiedeke}, {Pineda},
  {Stephens}, {Fern{\'a}ndez-L{\'o}pez}, {Looney}, {Caselli}, {Li}, {Mundy},
  {Kwon}, \& {Harris}}]{SeguraCox2020}
{Segura-Cox}, D.~M., {Schmiedeke}, A., {Pineda}, J.~E., {et~al.} 2020, \nat,
  586, 228

\bibitem[{{Sheehan} {et~al.}(2020){Sheehan}, {Tobin}, {Federman}, {Megeath}, \&
  {Looney}}]{Sheehan2020}
{Sheehan}, P.~D., {Tobin}, J.~J., {Federman}, S., {Megeath}, S.~T., \&
  {Looney}, L.~W. 2020, \apj, 902, 141

\bibitem[{{Sierra} \& {Lizano}(2020)}]{Sierra2020}
{Sierra}, A. \& {Lizano}, S. 2020, \apj, 892, 136

\bibitem[{{Sierra} {et~al.}(2021){Sierra}, {P{\'e}rez}, {Zhang}, {Law},
  {Guzm{\'a}n}, {Qi}, {Bosman}, {{\"O}berg}, {Andrews}, {Long}, {Teague},
  {Booth}, {Walsh}, {Wilner}, {M{\'e}nard}, {Cataldi}, {Czekala}, {Bae},
  {Huang}, {Bergner}, {Ilee}, {Benisty}, {Le Gal}, {Loomis}, {Tsukagoshi},
  {Liu}, {Yamato}, \& {Aikawa}}]{Sierra2021}
{Sierra}, A., {P{\'e}rez}, L.~M., {Zhang}, K., {et~al.} 2021, \apjs, 257, 14

\bibitem[{{Teague} {et~al.}(2018){Teague}, {Henning}, {Guilloteau}, {Bergin},
  {Semenov}, {Dutrey}, {Flock}, {Gorti}, \& {Birnstiel}}]{Teague2018}
{Teague}, R., {Henning}, T., {Guilloteau}, S., {et~al.} 2018, \apj, 864, 133

\bibitem[{{Testi} {et~al.}(2014){Testi}, {Birnstiel}, {Ricci}, {Andrews},
  {Blum}, {Carpenter}, {Dominik}, {Isella}, {Natta}, {Williams}, \&
  {Wilner}}]{Testi2014}
{Testi}, L., {Birnstiel}, T., {Ricci}, L., {et~al.} 2014, in Protostars and
  Planets VI, ed. H.~{Beuther}, R.~S. {Klessen}, C.~P. {Dullemond}, \&
  T.~{Henning}, 339

\bibitem[{{Testi} {et~al.}(2016){Testi}, {Natta}, {Scholz}, {Tazzari}, {Ricci},
  \& {de Gregorio Monsalvo}}]{Testi2016}
{Testi}, L., {Natta}, A., {Scholz}, A., {et~al.} 2016, \aap, 593, A111

\bibitem[{{Toci} {et~al.}(2021){Toci}, {Rosotti}, {Lodato}, {Testi}, \&
  {Trapman}}]{Toci2021}
{Toci}, C., {Rosotti}, G., {Lodato}, G., {Testi}, L., \& {Trapman}, L. 2021,
  \mnras, 507, 818

\bibitem[{{Toon} \& {Ackerman}(1981)}]{Toon1981}
{Toon}, O.~B. \& {Ackerman}, T.~P. 1981, \ao, 20, 3657

\bibitem[{{Trapman} {et~al.}(2019){Trapman}, {Facchini}, {Hogerheijde}, {van
  Dishoeck}, \& {Bruderer}}]{Trapman2019}
{Trapman}, L., {Facchini}, S., {Hogerheijde}, M.~R., {van Dishoeck}, E.~F., \&
  {Bruderer}, S. 2019, \aap, 629, A79

\bibitem[{{Tychoniec} {et~al.}(2020){Tychoniec}, {Manara}, {Rosotti}, {van
  Dishoeck}, {Cridland}, {Hsieh}, {Murillo}, {Segura-Cox}, {van Terwisga}, \&
  {Tobin}}]{Tychoniec2020}
{Tychoniec}, {\L}., {Manara}, C.~F., {Rosotti}, G.~P., {et~al.} 2020, \aap,
  640, A19

\bibitem[{{Ueda} {et~al.}(2020){Ueda}, {Kataoka}, \& {Tsukagoshi}}]{Ueda2020}
{Ueda}, T., {Kataoka}, A., \& {Tsukagoshi}, T. 2020, \apj, 893, 125

\bibitem[{{van Boekel} {et~al.}(2017){van Boekel}, {Henning}, {Menu}, {de
  Boer}, {Langlois}, {M{\"u}ller}, {Avenhaus}, {Boccaletti}, {Schmid},
  {Thalmann}, {Benisty}, {Dominik}, {Ginski}, {Girard}, {Gisler}, {Lobo Gomes},
  {Menard}, {Min}, {Pavlov}, {Pohl}, {Quanz}, {Rabou}, {Roelfsema}, {Sauvage},
  {Teague}, {Wildi}, \& {Zurlo}}]{vanBoekel2017}
{van Boekel}, R., {Henning}, T., {Menu}, J., {et~al.} 2017, \apj, 837, 132

\bibitem[{{van der Marel} {et~al.}(2013){van der Marel}, {van Dishoeck},
  {Bruderer}, {Birnstiel}, {Pinilla}, {Dullemond}, {van Kempen}, {Schmalzl},
  {Brown}, {Herczeg}, {Mathews}, \& {Geers}}]{vanderMarel2013}
{van der Marel}, N., {van Dishoeck}, E.~F., {Bruderer}, S., {et~al.} 2013,
  Science, 340, 1199

\bibitem[{{van der Plas} {et~al.}(2016){van der Plas}, {M{\'e}nard},
  {Ward-Duong}, {Bulger}, {Harvey}, {Pinte}, {Patience}, {Hales}, \&
  {Casassus}}]{vanderPlas2016}
{van der Plas}, G., {M{\'e}nard}, F., {Ward-Duong}, K., {et~al.} 2016, \apj,
  819, 102

\bibitem[{{Villenave} {et~al.}(2020){Villenave}, {M{\'e}nard}, {Dent},
  {Duch{\^e}ne}, {Stapelfeldt}, {Benisty}, {Boehler}, {van der Plas}, {Pinte},
  {Telkamp}, {Wolff}, {Flores}, {Lesur}, {Louvet}, {Riols}, {Dougados},
  {Williams}, \& {Padgett}}]{Villenave2020}
{Villenave}, M., {M{\'e}nard}, F., {Dent}, W.~R.~F., {et~al.} 2020, \aap, 642,
  A164

\bibitem[{{Wilking} {et~al.}(2005){Wilking}, {Meyer}, {Robinson}, \&
  {Greene}}]{Wilking2005}
{Wilking}, B.~A., {Meyer}, M.~R., {Robinson}, J.~G., \& {Greene}, T.~P. 2005,
  \aj, 130, 1733

\bibitem[{{Williams} \& {Best}(2014)}]{Williams2014}
{Williams}, J.~P. \& {Best}, W. M.~J. 2014, \apj, 788, 59

\bibitem[{{Williams} \& {Cieza}(2011)}]{Williams2011}
{Williams}, J.~P. \& {Cieza}, L.~A. 2011, \araa, 49, 67

\bibitem[{{Woitke} {et~al.}(2016){Woitke}, {Min}, {Pinte}, {Thi}, {Kamp},
  {Rab}, {Anthonioz}, {Antonellini}, {Baldovin-Saavedra}, \&
  {Carmona}}]{Woitke2016}
{Woitke}, P., {Min}, M., {Pinte}, C., {et~al.} 2016, \aap, 586, A103

\bibitem[{{Wolf} {et~al.}(2008){Wolf}, {Schegerer}, {Beuther}, {Padgett}, \&
  {Stapelfeldt}}]{Wolf2008}
{Wolf}, S., {Schegerer}, A., {Beuther}, H., {Padgett}, D.~L., \& {Stapelfeldt},
  K.~R. 2008, \apjl, 674, L101

\bibitem[{{Zhu} {et~al.}(2019){Zhu}, {Zhang}, {Jiang}, {Kataoka}, {Birnstiel},
  {Dullemond}, {Andrews}, {Huang}, {P{\'e}rez}, {Carpenter}, {Bai}, {Wilner},
  \& {Ricci}}]{Zhu2019}
{Zhu}, Z., {Zhang}, S., {Jiang}, Y.-F., {et~al.} 2019, \apjl, 877, L18

\bibitem[{{Zubko} {et~al.}(1996){Zubko}, {Mennella}, {Colangeli}, \&
  {Bussoletti}}]{Zubko1996}
{Zubko}, V.~G., {Mennella}, V., {Colangeli}, L., \& {Bussoletti}, E. 1996,
  \mnras, 282, 1321

\end{thebibliography}

\begin{appendix}

\section{Underestimation factors for face-on and edge-on disks}
\label{sec:extremecase}

\begin{figure*}[!h]
\centering
\includegraphics[width=0.7\textwidth]{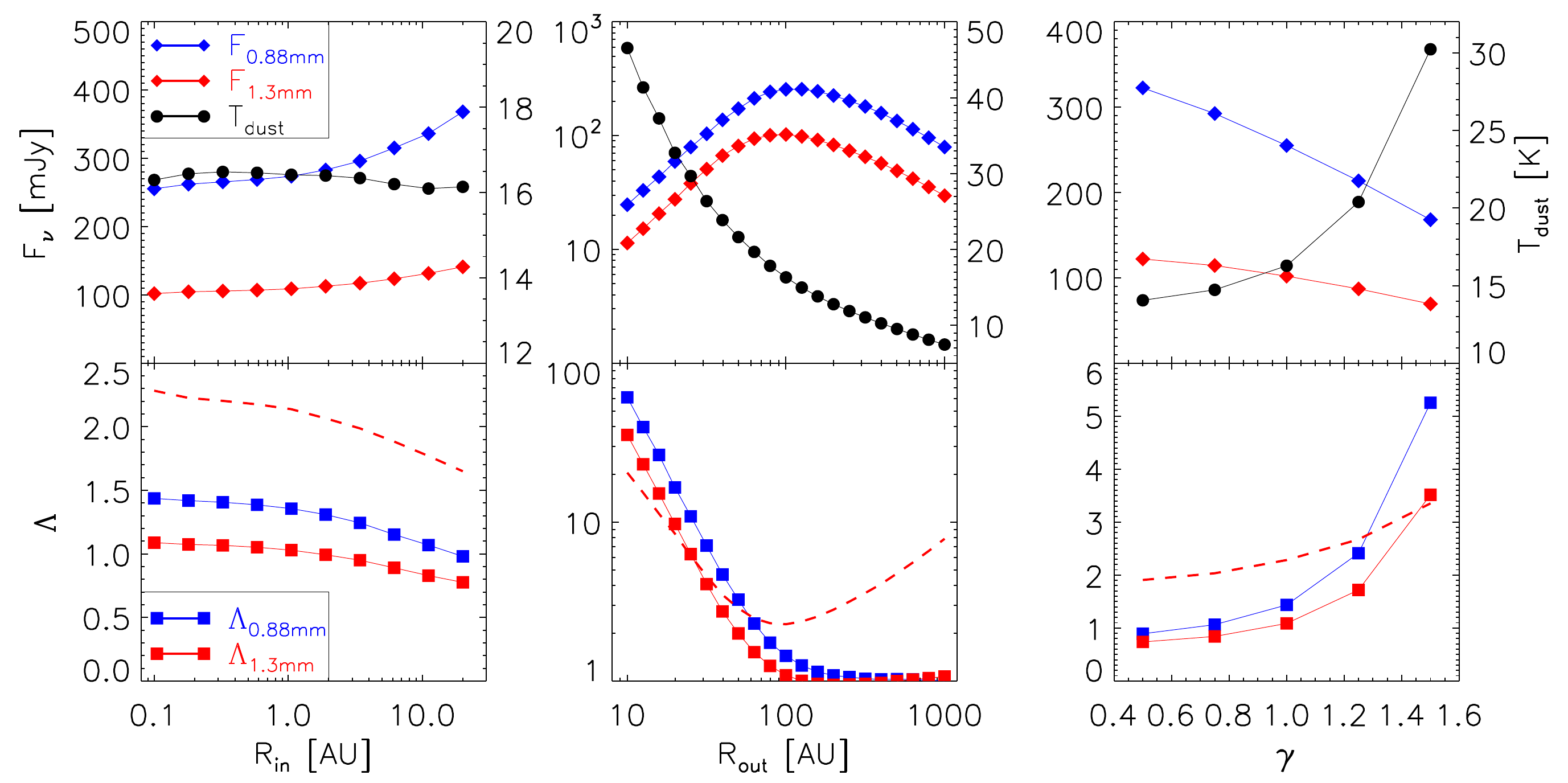}
\includegraphics[width=0.7\textwidth]{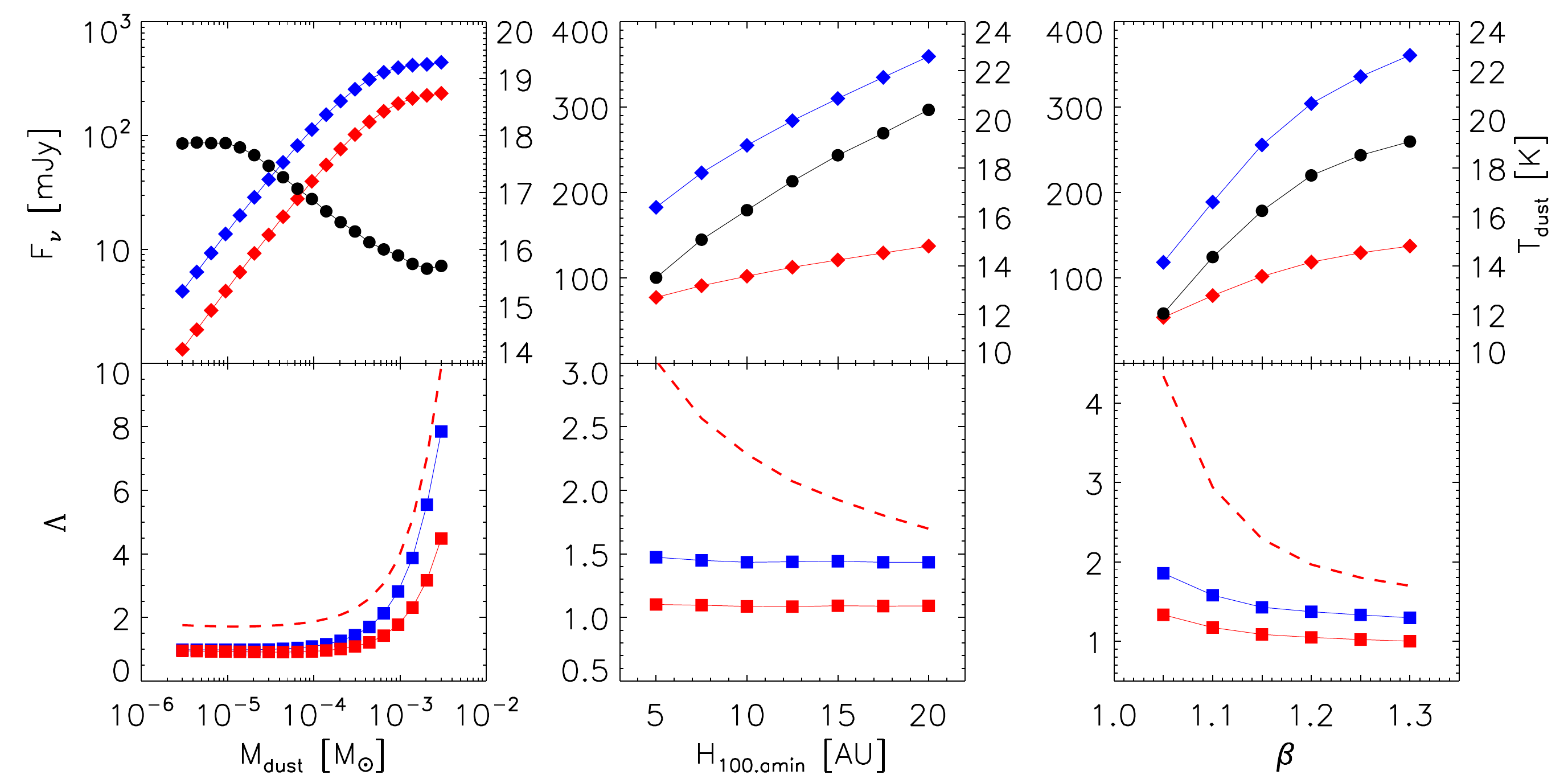}
\includegraphics[width=0.7\textwidth]{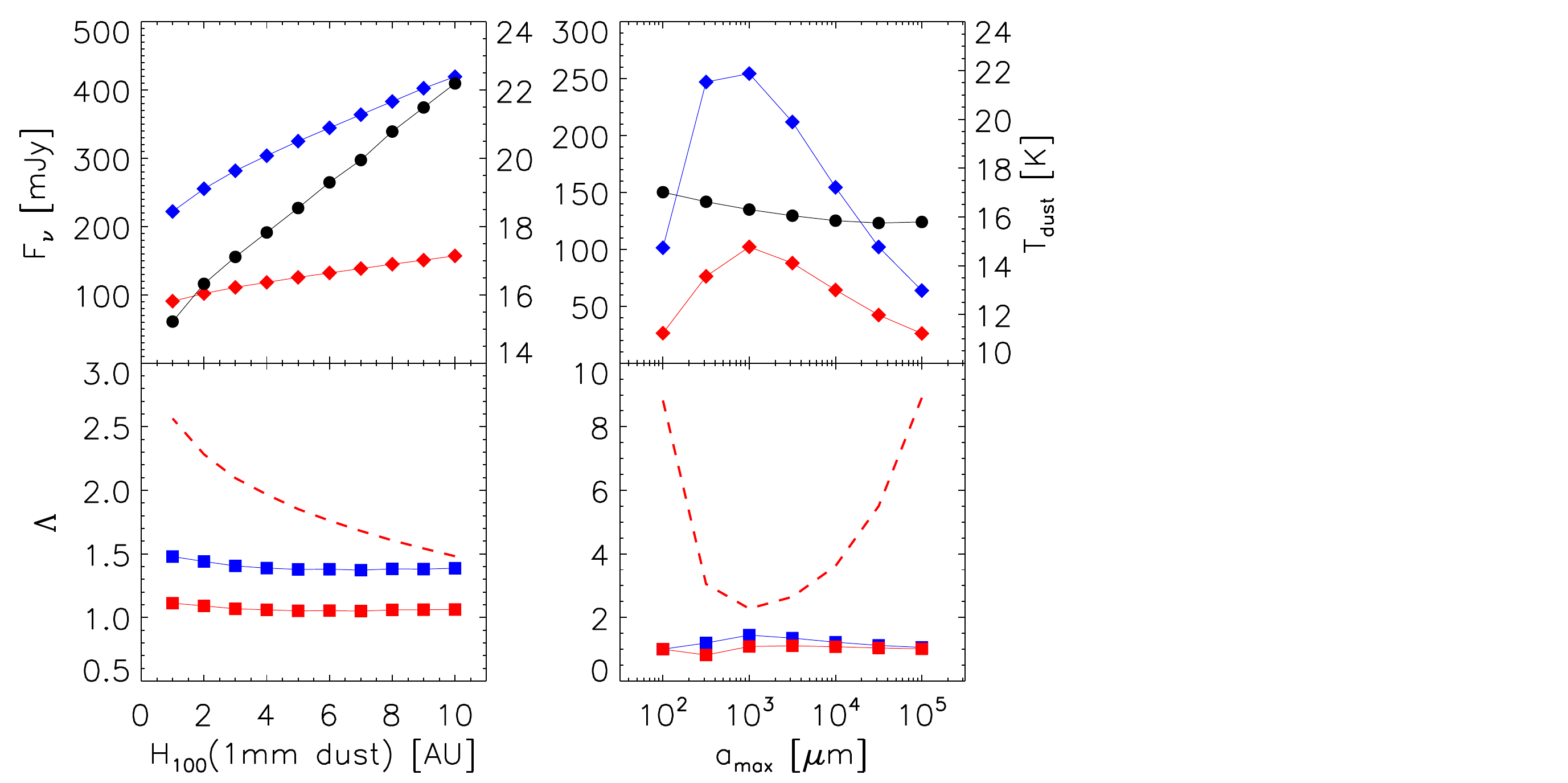}
\caption{Same as in Figure~\ref{fig:paraeff}, but the disk inclination is fixed to $i\,{=}\,0^{\circ}$ for all models.}
\label{fig:lambdafaceon}
\end{figure*}

\begin{figure*}[!h]
\centering
\includegraphics[width=0.7\textwidth]{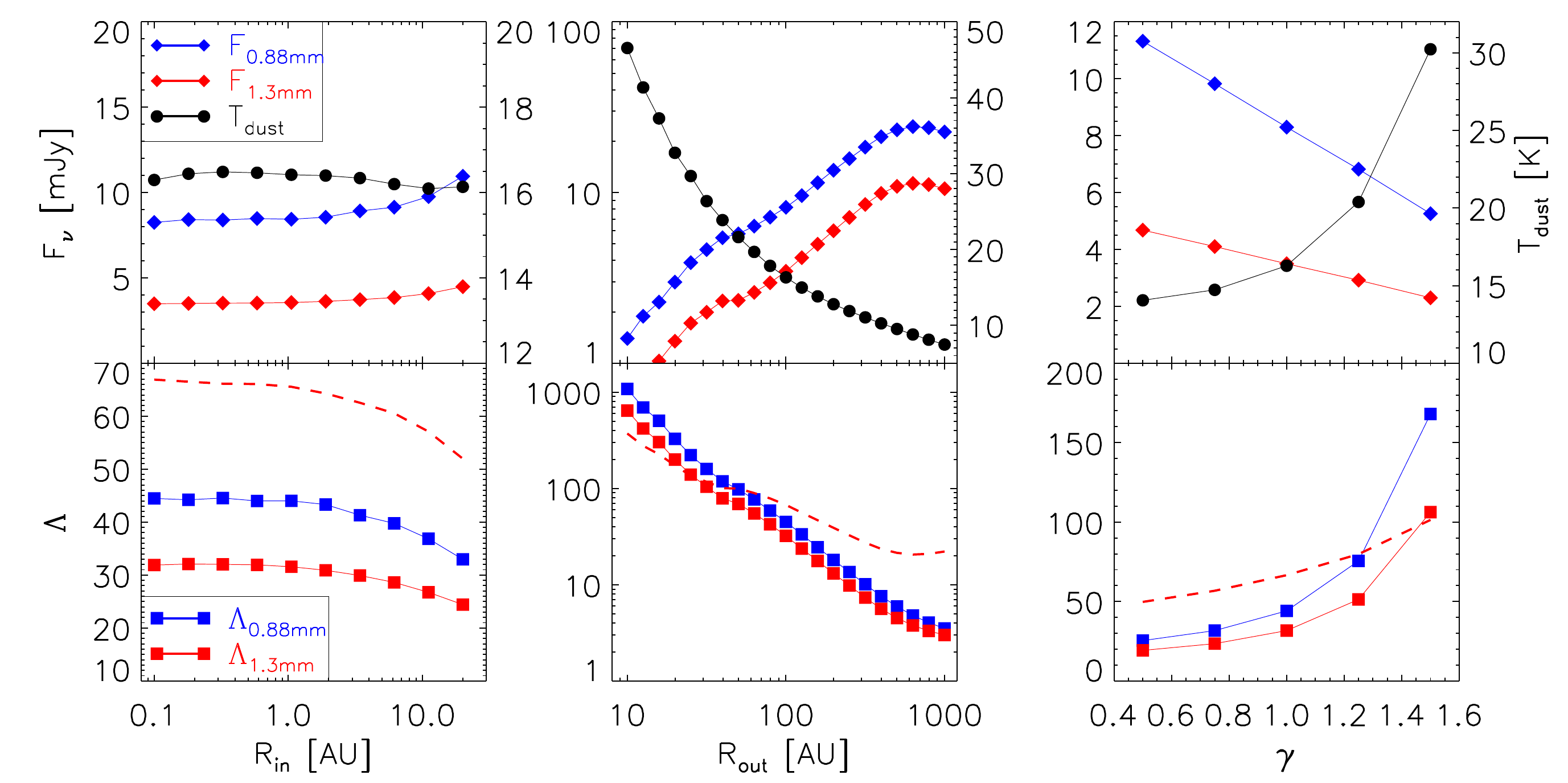}
\includegraphics[width=0.7\textwidth]{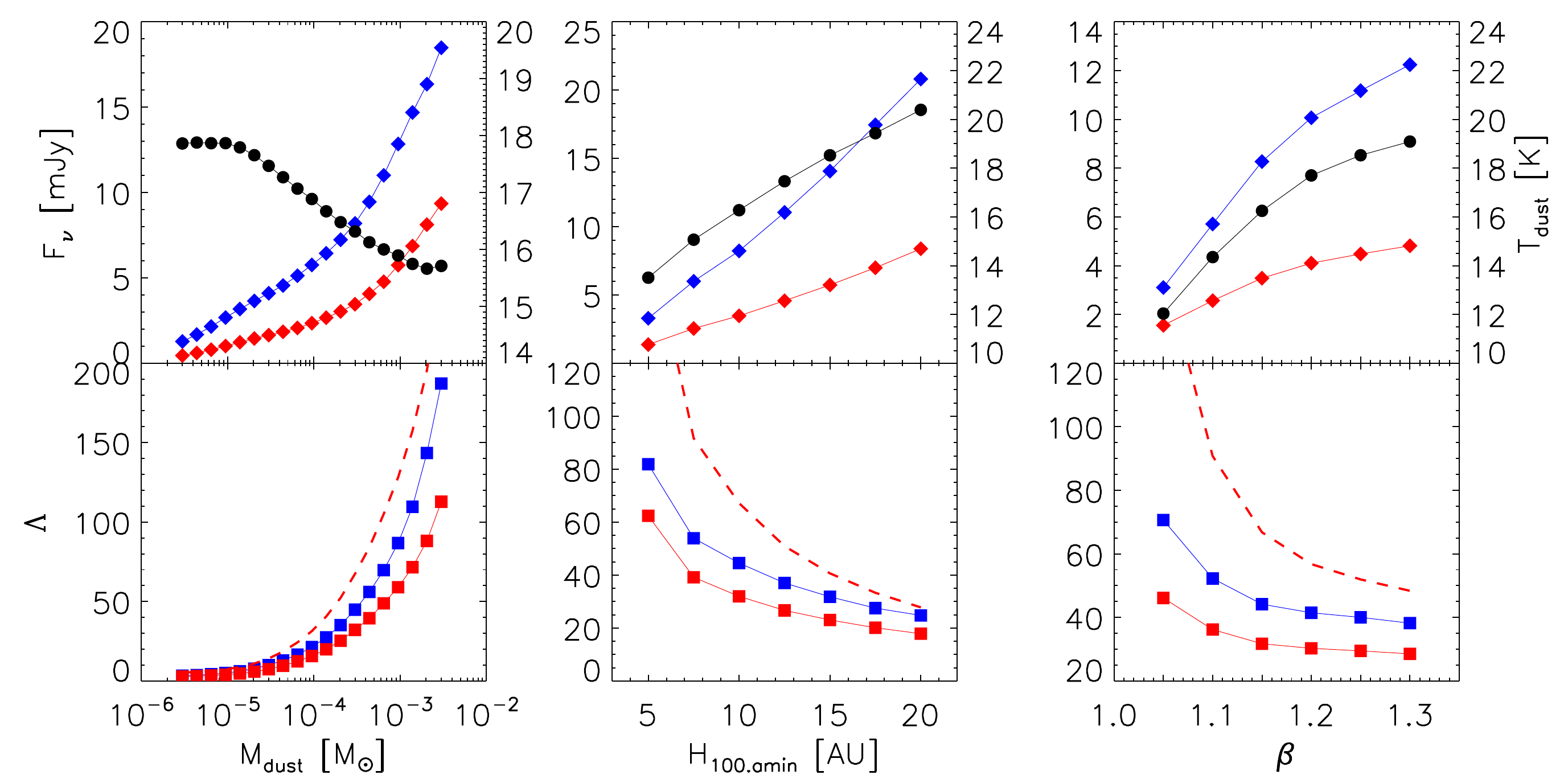}
\includegraphics[width=0.7\textwidth]{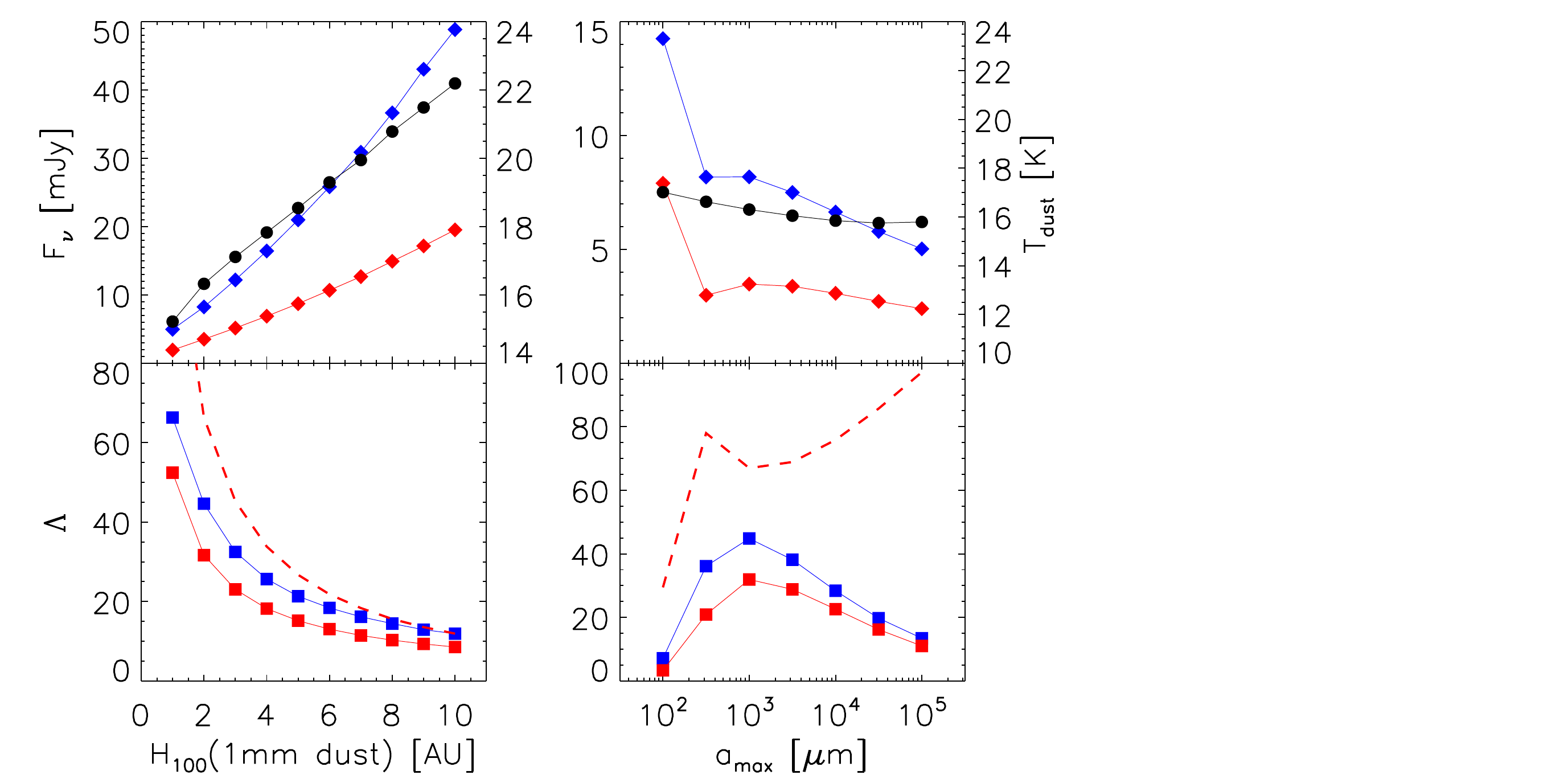}
\caption{Same as in Figure~\ref{fig:paraeff}, but the disk inclination is fixed to $i\,{=}\,90^{\circ}$ for all models.}
\label{fig:lambdaedgeon}
\end{figure*}

In this section, we explore how the underestimation factor changes with different model parameters for face-on ($i\,{=}\,0^{\circ}$) and edge-on 
($i\,{=}\,90^{\circ}$) disks. The results are shown in Figure ~\ref{fig:lambdafaceon} and \ref{fig:lambdaedgeon}. 

The degree of underestimation, and its variation with different mode parameters, are broadly comparable between the face-on and fiducial 
inclined ($i\,{=}\,41.8^{\circ}$) disks. However, the dependencies of $\Lambda$ on most of the explored parameters are found to be 
tighter for edge-on disks. For instance, when the disk is viewed face-on, $\Lambda$ decreases with increasing disk outer radius $R_{\rm out}$, 
and it becomes insensitive to $R_{\rm out}$ when $R_{\rm out}\,{\gtrsim}\,100\,{\rm AU}$. However, for edge-on disks, a strong correlation 
between $\Lambda$ and $R_{\rm out}$ exists for a wide range of $R_{\rm out}$, i.e., $10\,{\leq}\,R_{\rm out}\,{\leq}\,100\,{\rm AU}$.
Moreover, different dust scale heights ($H_{100.a_{\rm min}}$ or $H_{100}({\rm 1\,mm\,\,dust})$) generally result in a similar underestimation for 
disks with low inclinations. But, edge-on disks show clear dependencies between $\Lambda$ and both parameters. 

\section{CSO/SHARC\,II observations}
\label{sec:cso}

Through the Chinese Telescope Plan, DoAr\,33 was observed with the SHARC\,II bolometer array \citep{dowell2003} on the 10.4\,m Caltech Submillimeter 
Observatory (CSO) on Mauna Kea. SHARC\,II contains a 12 $\times$ 32 array of pop-up bolometers with a $>90\%$ filling factor over the field. 
The field-of-view of the SHARCII maps was ${\sim}2^{\prime} \times 0.6^{\prime}$, and the beam size at $350\,\mu{\rm{m}}$ is about $8.5^{\prime\prime}$.

\begin{figure}[!t]
\centering
\includegraphics[width=3.2in]{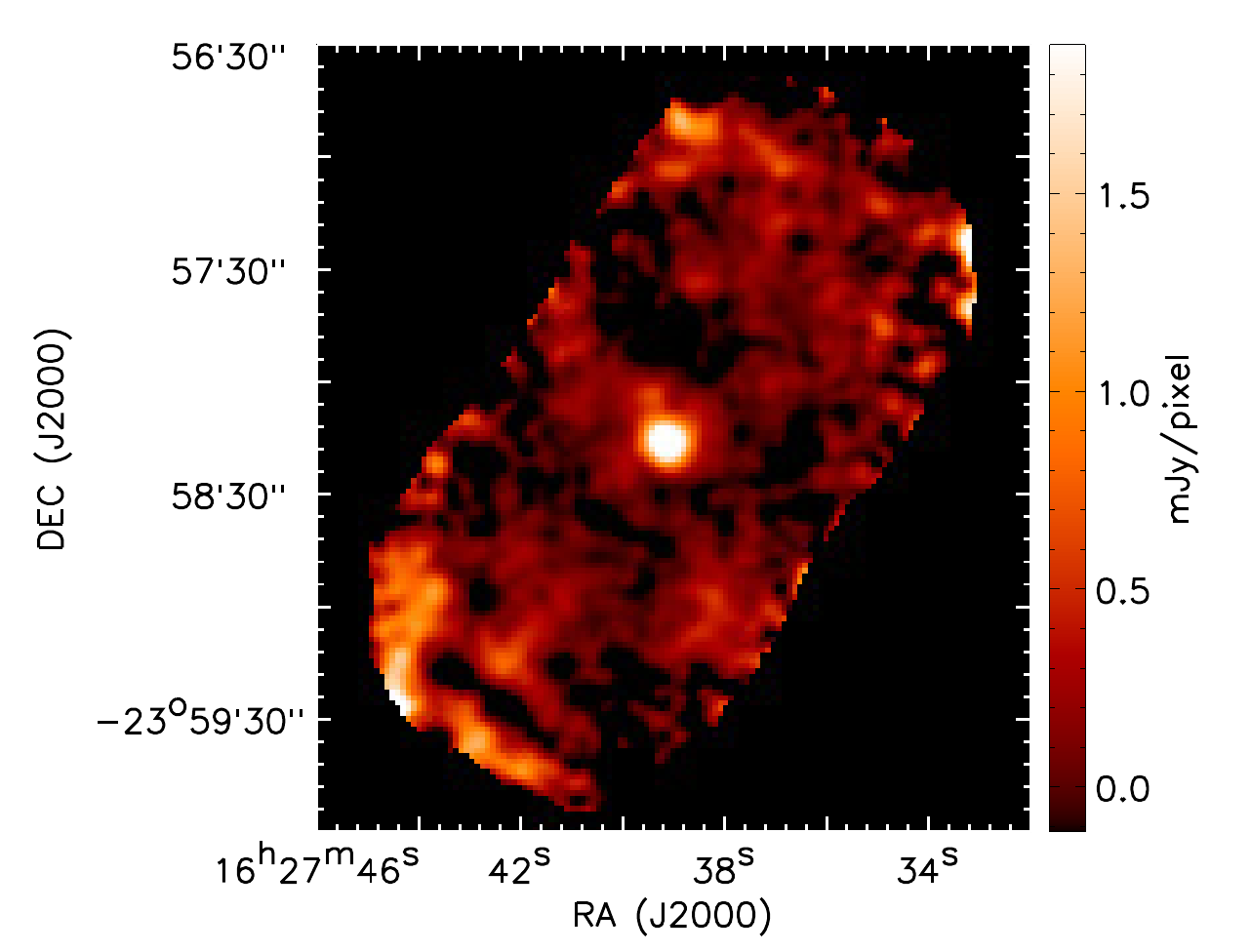}
\caption{The CSO/SHARC\,II image of DoAr 33.}
\label{fig:csoimg}
\end{figure}

The data at $350\,\mu{\rm{m}}$ were acquired on June 05 2015. The conditions during the observations were excellent with $\tau_{225\,\rm{GHz}} = 0.03$. 
The observing sequence consists of a series of target scans bracketed by scans of Pollas and \texttt{cal\_1629m2422}, serving as both absolute 
flux calibration and pointing calibration measurements. We used the sweep mode for the observations, in which the telescope moves in a Lissajous 
pattern that keeps the central regions of the maps fully sampled. Typical individual target scan time was 300\,s, while 120\,s was sufficient for 
the calibrators. We obtained a total of 15 scans on the science target. The data analysis consists of reduction of raw data with the CRUSH pipeline \citep{kovacs2008}, 
flux calibration and aperture photometry. Figure~\ref{fig:csoimg} shows the reduced image. The flux density of DoAr\,33 is measured to be 
0.212\,Jy. The total uncertainty on the flux ranges from ${\sim}10{-}20\%$ mainly due to the uncertainties of the calibrator's flux.

\section{Revisiting the Herschel data}
\label{sec:herschel}

DoAr\,33 was observed with \Herschel within the framework of a large survey of circumstellar disks in the Ophiuchus cloud \citep{andre2010}. 
We re-reduced the data and performed the photometry based on point-spread function (PSF) fitting although literature aperture photometry have 
been reported on \PACS and \SPIRE data \citep{2015A&A...581A..30R}. We work on the same observations as \citet{2015A&A...581A..30R}. The \PACS 70~$\mu$m and 160~$\mu$m data are from program 
{\tt OT1\_pabraham\_3}, while the \SPIRE data at 250, 350, and 500~$\mu$m are from the program {\tt KPGT\_pandre\_1}. 
In addition, we newly report on \PACS 100~$\mu$m photometry, also from the  {\tt KPGT\_pandre\_1} program. We 
downloaded Level2.5 data (scan and cross-scan combined) from the Herschel Science Archive (HSA), based on the 
recent bulk reprocessing which had employed the pipeline versions SPG 14.2.0 for the \PACS data, and SPG v14.2.1
for the \SPIRE data. This iteration of the data reduction includes improved versions of the pointing products which can
refine the sharpness of the PSFs especially at short wavelengths (70 and 100 $\mu$m). The maps for \PACS are 
{\sc JScanam} maps \citep{2015arXiv151203252G}, which are an implementation of the {\sc Scanamorphos} suite of algorithms 
devised by \citet{2013PASP..125.1126R}. These standard maps have been constructed using the pixfrac=0.1 setting 
for the {\sc Drizzle} algorithm \citep[see][]{2002PASP..114..144F}. For the \SPIRE data, we used the maps from the 
standard mapper, with the point-source calibration intact (map units Jy/beam). 

For PSF references, we used the data released by the \PACS and \SPIRE instrument teams, i.e., maps of asteroid 
Vesta\footnote{ftp://ftp.sciops.esa.int/pub/hsc-calibration/PACS/PSF/ PACSPSF\_PICC-ME-TN-033\_v2.2.tar.gz} 
and of planet Neptune\footnote{http://herschel.esac.esa.int/twiki/bin/view/Public/SpirePhotometer BeamProfile2}, 
respectively. For the \PACS PSFs, appropriate files were available for the pixfrac, scan instrument angle, and scan speed 
values of the science data. We just had to regrid the fine PSF data to the somewhat coarser pixel scale of the 
science maps. The \SPIRE PSFs were available in a pixel scale adapted to the science data. These reference PSFs were obtained 
using the standard scan speed for \SPIRE (30$''$/s, sampling 18.2 Hz), while the science data have been taken with the fast scan 
speed of the \SPIRE/\PACS parallel mode (60$''$/s, sampling 10 Hz). Potential differences in the PSF shape would remain a 
second-order effect \citep[see][and especially the Parallel-Mode Observers'
 Manual\footnote{http://herschel.esac.esa.int/Docs/PMODE/html/parallel\_om.html}, Section 2.1]{2010A&A...518L...3G}. 
All reference PSFs were rotated in order to match the PSF orientation in the individual science maps, in dependence 
of the satellite position angles, taken from the science map fits headers. 

\begin{table*}[!t]
\caption[]{Observational details of \Herschel data and derived photometry}
\centering          
\begin{tabular}{l | r r r r r r }     
\hline     
Wavelength [$\mu$m] 	& 70 & 100 & 160 & 250 & 350 & 500 \\ 
\hline  
                  
\multirow{2}{*}{obsids}				& 1342238816 + 	& 1342227148 + 	& 1342238816 + 	& 1342205093 + 	& 1342205093 +		& 1342205093 +\\
						& 1342238817 		& 1342227149 		& 1342238817   		& 1342205094 		& 1342205094 		& 1342205094 \\  
scan speed [$''$/s]		& 20 				& 20    				& 20 				& 60 				& 60 			& 60 \\
parallel mode [y/n]   		& n 				& n    		 		& n 				& y				& y 				& y   \\
map pixel size [$''$]		& 1.6				& 1.6				& 3.2				& 6.0				& 10.0				& 14.0  \\
PSF correlation coeff.		& 0.991	 			& 0.985    			& 0.984				& 0.982				& 0.945 			& 0.892 \\
PSF radius used [$''$]	        & 32				& 32				& 57.6				& $-$		& $-$		& $-$ \\
PSF integrated flux [Jy] 	& 0.407 			& 0.474    			& 0.495 			& $-$ 		& $-$		& $-$ \\
Aperture corrected		& \multirow{2}{*}{0.455}	& \multirow{2}{*}{0.531}	& \multirow{2}{*}{0.543}	&  \multirow{2}{*}{$-$}		& \multirow{2}{*}{$-$}		& \multirow{2}{*}{$-$} \\
\, \, \,  flux density [Jy]	&	\\
PSF peak flux [Jy/beam]	        & $-$		& $-$		& $-$		& 0.422				& 0.269				& 0.130 \\	
Colour-corrected		& \multirow{2}{*}{0.469$\pm$0.033}	& \multirow{2}{*}{0.537$\pm$0.038}	& \multirow{2}{*}{0.539$\pm$0.038}	& \multirow{2}{*}{0.416$\pm$0.023}	& \multirow{2}{*}{0.260$\pm$0.014}	& \multirow{2}{*}{0.121$\pm$0.007}   \\
\, \, \,  flux density [Jy]	&  \\
\hline                  
\end{tabular}
\label{Tab:PSF-photometry}
\end{table*}

The actual PSF photometry was performed employing the {\sc Starfinder} package \citep{2000A&AS..147..335D}, which has proven 
to successfully cope with the circumstances in \PACS maps with extended emission \citep[e.g.][]{2012A&A...547A..49R}. For \SPIRE,
several photometry algorithms are implemented in the \Herschel data reduction environment {\sc HIPE}, but their performance has
been mainly tested on clean stellar or extragalactic fields \citep{2014ExA....37..175P}. For our circumstances -- a point source 
embedded in rivalling extended emission especially at the longest \Herschel wavelengths -- we also employ the {\sc Starfinder}
PSF photometry for the \SPIRE maps. DoAr 33 was detected with high confidence in all six bands (see also the high correlation coefficients 
mentioned in Table~\ref{Tab:PSF-photometry}). During the photometry setup, we explicitly truncated the PSF maps radially. 
For PACS, we work on Jy/pixel maps and are interested in the pixel-integrated flux densities. Therefore, we applied an 
appropriate aperture correction as tabulated in the encircled-energy-fraction files coming with the aforementioned PSF release 
\citep[see also][]{2014ExA....37..129B}. For the \SPIRE maps (in Jy/beam) we are interested in the peak fluxes of the fitted PSFs when the
extended background is eventually removed. {\sc Starfinder} produces such maps as a side product which just contain the
appropriately scaled copies of the PSFs, depending on the brightness of the identified sources in the field. We took these 
maps and fitted a 2-dimensional Gaussian to the source representing DoAr 33 (without extended background emission) 
in order to better recover the true peak flux, partly mitigating the relatively coarse pixel resolution of the SPIRE data. For a
true point source, the value of the peak flux (in Jy/beam) will be identical to the integrated flux of the source.

Finally, a colour correction has to be applied. The \PACS and \SPIRE instrument teams provide correction factors for different
types of SEDs, in particular for the situation of modified black-body SEDs ($F_\nu\,{\sim}\,{B(\rm{T})}{\cdot}\nu^{\beta}$). 
We therefore fitted the 70--500~$\mu$m SED with a simple but robust SED fitting tool \citep{2007ApJ...656L..85B}. The SED of DoAr 33
in the FIR cannot be approximated with one temperature component, but the inclusion of two temperature components (fit values: 12.6 K 
and 32.5 K) gave satisfactory results when adopting $\beta=+2$ (a case explicitly listed in the colour correction tables).  
For all six \Herschel wavelengths, the tool gave the relative flux contributions for both temperature components. We applied the 
colour correction terms (interpolated to the above-mentioned temperatures) to the respective fractions of the measured flux values, 
and added up the two corrected values per wavelength. The finally derived flux densities for DoAr 33 are given in Table~\ref{Tab:PSF-photometry}.
The formal fitting errors from the PSF photometry are quite small. As overall uncertainties we therefore quote the sum of instrument 
calibration uncertainties and a second term arising from uncertainties in the underlying models of the flux calibration standards. These uncertainties are 
(2.0 + 5.0)\% for \PACS and (1.5 + 4.0)\% for \SPIRE, respectively. 

On the whole, we can confirm the \PACS 70~$\mu$m and 160~$\mu$m photometry derived by \citet[][cf.~our Table~\ref{Tab:PSF-photometry}]{2015A&A...581A..30R}. 
However, we derive \SPIRE fluxes 3--4 times higher than reported by their study. To assume that the photometry for the wrong \SPIRE 
source was given in the paper (and in the associated Vizier catalogue) remains speculation. Therefore, we have to conclude that their 
automatised aperture photometry did not handle the subtraction of the extended background well in this case and led to a strong over-subtraction of flux.

\end{appendix}

\end{document}